\documentclass[acmlarge, authordraft, review=false]{acmart}

\usepackage[utf8]{inputenc}
\usepackage[T1]{fontenc}
\usepackage{graphicx}
\usepackage{adjustbox}
\usepackage{tabularx}
\usepackage[english]{babel}
\usepackage[figuresright]{rotating}
\usepackage{makecell}
\usepackage{array}
\usepackage{caption}
\usepackage{multirow}
\usepackage{url}
\usepackage{float}
\usepackage{hyperref}
\usepackage{adjustbox}
\usepackage{threeparttable}
\usepackage{rotating} 
\usepackage{array}
\usepackage{makecell}
\usepackage{tikz}
\usepackage{afterpage}
\usepackage{longtable,lscape}
\usepackage{enumitem}
\usepackage{color, colortbl}
\usepackage{makecell}
\usepackage{wrapfig}
\usepackage[linesnumbered]{algorithm2e}
\usepackage[most]{tcolorbox}
\hypersetup{colorlinks, citecolor={black}, filecolor=black, linkcolor=black, urlcolor=black}
\usepackage{multirow}
\usepackage{hyperref}
\usepackage{enumitem}
\usepackage{subcaption}
\usepackage{graphicx}
\usepackage{multirow}
\usepackage[per-mode=fraction]{siunitx}

\definecolor{Gray}{gray}{0.9}
\definecolor{LightGray}{gray}{0.6}
\definecolor{greenv2}{HTML}{7AFF72}
\definecolor{redv2}{HTML}{FF4D42}

\makeatletter
\def\@fnsymbol#1{\ensuremath{\ifcase#1\or \dagger\or 
   \mathsection\or \mathparagraph\or \|\or **\or \dagger\dagger
   \or \ddagger\ddagger \else\@ctrerr\fi}}
\makeatother

\AtBeginDocument{%
  \providecommand\BibTeX{{%
    \normalfont B\kern-0.5em{\scshape i\kern-0.25em b}\kern-0.8em\TeX}}}

\acmVolume{0}
\acmNumber{0}
\acmArticle{0}
\acmMonth{0}

\setcopyright{none}

\begin{document}


\title{Quantified Canine: Inferring Dog Personality From Wearables}

\author{Lakmal Meegahapola}
\authornote{work done during the research internship at Nokia Bell Labs, Cambridge, UK}
\email{lakmal.meegahapola@epfl.ch}
\affiliation{\institution{Idiap Research Institute \& EPFL, Switzerland}}

\author{Marios Constantinides}
\affiliation{
  \institution{Nokia Bell Labs, UK}
}

\author{Zoran Radivojevic}
\affiliation{
  \institution{Nokia Bell Labs, UK}
}

\author{Hongwei Li}
\affiliation{
  \institution{Nokia Bell Labs, UK}
}

\author{Daniele Quercia}
\affiliation{
  \institution{Nokia Bell Labs, UK}
}

\author{Michael S. Eggleston}
\affiliation{
  \institution{Nokia Bell Labs, USA}
}

\renewcommand{\shortauthors}{Meegahapola et al.}

\begin{abstract}

Being able to assess dog personality can be used to, for example, match  shelter dogs with future owners, and personalize dog activities. Such an assessment typically relies on experts or psychological scales administered to dog owners, both of which are costly. To tackle that challenge, we built a device called ``Patchkeeper'' that can be strapped on the pet's chest and measures  activity through an accelerometer and a gyroscope. In an in-the-wild deployment involving 12 healthy dogs, we collected 1300 hours of sensor activity data and dog personality test results from two validated questionnaires. By matching these two datasets, we trained ten machine-learning classifiers that predicted dog personality from activity data, achieving AUCs in [0.63-0.90], suggesting the value of tracking the psychological signals of pets using wearable technologies.
\end{abstract}




\maketitle

\section{Introduction}
\label{sec:introduction}
When it comes to dog adoption, breed may not be the only important factor to consider~\cite{bradshaw2021impact, personality_over_breed} as humans tend to favor, for example, a pet's looks (e.g., attractiveness~\cite{chersini2018dog} based on poses and facial areas~\cite{isgate2018makes, hecht2015seeing}) and perceived human-directed sociability~\cite{lamb2021role}. However, according to a study from the Animal Farm Foundation~\cite{animal_foundation}, one in every four pets that are chosen based on breed (or looks) end up in shelters and rescues. By contrast, personality traits tend to offer a more comprehensive behavioral description of a dog, which is consistent over time and context~\cite{gartner2015pet}. Dog personality has been described in the literature using a variety of traits including fearfulness, energy levels, aggression, excitability, motivation, and amicability~\cite{jones2005temperament, jones2008development, ley2008personality}.

Not only could dog personality assessment reduce the number of owner-dog mismatches, but it could also put an end to (unfortunate) cases in which dogs get flocked into shelters or destroyed by authorities when expelled from their homes~\cite{corsetti2021different}. In fact, a few dog agencies and shelters are already experimenting with the use of dog personality traits for matchmaking dogs with future owners~\cite{nhhumane2022, knowsley2021meet}. Further, like humans, dogs also need different levels and types of companionship, activities, and emotional connection depending on their inherent personality traits~\cite{burgesspetcare2021tailor, knowsley2021meet}. It is therefore extremely important to identify a set of activities that `work' for a dog, and to find the right companion dogs for socializing~\cite{caroline2022do}, not least because inadequate socialization may escalate the pet's fear levels and may lead to aggression~\cite{shibashake2017do}.

Dog personality assessment is typically done through observational assessments by experts or psychological scales administered to the dog's owner \cite{cox2020understanding, jones2008development}. The former is expensive, time-consuming, and requires highly specialized facilities, and the latter is time-consuming, is prone to biases, and requires knowledge of someone who already knows the dog very well~\cite{jones2005temperament, jones2008development, rowan1992shelters}. That is why we set out to computationally assess dog personality in everyday settings (compared to highly specialized facilities or laboratory settings) with wearables. In the wearable sensing literature, studies used devices for monitoring dog activity~\cite{weiss2013wagtag, ladha2013dog, chambers2021deep, byrne2018predicting, ladha2013dog}, detecting pruritic behaviors (i.e., scratching, head  shaking)~\cite{griffies2018wearable}, and tracking breathing patterns~\cite{cotur2022bioinspired}. This stream of research recently inspired the fast-growing market of pet wearables~\cite{zamansky2019log} with a number of consumer-grade platforms readily available such as FitBark\footnote{\url{https://www.fitbark.com/}} (location, activity, and sleep tracking), PetPace\footnote{\url{https://petpace.com/}} (vital signs and behavior tracking), and PitPat\footnote{\url{https://www.pitpat.com/}} (activity tracking with gamified social elements).

Similar to how passive sensing of human personality drives the design of personalized apps~\cite{khwaja2019modeling}, sensor-based modeling of dog behavior through activity trackers has the potential to benefit both dogs and owners~\cite{zamansky2019log}. It has been found to impact owners' motivation to increase their mutual physical activities with their dogs, and increased human awareness to animals' needs \cite{jones2014use}. 
However, computational personality assessment techniques for dogs are non-existent. Therefore, we set out to develop and test an automatic way of operationalizing dog personality through passively sensed data from wearables. In so doing, we made three sets of contributions:
 
\begin{itemize}[leftmargin=*,align=left]
    \item We developed a wearable device, called ``Patchkeeper'', which can be easily strapped on a dog's chest (Section~\ref{sec:patchkeeper}). The device is equipped with accelerometer and gyroscope sensors. Since its processing pipeline was initially developed for wearable data  obtained from human subjects, we conducted a validation study of our device and the pipeline on dogs, together with a consumer-grade dog activity monitor. We found that our device is capable of determining four activity levels: moderate-vigorous activity with an accuracy of 92\%; light and sedentary activity with an accuracy of 96\%; and sleep with an accuracy of 98\%. 
    \item We launched a data collection campaign to recruit dog owners whose pets participated in a one-week study. The campaign was launched on four social media platforms (i.e., posts were made on Twitter, Facebook, Instagram, and NextDoor) and was also spread via word of mouth (Section~\ref{sec:animal_study}), resulting in a total of 22 dogs being successfully recruited and monitored for one week (i.e., the entire period of study).
    Dog owners answered two validated questionnaires (the Dog Personality Questionnaire (DPQ)~\cite{jones2008development} and the Refined Monash Canine Personality Questionnaire (MCPQ-R)~\cite{ley2009refinement}), and provided self-reports about their dog's activities (e.g., images were taken when walking the dog). Using the passively sensed data, we developed a data processing pipeline and extracted two types of features: (a) activity-level features (e.g., \% of sleep in the morning,  \% of sedentary activity in the afternoon) and (b) statistical features (e.g., acceleration histogram) (Section~\ref{sec:dataset}). We statistically analyzed the extracted features along with the self-reports from the two questionnaires and found that both types of features could discriminate dog personality traits (e.g., high or low fearfulness), with features capturing dog activity between 6am and 12pm (morning) being more informative for personality trait inferences than features capturing activity in the rest of the day. 
    This is expected as most dogs will be the most active and full of energy in the mornings after a dedicated sleep, and that was reflected in the signal captured from our device's sensors.
    \item We set up an inference task to predict dog personality traits using both activity-level and statistical features (Section~\ref{sec:statistical_analysis}). Our models achieved AUC scores in the range of 0.63-0.90 with a time-window-based setup (i.e., using the same features computed at different times of the day) (Section~\ref{sec:inference}). Interestingly, statistical features (e.g., acceleration histogram) were more informative than activity-level features (e.g., sedentary); yet, despite explaining more variance in personality traits, the former set of features is less interpretable than the latter one, opening up the need for Explainable AI in this kind of wearables too. When it comes to the usability of dog monitoring wearables, dog owners had split opinions about battery life (some found a day of battery life to be sufficient, while others expressed the opposite). For the development of future monitoring wearables, the majority of dog owners stressed their immense value, echoing a dog owner's statement: ``as dogs cannot speak, a device that allows my dog to `speak' and `express her feelings' is worth everything''.
\end{itemize}{}
\section{Background and Related Work}\label{sec:related_work}

Next, we surveyed various lines of research that our work draws upon, and grouped them into four main areas: \emph{i)} dog personality research; \emph{ii)} psychological scales for assessing dog personality; \emph{iii)} monitoring dog activity with wearable sensing; and \emph{iv)} activity levels and dog personality.

\subsection{Dog Personality Research}\label{subsec:dog_personality_research}
Dogs have personality~\cite{jones2008development,ley2009refinement,ley2008personality}, which refers to a set of dog behaviors and traits that are consistent over time and context~\cite{gartner2015pet, gosling2008personality}. These traits stem from the Five-Factor Model of personality, a.k.a. the Big-Five Traits~\cite{costa1992four}. As with personality, temperament is also being used in literature to describe both human and animal behavior. Researchers on animals and human infants tend to use the term temperament, while those studying human children and adults tend to use the term personality, with the two terms often being used interchangeably~\cite{mccrae2000nature}. On the one hand, temperament has been defined as the inherited, early appearing tendencies that continue throughout life and serve as the foundation for personality~\cite{goldsmith1987roundtable, jones2005temperament}; a definition that has not been widely adopted by animal researchers~\cite{gosling2001mice}. On the other hand, personality psychologists often study phenomena including temperament and character traits, attitudes, physical and bodily states, moods, and life stories~\cite{john2000personality}. Therefore, a broad definition includes characteristics of individuals
that describe and account for consistent patterns of feeling, thinking, and behaving~\cite{pervin1997}. As the distinction between temperament and personality has not been maintained consistently in the literature, we echo the statement by Jones and Gosling \cite{jones2005temperament, jones2008development}, that is, \emph{the term ``temperament'' is used whenever possible while the term ``personality'' is more appropriate when, for example, referring to work that explicitly discusses personality research. Hence, we use the term personality throughout the paper.}

In the scientific literature, Elliott Humphrey first hinted at the idea of dogs having personality in 1934~\cite{humphrey1934mental}. He described German Shepherd dogs with the traits of jealousy, apport, wildness-tameness, affection, initiative, attentiveness, curiosity, alertness, fighting and protection instincts, willingness to bite humans, confidence, self-right, energy, willingness, and intelligence. Seventy years later, by reviewing more than 50 scientific articles on dog personality, Jones and Gosling~\cite{jones2005temperament} found several inconsistencies, and proposed the first five-factor dog personality instrument, covering the dimensions of reactivity, fearfulness, responsiveness to training, submissiveness, and aggression. Building on Jones and Gosling's seminal work, researchers have incrementally added other dimensions such as calmness, boldness, trainability, and sociability~\cite{kubinyi2009dog}; extraversion, neuroticism, self-assuredness (motivation), training focus, and amicability~\cite{ley2009refinement}; stranger-directed sociability, activity, aggressiveness, and trainability~\cite{mirko2012preliminary}; and playfulness, chase-proneness, curiosity/fearlessness, sociability, and aggressiveness~\cite{svartberg2006breed}. 
Researchers, however, have split views when it comes to predictors of dog personality. Some studies found that different breeds have similar personalities~\cite{ley2009refinement, schneider2013temperament, svartberg2006breed}, while others reported the lack of evidence for it~\cite{mirko2012preliminary, sinn2010personality}. Two other attributes linked to personality traits are whether the dog is neutered or not (neutering is a surgical procedure to prevent a dog from reproducing) and its sex. Kubinyi et al.~\cite{kubinyi2009dog} found that not neutered dogs are more calm, while Lofgren et al.~\cite{lofgren2014management} found that neutered female dogs were less excitable and sought lower levels of attention. There is also evidence that older dogs are more calm~\cite{kubinyi2009dog} with lower amounts of fear~\cite{lofgren2014management} compared to their younger counterparts. Hence, as mentioned above, even though not conclusive, there is evidence that static attributes such as sex, age, and neutering could be associated with dog personality~\cite{kubinyi2009dog,lofgren2014management}. 

\subsection{Psychological Scales for Assessing Dog Personality}\label{subsec:psychological_scale_dog_personality}

While there are many dog personality measurement questionnaires~\cite{ani11051234, posluns2017comparing}, two widely established and validated psychological scales are: \emph{a)} the Dog Personality Questionnaire (DPQ)~\cite{jones2008development}, and \emph{b)} the Refined Monash Canine Personality Questionnaire (MCPQ-R)~\cite{ley2009refinement}. Next, we explain each scale.

\begin{itemize}[leftmargin=*,align=left]
    \item \textbf{DPQ:} Building on the work of Jones and Gosling~\cite{jones2005temperament}, the development of this scale aimed at reducing the time and resources (i.e., trained assessors, money, facilities) for dog personality assessment. Amanda Jones started from 1200 dog descriptors (i.e., statements describing dog behavior) identified in the literature and narrowed them down to 360 statements~\cite{jones2008development}. Then, in two studies with over 6000 participants, they narrowed these statements down to 75 items, grouped in five factors of \emph{Fearfulness}, \emph{Aggression towards People}, \emph{Excitability}, \emph{Responsiveness to Training}, and \emph{Aggression towards Animals}. 
    Scores for these traits can be derived using a list of statements marked by the dog owner on a Likert scale from 1 to 7 (1: disagree strongly; 7: agree strongly).
    \item \textbf{MCPQ-R:} This is the refined version of the original MCPQ questionnaire~\cite{ley2008personality}. The original questionnaire was developed using an adjective-based technique similar to the Big-Five Model of personality~\cite{john1990big}. Ley et al. \cite{ley2009refinement} revised the original MCPQ in a study with more than 450 participants. This led to the development of MCPQ-R, which consists of five factors: \emph{Extraversion} (perceived energy level of the dog), \emph{Motivation} (perceived persistence in the face of distractions---e.g., begging for food, finding a particular toy), \emph{Training Focus} (perceived trainability of the dog), \emph{Amicability} (perceived tolerance of the dog while being around humans and animals), and \emph{Neuroticism} (perceived nervous or cautious behavior of the dog). To assess these traits, dog owners rate 26 words (e.g., friendly, obedient, hyperactive) that describe their dog's personality by marking each word with the appropriate number from 1 to 6 (1 = really does not describe my dog; 6 = really describes my dog).
\end{itemize}

Even though the two scales come with different constructs, a fair amount of convergence has been observed~\cite{posluns2017comparing} between neuroticism (MCPQ-R) and fearfulness (DPQ); excitability (DPQ) and extraversion (MCPQ-R); responsiveness to training (DPQ) and training focus (MCPQ-R). While other widely used questionnaires such as the Canine Behavioral Assessment and Research Questionnaire (C-BARQ)~\cite{hsu2003development} were developed, recent research suggested that it is not suitable for general research use because it was designed to identify specific dog behavioral problems~\cite{cox2020understanding}. Hence, in the current study, we focused on DPQ and MCPQ-R questionnaires that capture a total of ten personality traits (factors).

\subsection{Dog Monitoring with Wearable Sensing}
Dog tracking and activity detection have gained much popularity due to advancements in sensor technology~\cite{hussain2022activity}, which led to a number of commercial dog monitoring products (e.g., FitBark, PetPace, PitPat). However, tying wearable sensing to behavioral tests (like dog personality in our case) is just starting to gain traction. In Animal-Computer Interaction research, prior studies focused on systems that facilitate better communication and interaction between dogs and owners~\cite{hirskyj2021forming} as well as among dogs~\cite{hirskyj2019internet}. Personality and dog behavior were also studied as part of certain games such as the spin-the-bottle~\cite{cox2020understanding}, concluding that dogs' preferences for human involvement were likely attributed to subtle differences in personality traits or prior training experiences.

Brugarolas et al.~\cite{brugarolas2015wearable} developed a non-invasive wearable sensor system for measuring dogs' vital signs using electrocardiogram (ECG), photoplethysmogram (PPG), and inertial measurement units (IMU). In a longitudinal study of monitoring puppies' cardiac changes, Foster et al.~\cite{foster2020preliminary} developed machine learning models for predicting puppies' Behavior Checklist (BCL) scores (including changes in energy and smoothness of movement, vocalization, tongue flicking, use of coping strategies, body language, and
changes in responsiveness to the handler), achieving up to 90\% of accuracy. Weiss et al.~\cite{weiss2013wagtag} developed WagTag that infers three dog activity levels (i.e., walk, run, and minimal), and concluded that personal models for predicting activity levels are better than universal models. Ladha et al.~\cite{ladha2013dog} also demonstrated that 17 dog activities (e.g., barking, running, chewing, digging) can be inferred with an accuracy of 70\% from a collar-worn wearable with accelerometers. More recently, Chambers et al.~\cite{chambers2021deep} used deep-learning models to infer dog activities with a collar-worn accelerometer, and showed that activities such as eating and drinking could be inferred with high accuracy, while behaviors such as licking, petting, rubbing, and sniffing were harder to identify. Beyond activity tracking, Griffies et al.~\cite{griffies2018wearable} used wearables to detect pruritic behaviors (i.e., scratching, head shaking). In a laboratory study with over 360 dogs, they showed that algorithms could be trained to infer head shaking and scratching with sensitivities over 70\% and specificities over 90\%. Wearable devices have also been used to monitor dog breathing patterns with reasonable accuracies~\cite{cotur2022bioinspired}.

\subsection{Dog Activity Levels and Personality Traits}\label{subsec:actvity_levels_personality}
Prior work in animal-computer interaction and canine behavior has highlighted certain relationships between personality traits and activity levels. For example, previous studies found that more extroverted dogs showed higher activity levels in the park~\cite{carrier2013exploring}, higher energy levels~\cite{gosling2003dog}, with significantly greater proportions of time spent with other dogs. Amicable dogs showed frequent behaviors indicative of play (high activity level), while neurotic dogs showed higher frequencies of hunched posture (low activity level)~\cite{carrier2013exploring}. Hence, extraversion, amicability, and neuroticism (traits that come from MCPQ-R) can be directly linked to activity levels. Further, even though not directly studied, prior studies linked psychological aspects such as fearfulness and aggression (corresponding to the three traits of DPQ--- fearfulness, aggression towards people, and aggression towards animals) to activity levels. For example, in domestic dogs, it has been found that a higher degree of impulsivity correlates with high activity levels~\cite{wan2013drd}, poor attention span~\cite{vas2007measuring}, and human-directed aggression~\cite{rayment2015applied, peremans2003estimates}. Further, previous studies linked activity levels to negative emotions and stress~\cite{beaudet1994predictive, rayment2015applied, jones2014use}, which, in turn, can be seen as the roots of fearfulness and aggression~\cite{beaudet1994predictive}. Moreover, neuroticism has been directly linked to activity levels in some studies \cite{carrier2013exploring}, but it has also been observed to be converging with fearfulness according to other studies \cite{posluns2017comparing}, hence providing evidence on how activity-levels could be indirectly informative of fearfulness. Studies have also found that excessively high or low activity levels are predictive of successful dog training (i.e., trainability and certain levels of fearfulness~\cite{weiss2002selecting}; traits captured from DPQ and MCPQ-R).

In summary, previous wearable sensing literature explored aspects such as monitoring dog activity, detecting pruritic behavior, and tracking breathing patterns. While previous literature explored a few aspects concerning the relationship between dog activity and personality traits, this relationship still represents an under-explored area. Our study aims to partly fill this gap by exploring the relationship between ten personality traits captured from two canine personality questionnaires and dog activity.
\section{Research Questions}
\label{sec:rqs}

We set out to explore whether dog personality can be automatically inferred from wearable data in everyday settings by answering three questions:

\begin{itemize}[wide, labelwidth=!, labelindent=0pt]
    \item[\textbf{RQ\textsubscript{1}:}] Which dog activity-level features and statistical ones can be extracted from wearable data?
     \item[\textbf{RQ\textsubscript{2}:}]  Which dog activity-level features and statistical ones are associated with dog personality?
    \item[\textbf{RQ\textsubscript{3}:}]  To what extent activity-level, statistical, and demographic features are predictive of dog personality? 
\end{itemize}{}

\section{Patchkeeper}
\label{sec:patchkeeper}

\begin{figure*}
\captionsetup{labelfont=normalfont}
\centerline{\includegraphics[width=\linewidth]{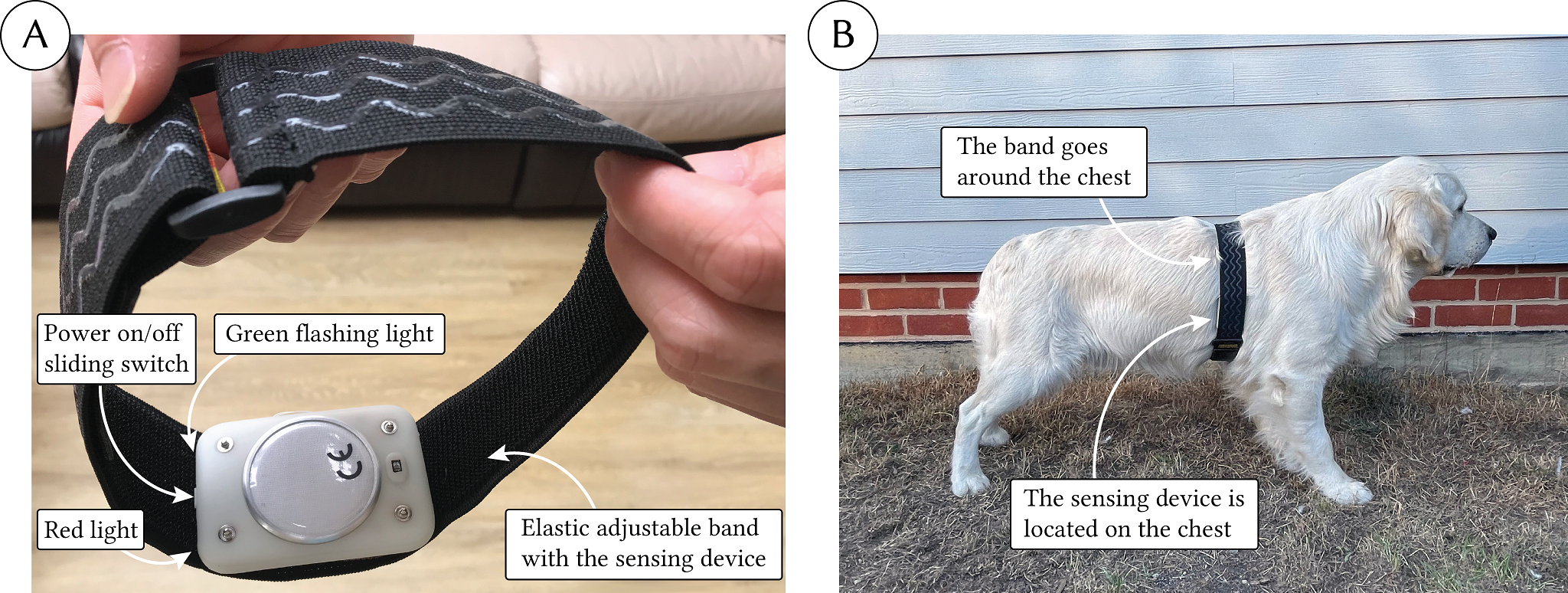}}
\caption{\emph{A:} The Patchkeeper device is attached to an elastic adjustable band. The device comes with two lights: a \emph{green light} indicating whether the device is ON; and a \emph{red light} indicating whether the device is in charging mode. \emph{B:} The band can be strapped on the dog's chest. }
\label{fig:device_and_dogs}
\end{figure*}

Patchkeeper (Figure~\ref{fig:device_and_dogs}a) is a wearable device developed at {Nokia Bell Labs} for behavioral monitoring of both humans and animals. It contains a photoplethysmography (PPG) sensor, an electrocardiogram (ECG) sensor, an accelerometer, a gyroscope, and a microphone. In the current study, only the inertial measurement unit (IMU) sensors (i.e., accelerometer and gyroscope) were used, and PPG, ECG, and microphone were not used due to the dog hair and privacy concerns (more details in Section~\ref{subsec:device_sensors_used}). The IMU sensor is a BMI160 from Bosch Sensortec\footnote{\url{https://www.bosch-sensortec.com/products/motion-sensors/imus/bmi160/}}. It is a small, low-power, low-noise 16-bit chip designed for mobile applications. It provides highly accurate gyroscope and accelerometer data in real time. The IMU's sampling rate was set to 50 samples per second. This sampling rate allowed for striking the right balance between obtaining reasonably fine-grained data for our analysis and storage capacity requirements. The microcontroller unit (MCU) is an nRF52840 from Nordic Semiconductor\footnote{\url{https://www.nordicsemi.com/products/nrf52840}}, which contains a 64 MHz Cortex-M4 processor with floating point unit (FPU). All data was saved in a micro-SD card on the printed circuit board (PCB). 

The device dimensions are 76x52x15mm with a weight of 56 grams. It contains a 400mAh lithium polymer battery, which can last more than 24 hours while continuously recording data. The battery takes around two hours to be fully charged and comes with a USB-C charging port for hassle-free charging with any commercially available charger. With a two-hour daily charge, the device runs continuously without any loss of data. The device has a switch with ON and OFF sides marked with red and green colors. For better user experience, we included different lights on the device (Figure~\ref{fig:device_and_dogs}a): \emph{(i)} a green light flashing every 10 seconds indicates that the device is ON, it is working properly, and data is being recorded; \emph{(ii)} a static red light indicates that the device is fully charged, and \emph{(iii)} a flashing red light indicates an issue with the device or the memory card.

\section{Animal Study}
\label{sec:animal_study}

Having developed our custom-made wearable device to collect dog activity data, we conducted a one week in-the-wild study to understand the link between dog behavior and personality.

\subsection{Materials and Apparatus}\label{subsec:materials}
Each dog owner received a package, fitted in a medium-sized letter envelope, weighing approximately 500 grams. The package contained: a Patchkeeper device, a charging cable, three black elastic straps, a consent form, an information sheet, questionnaires (i.e., DPQ, MCPQ-R, and a post-study questionnaire), and a pre-paid return envelope. Upon completion of the study, the owner shipped back the package using the pre-paid return package. 

\subsubsection{{Patchkeeper and Elastic Straps.}}\label{subsec:device_sensors_used} As the device can be used on both human and animal subjects (Section~\ref{sec:patchkeeper}), and given the requirement of continuous monitoring for one week, we decided to deactivate the ECG and PPG sensors, and the audio microphone. ECG and PPG were disabled for two reasons: first, they relied on skin conductance, which is made difficult by dog hair; and second, they required additional straps, which would place additional effort on the owners, making it more likely for them to drop out. Audio was also deactivated due to privacy reasons as the device would otherwise continuously capture audio throughout the day. It would be extremely awkward to listen to intimate moments or any audio conversation that creeps into the device due to the pet's movements. To ensure that the device would fit various dog sizes, we used an adjustable elastic band that can be strapped to the pet's chest (Figure~\ref{fig:device_and_dogs}b). These are off-the-shelf straps that can be found on Amazon and are comfortable to wear. The device can be simply attached to the strap using a sticky patch. We also considered alternative areas (e.g., neck) to place the device, weighing various aspects. First, some breeds have a more pronounced dewlap (loose, saggy skin around the neck/throat) than others, whilst the chest is generally not affected in such a way. Second, it has been found that the skin near the axilla (armpit) and ventral abdomen (lower chest/thorax, top of the belly) is significantly thinner than that in the dorsal (top of the dog) areas~\cite{theerawatanasirikul2012histologic}. Third, double coated breeds (e.g., the golden retriever, Samoyed, and German shepherd included in this study) have coarse guard hairs and dense undercoats, with this being particularly pronounced in the dorsal areas but less so near the axilla and ventral abdomen areas. Taking all these aspects into consideration, the chest area (behind the forelegs) has the benefits of thinner skin whilst removing breed-to-breed variation in our sample (previous work also favored the chest area~\cite{foster2020preliminary}). To ensure that the position of the device did not impact the results, we intentionally used wider straps with 5cm wide rubbered features and strong tensioning force to make the device intimately connected to the fixed location at the pet's chest. During pilot studies in the design of the device, we estimated an approximate sensor dislocation of $\pm$2cm, which was sufficient to guarantee fixed sensor location over a long period of use.

\subsubsection{{Questionnaires.}} Dog owners answered two types of questionnaires. A Pre-Study Questionnaire (Q1) and a Post-Study Questionnaire (Q2). Q1 was completed before the study and had two sections. The first section captured demographic information of the owner (i.e., age, sexual identity, occupation status, and ethnicity), followed up with the Personality Inventory (TIPI)~\cite{gosling2003very}, which is a 10-item measure of the Big Five (or Five-Factor Model) dimensions. The second section captured basic information about the dog (i.e., the dog's age, breed, sexual identity, weight, typical activity levels, disease conditions, and whether it is neutered or not). Q2 was completed after the study and had two sections as well. The first section captured user experience and dog owners' perceived utility of wearable platforms for dog monitoring. The user experience of the Patchkeeper was captured by a Likert scale of 1 to 7 (1 = very bad, 7 very good) and corresponding feedback. In a similar vein, we captured the perceived utility of commercial wearable pet monitoring devices in general by a Likert scale of 1-7 (1 = not very important; 7 = very important) and corresponding feedback. Additionally, we asked dog owners to rate on a scale of 1-7 (1 = strongly not preferred, 7 = strong preferred) their likelihood of adopting a mobile app that uses Patchkeeper's data for dog monitoring. We provided sample options including: monitoring activity types, identifying when dogs are not in a healthy state, finding a community of dogs with a similar personality, or monitoring the mood and stress of dogs. The second section of Q2 asked owners to complete the Dog Personality Questionnaire (DPQ)~\cite{jones2008development} and the Refined Monash Canine Personality Questionnaire (MCPQ-R)~\cite{ley2009refinement}.

\begin{table*}
\captionsetup{labelfont=normalfont}
\caption{Overview of dog demographics.}
\label{tab:participants}
  \centering
  \begin{tabular}{lccccc}
    
    \textbf{\cellcolor[HTML]{FFFFFF} Dog ID} & 
    \textbf{\cellcolor[HTML]{FFFFFF} Breed} & 
    \textbf{\cellcolor[HTML]{FFFFFF} Sex} & 
    \textbf{\cellcolor[HTML]{FFFFFF} Weight} & 
    \textbf{\cellcolor[HTML]{FFFFFF} Neutered?} &
    \textbf{\cellcolor[HTML]{FFFFFF} Birth Year}
    \\
    
    \hline
    
    \#1 &
    Golden Retriever &
    Female &
    30 kg &
    Yes &
    2018
    \\
    
    \#2 &
    Golden Retriever &
    Male &
    35 kg &
    No &
    2021
    \\
    
    \#3 &
    Poodle (Toy) &
    Male &
    8 kg &
    No &
    2020
    \\
    
    \#4 &
    Dalmadoodle - 75\% Poodle, 25\% Dalmatian &
    Female &
    13 kg &
    Yes &
    2020
    \\
    
    \#5 &
    Golden Retriever &
    Male &
    40 kg &
    No &
    2021
    \\
    
    \#6 &
    Working English Setter &
    Male &
    29 kg &
    Yes &
    2017
    \\
    
    \#7 &
    Boxer &
    Female &
    25 kg &
    No &
    2020
    \\
    
    \#8 &
    Samoyed &
    Male &
    25 kg &
    No &
    2018
    \\
    
    \#9 &
    Cockapoo &
    Female &
    10 kg &
    Yes &
    2011
    \\
    
    \#10 &
    Working English Setter &
    Female &
    31 kg &
    No &
    2016
    \\
    
    \#11 &
    Mixed &
    Male &
    15 kg &
    Yes &
    2021
    \\
    
    \#12 &
    Cavalier King Charles Spaniel &
    Female &
    8.5 kg &
    Yes &
    2019
    \\
    
  \bottomrule
\end{tabular}
\end{table*}   

\subsubsection{{Information Sheet and Consent Form.}} 
The information sheet described the study protocol (Section~~\ref{subsec:study_protocol}). The consent form highlighted two aspects: confidentiality and voluntary participation. In terms of confidentiality, the form explained that all data would be kept confidential except in cases where the researchers were legally obligated to report specific incidents (e.g., dog abuse). The collected phone numbers and email addresses will not be used in any scientific output, and that confidentiality will be preserved by: \emph{a)} assigning code numbers for dog owners in all research documents; and \emph{b)} keeping notes, data, and any other dog owner identifiers in a password-protected hard drive, securely stored at the facilities of the {Nokia Bell Labs}. In terms of voluntary participation, the form explained that a signature was required to participate. Additionally, withdrawal from the study was allowed at any time and without giving a reason, even after signing the consent form. Upon withdrawal, all data will be deleted.    

\subsubsection{{Pre-Paid Letter Cover and WhatsApp Hotline.}} To ease dog owners participation, we included in our package a pre-paid return package. Upon completion, they placed all materials received into the return package and posted it. To have effective communication with the dog owners throughout the study, we used a dedicated WhatsApp number as a hotline. This number was used by the first author to deal with matters related to the study (e.g., unable to place the strap or charge the device). 

\begin{figure*}
\captionsetup{labelfont=normalfont}
\centerline{\includegraphics[width=\linewidth]{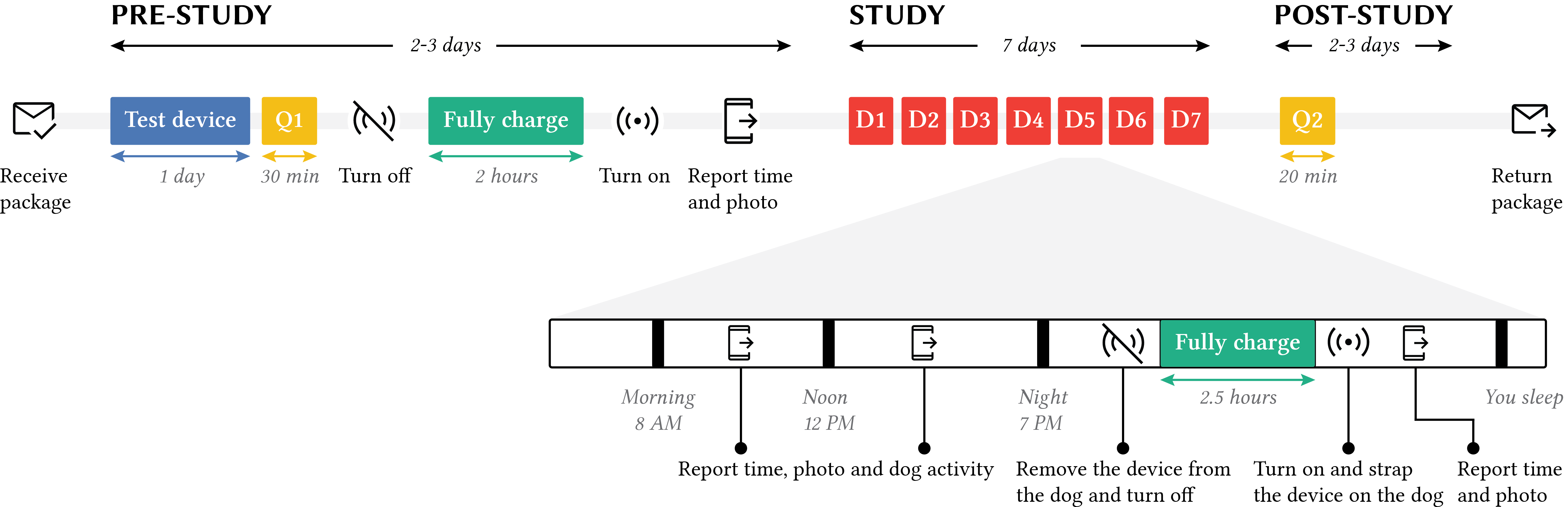}}
\caption{Our study protocol has three phases. In the \emph{pre-study} phase, dog owners received the study package, including the PatchKeeper device and questionnaires; in the \emph{study} phase, data collection for seven days took place; and, in the \emph{post-study} phase, dog owners returned the package and answered a follow-up survey about their experience using the device.}
\label{fig:protocol}
\end{figure*}

\subsection{Study Protocol}\label{subsec:study_protocol}
The study protocol has three periods: pre-study, study period, and post-study (Figure~\ref{fig:protocol}).

\subsubsection{Pre-Study} Once the dog owners received the package, they familiarized themselves with the device and answered the pre-study questionnaire. During that period, they were encouraged to ask questions via the hotline, and were instructed to fully charge the Patchkeeper and send a picture of the dog with the device turned on, after a full charge every day (this was a preemptive measure to ensure compliance, but, at the same time, to guarantee data quality).

\subsubsection{Study Period.} During the seven-day period, the device captured sensor data between 12am and 6pm (continuously for 18 hours), and it could be charged for two hours between 6pm and 12am. Enforcing the same charging schedule across all dog owners enabled us to obtain comparable data across dogs. Of course, this comes with the caveat that we might not have 2-3 hours of data between evening hours; a drawback that we were willing to accept to ensure high quality data during other time slots. In summary, each evening, the dog owners would remove the strap from the dog, turn the device off, fully charge it, turn it back on, and put it back on the dog. Afterward, they would send a message with a picture of the dog wearing the device via the hotline. In the morning, they would again be asked to check that the device was working and was correctly positioned around the dog's chest. During all other time periods, no interaction was required from dog owners as the device would automatically capture all data. Furthermore, we encouraged the dog owners to voluntarily send us in-situ self-reports (in the form of images or short video clips) of various dog activities throughout the day.

\subsubsection{Post-Study.}\label{subsec:post_study} During that period, dog owners answered the post-study questionnaire. They placed all materials and apparatus in the pre-paid package, and shipped it back to the return address. Upon successful completion of the study, dog owners received a \$25 Amazon gift voucher and a report summarizing their dog activity profile over the seven days of the study.

\subsection{Recruitment}
Recruitment for in-the-wild human studies is typically difficult~\cite{ellard2015finding}, and so it is for animal studies. We employed two techniques that were proven (un)successful to varying degrees.

\begin{itemize}[leftmargin=*,align=left]
    \item Social media and local communities (Twitter, Facebook, Instagram, and NextDoor): Twitter and Facebook are used to advertise scientific studies~\cite{whitaker2017use, sibona2012purposive}---both channels were not very successful in this study. Instead, Instagram posts on profiles dedicated to dogs with 1000s of followers were successful to some extent. Finally, Nextdoor\footnote{\url{https://nextdoor.com/}}, a social media site for local communities, was the most successful recruitment strategy (40\% of the dogs were recruited through it). A banner of the study was also shared within the communities of Cambridge Dog Meetup.
    \item Word of Mouth: One researcher from Nokia Bell Labs, who is not part of conducting the study, participated in the study with his dog. He spoke to his neighbors about the study, who also signed up. Shortly after, this created a snowball effect (30\% of the dogs were recruited through word of mouth). 
\end{itemize}

Having a variety of recruitment techniques, we were able to reach out to 31 dog owners in Cambridge, United Kingdom. Of these, 22 signed up for the study and received the package. 10 of them withdrew during the study for various reasons: high temperature, including a heatwave, making it difficult for the dog to wear the strap continuously (2/10), owners going away for summer holidays (2/10), strap not holding to the body of the dog due to its curvy shape (1/10), dogs not being in healthy conditions (i.e., leg injury after a run, wound on the neck, bug bites) during the time of the experiment (3/10), and dogs not appearing to feel happy about wearing the strap (2/10). This left us with 12 healthy dogs that successfully completed the study. Note that these 12 dogs were all healthy (as reported by their owners), and every morning the first author checked with the owners whether any of the dogs displayed peculiar behavior (e.g., snagging on objects, appearing to feel uncomfortable) due to the wearable. No such incident was reported. However, we had an incident wherein a dog jumped into a body of water, destroying the device. This dog continued the study later with a replacement device. 
The recruitment took place during the summer period, with starting dates ranging from July to August. The study was approved by Nokia Bell Labs, and the study protocol stated that the collected data will be analyzed for research purposes only. In accordance to GDPR, no researcher involved in the study could have tracked the identities of the dog owners after the end of the study, and all responses were analyzed after anonymization at an aggregated level.

\section{Dataset}
\label{sec:dataset}
Having successfully deployed Patchkeeper in an in-the-wild study and collected more than 1300 hours of accelerometer and gyroscope data, we then applied a processing pipeline to that data.

\begin{figure*}
\captionsetup{labelfont=normalfont}
\centerline{\includegraphics[width=0.8\linewidth]{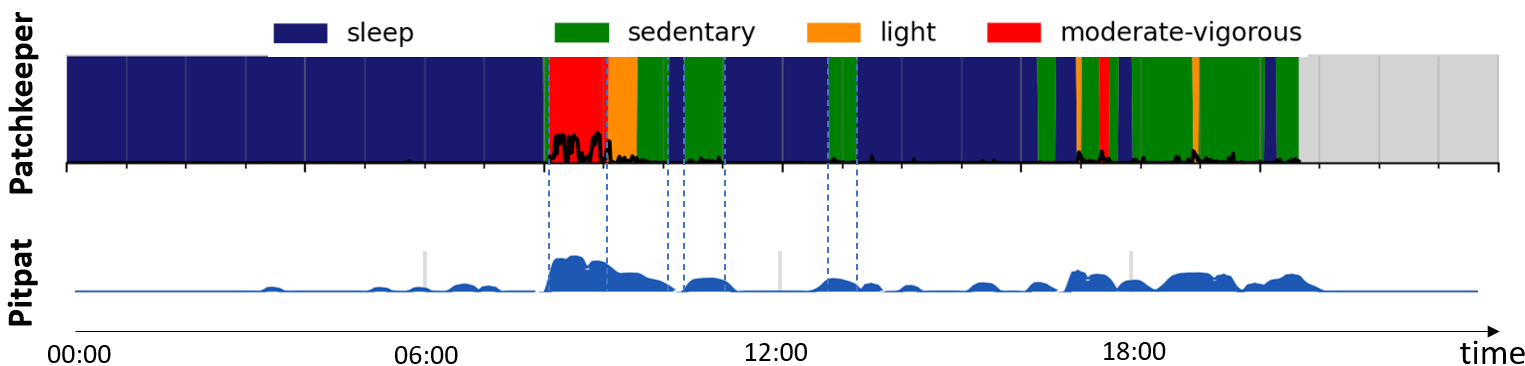}}
\caption{Comparison of dog\#3 is monitored by activity levels generated from Patchkeeper (top) and Pitpat (bottom) for 24 hours.}
\label{fig:validation_study}
\end{figure*}

\begin{figure*}
\captionsetup{labelfont=normalfont}
\centerline{\includegraphics[width=0.8\linewidth]{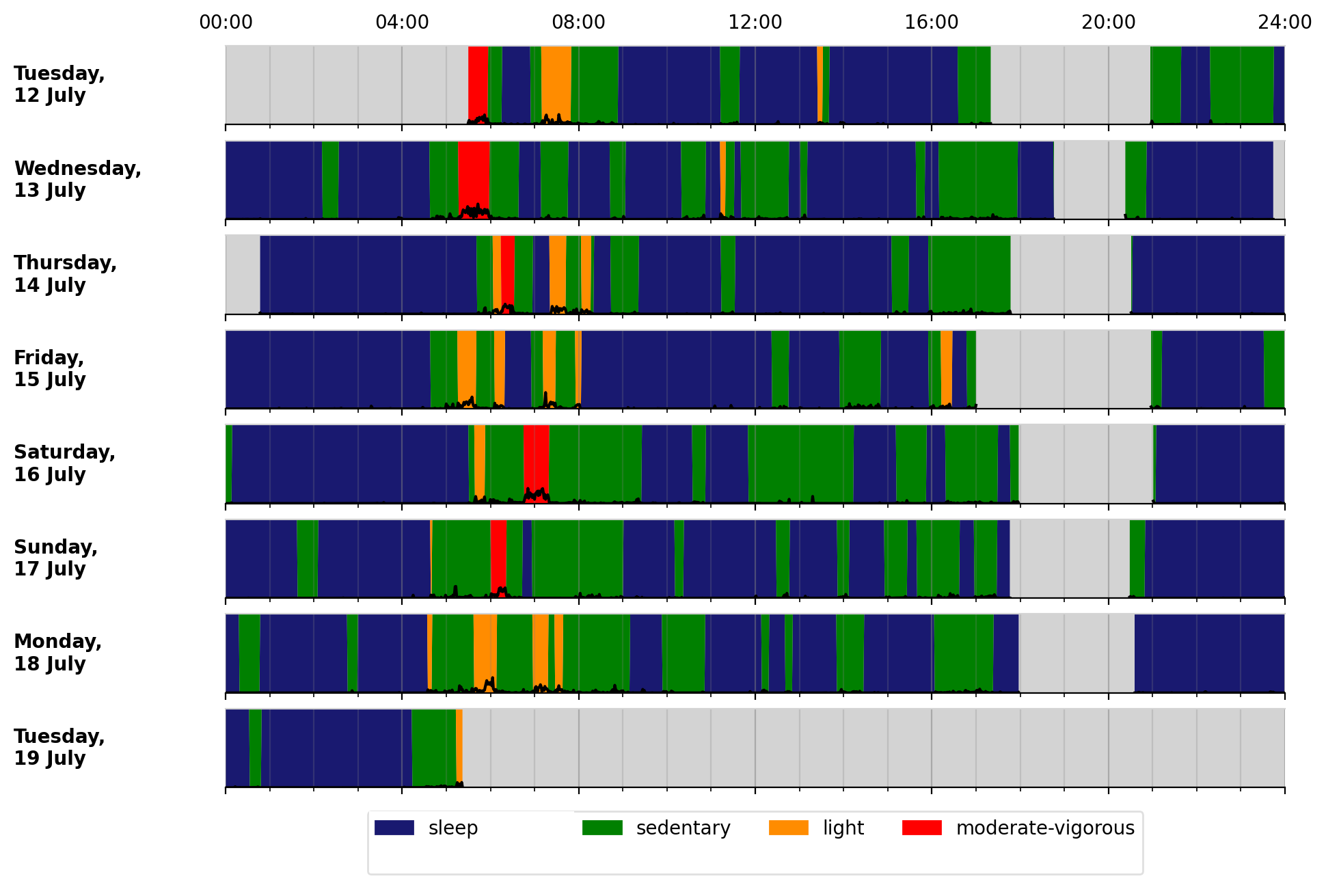}}
\caption{Example of four types of activity levels of dog\#1 (i.e., sleep, sedentary, light, and moderate-vigorous) generated from Patchkeeper.}
\label{fig:activity_level_diagram}
\end{figure*}

\subsection{Data Processing Pipeline}\label{subsec:data_processing_pipeline}
\subsubsection{{Activity-Level features.}} 
\label{subsec:activity-level}
This set of features describes dog behaviors derived from accelerometer data, and is interpretable. To extract these features, we used a state-of-the-art data processing pipeline to convert the triaxial data to acceleration~\cite{doherty2017large, willetts2018statistical}. The processing included four steps: \emph{a)} ten-second samples from static sections (no movement) of accelerometer data were obtained to optimize the gain and offset for each of the X, Y, and Z axes to fit a unit gravity sphere~\cite{willetts2018statistical}; \emph{b)} data were re-sampled at 100Hz using linear interpolation, and acceleration was calculated using the euclidean norm of X, Y, and Z axis values; \emph{c)} a fourth-order butterworth filter was used to remove noise; and \emph{d)} one gravity (1G) unit was removed from the data, and the remaining negative values were truncated at zero.

Next, using non-overlapping time windows of 60 seconds, 126 time and frequency domain features such as mean, standard deviation, median, minimum, maximum, 25\textsuperscript{th} and 75\textsuperscript{th} percentiles of vector magnitude, kurtosis, and skewness were generated~\cite{willetts2018statistical, doherty2018gwas, walmsley2021reallocation}. Using these features, we used a pre-trained model based on Hidden Markov Models and Balanced Random Forests~\cite{walmsley2022reallocation} to classify acceleration into four different activity levels: \emph{sleep}, \emph{sedentary}, \emph{light}, and \emph{moderate-vigorous}. These activity levels are in line with prior studies on dog activity-levels~\cite{morrison2013associations, ortmeyer2018combining, weiss2013wagtag}.

As the data processing pipeline was initially developed for wrist wearables worn by humans, we conducted a validation step to ensure transferability to animals. To do that, we used a consumer-grade dog activity monitor called PitPat\footnote{\url{https://www.pitpat.com/}} on two dogs (dog\#3 and dog\#8) for three days. These two dogs also took part in the larger in-the-wild study. In total, we collected over 120 hours of sensor data from both devices, and a total of 83 self-reports (e.g., the dog is sleeping, running) from dog owners. A comparison of our 24-hour data processing pipeline and PitPat's output is shown in Figure~\ref{fig:validation_study}.  In terms of ground truth obtained from PitPat (in total, we analyzed 100 data points), our model performed with an accuracy of: 98\% in detecting sleeping (sections where PitPat showed no activity); 92\% in detecting high intense moderate-vigorous activities (sections where PitPat showed a peak in activity levels), and 96\% in detecting sedentary or light activity levels (sections where PitPat showed a medium level of activities). In terms of self-reported ground truth (including pictures), our model was 91\% accurate in determining the 83 activity levels provided by dog owners. This answered our \textbf{RQ\textsubscript{1}}, allowing us to conclude that activity levels can be extracted with accuracies over 90\%. 

Having established the reliability of our data processing pipeline, we first obtained how long a particular dog had been engaging in activities at different levels (i.e., percentage of time spent in sleep, sedentary, light, and moderate-vigorous activity levels), resulting in four features. We then used the acceleration time series to extract statistical features such as its minimum, maximum, mean, median, and standard deviation, resulting in five features. For simplicity, we call these nine features \emph{activity-level} features throughout the paper.

\subsubsection{{Statistical Features.}} 
This set of features was derived from complex associations in the time series of both accelerometer (x,y,z) and gyroscope (x,y,z) and, as such, is less interpretable compared to the activity-level features but computationally less expensive to obtain. To extract these statistical features, we used the tsfel library~\cite{barandas2020tsfel}. The library allowed us to extract 56 features (e.g., min, max, std, mean, median, kurtosis, skewness, absolute energy, zero crossing rate, histogram, and empirical cumulative distribution function) that describe temporal and statistical aspects of the time series nature of the data\footnote{\url{https://tsfel.readthedocs.io/en/latest/descriptions/feature_list.html}}. 

\subsubsection{{Unit of Analysis.}}
A typical way of capturing temporal dynamics in HCI and UbiComp studies is to use time windows at different times of day when calculating features~\cite{obuchi2020predicting, wang2020predicting, wang2017predicting, nepal2020detecting, wang2022first, constantinides2018personalized}. A large time window of eight hours, dividing the day into three periods, has been previously used in dog studies, and it has been found, for example, that studying night sleep separately from day sleep provided more meaningful insights about sleeping patterns~\cite{schork2022cyclic, bodizs2020sleep} than studying sleeping during the whole day. Drawing from this prior line of work, we resorted to three time windows for our analysis: \emph{a)} night (N): time from 12am to 5.59am; \emph{b)} morning (M): time from 6am to 11.59am.; and \emph{c)} afternoon (A): time from 12pm to 5.59pm. For example, at night, a dog could be showing activity levels as 60\% sleeping, 20\% sedentary, 15\% light, and 5\% moderate-vigorous. Hence, for each time window, we extracted a total of 65 features, including the nine activity-level and the 56 statistical features. 

\begin{figure*}[t]
\begin{center}

    \begin{subfigure}[t]{0.45\textwidth}
        \centering
        \includegraphics[width=\textwidth]{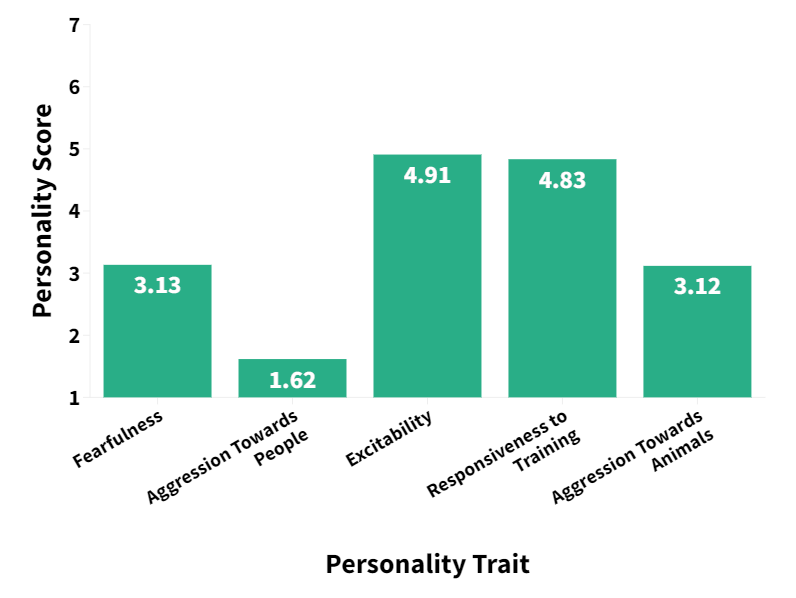}
        \caption{\centering DPQ}
        \label{fig:dpq}
    \end{subfigure}
    \hfill 
    \begin{subfigure}[t]{0.45\textwidth}
        \centering
        \includegraphics[width=\textwidth]{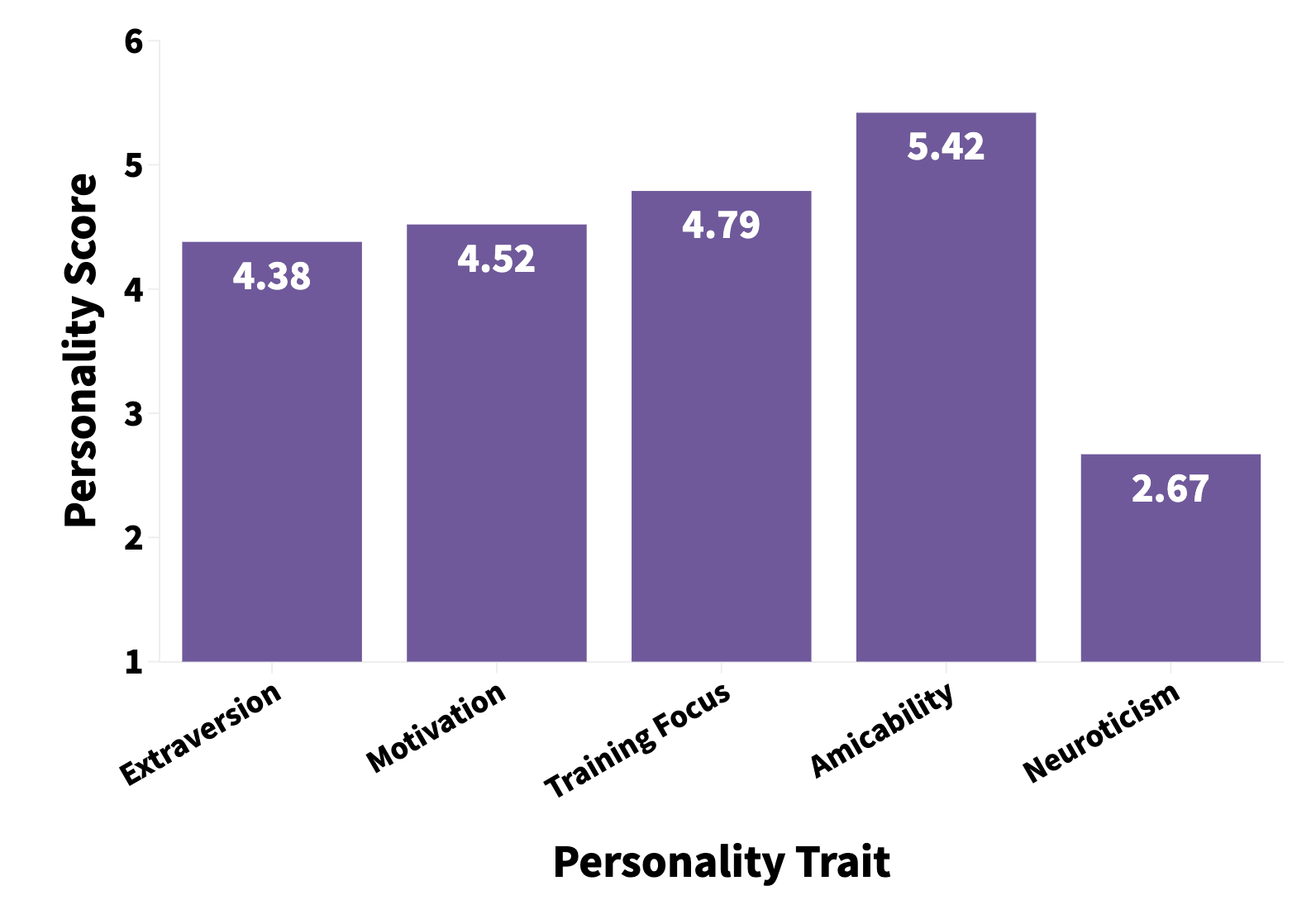}
        \caption{\centering MCPQ-R}
        \label{fig:mcpqr}
    \end{subfigure}
    \caption{Average personality trait scores for five personality factors measured through DPQ (in a Likert scale 1-7) and MCPQ-R (in a Likert scale 1-6), respectively. Scores for MCPQ-R were higher (except Neuroticism), whereas scores for DPQ were more spread out.
    }
    \label{fig:personality_scores}
\end{center}
\vspace{-0.2 in}
\end{figure*}

\subsection{Descriptive Statistics of Personality Traits and Activity Levels}
\begin{figure*}
\captionsetup{labelfont=normalfont}
\centerline{\includegraphics[width=0.8\linewidth]{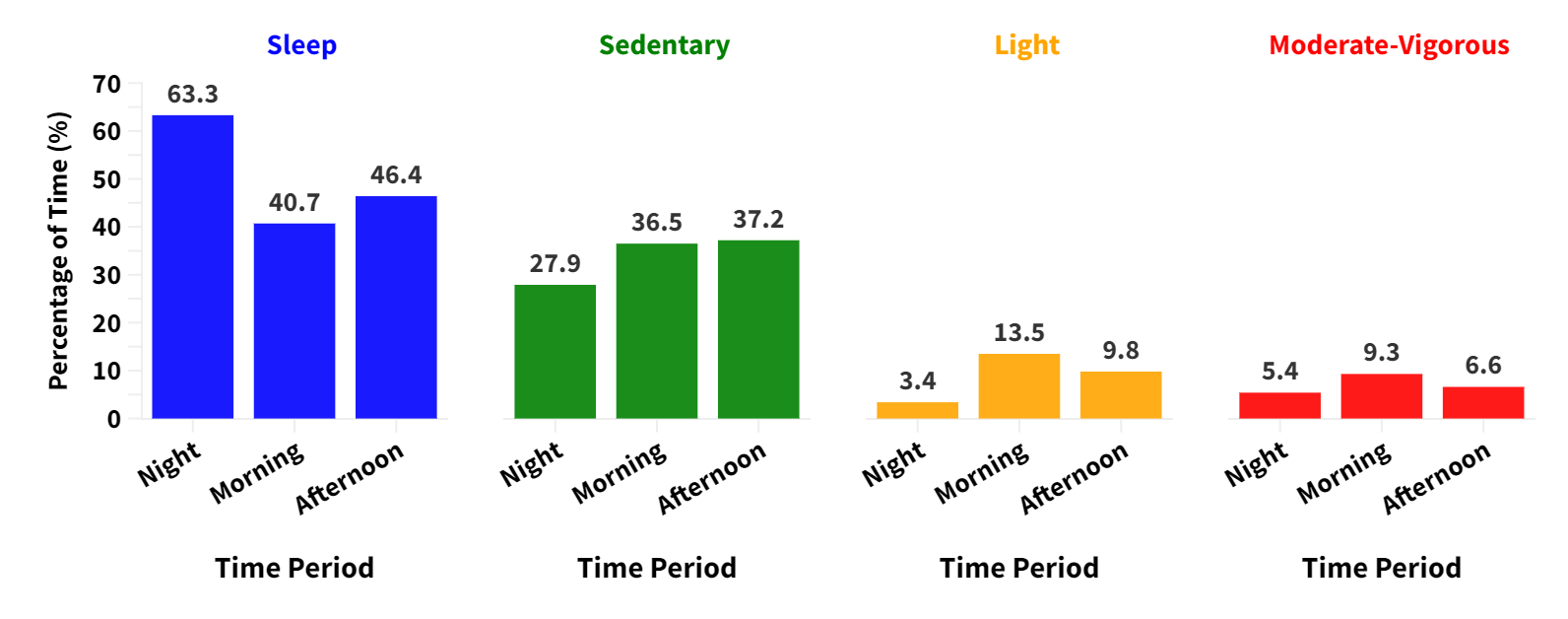}}
\caption{Percentage of time spent by all dogs on average in each of the four activity level for sleep, sedentary, light, moderate-vigorous at three different time of day (Night, Morning, Afternoon). Dogs slept more at night, engaged in sedentary activity in the morning and the afternoon, and engaged in physical activity (light or moderate-vigorous) during the morning.}
\label{fig:activity_time_periods}
\end{figure*}

The distributions of personality traits are shown in Figure~\ref{fig:personality_scores}, and a summary of statistics of the recruited dogs is in Table~\ref{tab:participants}. Recruited dogs were over one year old, with a mean age of three years and ten months, 65\% of them were female dogs, and were all medium to small dogs. In terms of dog personalities, the DPQ factors: \emph{Fearfulness}, \emph{Aggression Towards People}, and \emph{Aggression Towards Animals} had average scores below 3.2, whereas the \emph{Excitability} and \emph{Responsiveness to Training} had high average scores above 4.8. For MCPQ-R, \emph{Neuroticism} had an average score of 2.67, while the other four factors had average scores on or above 4.38. Overall, the mean percentages across the personality dimensions were comparable to previous studies~\cite{ley2009refinement, carrier2013exploring}. Further, Figure~\ref{fig:activity_time_periods} shows the average activity level across all dogs as a percentage of total time ($y$-axis) for different time periods of the day ($x$-axis). The night was predominantly spent sleeping (63.3\%) whereas morning was predominantly spent at other types of activity such as sedentary (36.5\%), light (13.5\%), or moderate-vigorous (9.3\%). 

\section{Methodology}
\label{sec:statistical_analysis}

Using the extracted features and self-reported personality (Section~\ref{sec:dataset}), we set out to understand which features are associated with dog personality (\textbf{RQ\textsubscript{2}}), and, to what extent these features are predictive of personality (\textbf{RQ\textsubscript{3}}). In so doing, we defined our dependent variables, conducted a series of statistical analyses, and developed machine learning classifiers to predict dog personality, which we describe next.

\subsection{Dependent Variables}
The ten personality traits of both DPQ and MCPQ-R (five each) served as our dependent variables for both statistical analyses and classification tasks. We binarized each personality trait (i.e., whether a dog scored high or low in a given trait ---for example, high or low in fearfulness) using the median value across all dogs in our dataset. In other words, we computed the median value across all dogs for each trait, and binarized each trait to be either high or low. The choice of binary traits was reinforced by previous literature on inferring human personality from mobile data~\cite{khwaja2019modeling, vinciarelli2014survey}. 

\subsection{Statistical Analyses}
These analyses allowed us to identify statistically significant features that help discriminate between high and low personality scores for each trait (Table~\ref{tab:tstatistics}). We report the top five features for each personality trait with: \emph{a)} the highest $t$-statistic\footnote{$p$-values~\cite{Greenland2016} are marked with an asterisk (*) after bonferroni correction~\cite{weisstein2004bonferroni}.}~\cite{Kim2015}, and \emph{b)} the highest Cohen's-d\footnote{the 95\% confidence intervals~\cite{Lakens2013} overlapping with zero are marked with an asterisk (*).}~\cite{Rice2005}. As a rule of thumb, a Cohen's-d of 0.2 illustrates a small effect size, 0.5 a medium effect size, while 0.8 a large effect size~\cite{kim2015statistical}. Results are presented in Section~\ref{subsec:statistical_analysis}.

\subsection{Classification and Cross-Validation Methods}
In the next set of experiments, we used Python with Keras~\cite{chollet2015keras} and scikit-learn~\cite{pedregosa2011scikit} frameworks. For dimensionality reduction, we used principle component analysis (PCA)~\cite{abdi2010principal} and retained features with a variance of 95\% (leaving us with 3-5 features to train models, depending on the set of features used---we discuss these features in a subsequent section). All experiments were done with the leave-k-dogs-out strategy (in a similar manner to the leave-one/K-out setting, which is typical cross-validation for human subjects \cite{meegahapola2021one}) in which data in training and testing splits do not come from the same dog. Hence, this is subject-independent~\cite{meegahapola2022sensing}. We conducted all experiments with five iterations and K = 4 --- that is, in each experiment, four dogs were left out for the testing set such that this set contains instances from both high and low scores of each personality trait. This allowed us to measure the mean and standard deviation of the models' performance across iterations. Given the small sample size, the choice of four dogs was a reasonable one. As for performance metric, we chose the area under the receiver operating characteristic curve (AUC), which is a holistic measure assessing how well a model performs for both classes (i.e., distinguishing high and low traits)~\cite{bradley1997use}. In total, we set up three experiments, and tested four types of models (S1) using a combination of features (S2 and S3).

\begin{itemize}[leftmargin=*,align=left]
    \item S1---Model Types: Given the small dataset size (typical in animal studies \cite{hirskyj2021forming, weiss2013wagtag, ladha2013dog}), we used four types of classifiers: \emph{(a)} Support Vector Machines (SVM)~\cite{noble2006support}, \emph{(b)} Light Gradient Boosting Machine (L-GBM)~\cite{ke2017lightgbm}, \emph{(c)} Naive Bayes (NB)~\cite{webb2010naive}, and \emph{(d)} Random Forest (RF)~\cite{cutler2011random}.
    \item S2---Feature Types: As previously mentioned (Section~~\ref{subsec:data_processing_pipeline}), we generated two main types of features from the inertial data: activity-level features ({ACT}) and statistical features ({STAT}). The former set of features is typically more interpretable but costly to obtain due to the power processing needed to generate the features, while the latter set of features is not computationally expensive but it is less interpretable. In addition, we used demographic attributes ({DEM}) such as sex of the dog, weight, age, neutered, and training rating as input to the model because prior work suggested connections between these attributes and dog personality (Section~\ref{sec:related_work}). As the dog owner's personality has been previously linked to the dog's personality and well-being~\cite{kubinyi2009dog, hoglin2021long}, we also used the dog owner's information ({O-INFO}) including their sex and personality traits captured from the Personality Inventory~\cite{gosling2003very}.
    \item S3---Time of Day: These features were computed for three time periods of the day (i.e., night, morning, and afternoon). In S1 and S2, we used all the features. As mentioned in Section~\ref{sec:statistical_analysis}, the same feature captured at different times of day may have differences in statistical significance values. For example, sleeping time in the morning (M) could be informative to discriminate high and low levels of Extraversion, while sleeping time in the afternoon (A) might not be. Hence, in this set of experiments, we incorporated the time period of the day, and sought to understand whether developing separate models for different time periods yield better performance. For example, if a model trained with data from only morning features performs better, it would mean that we only need six hours of data from a dog to perform the inference. 
\end{itemize}
\section{Results}
\label{sec:inference}

\begin{table*}
    \caption{{t-statistic (p-value=<0.05:*; =<0.5:**; >0.5: ***) and Cohen's-d (95\% confidence intervals do not overlapping with zero: *) of activity-level and statistical features for personality traits in DPQ and MCPQ-R:} Night (12am-6am), Morning (6am-12pm), and Afternoon (12pm-6pm) are denoted by N, M, and A, respectively; t-statistic and Cohen's-d values are sorted in descending order, with the highest value of each trait in boldface. Results for activity-level features (left) and statistical features (right) are shown separately. Standard notation min, max, mean, median, std were used for minimum, maximum, mean, median, and standard deviation of the signal, respectively. Other notations include: acceleration -- acceleration value was calculated in Section~\ref{subsec:data_processing_pipeline}, acc -- accelerometer, gyro -- gyroscope, x,y,z -- axes of the accelerometer and gyroscope, \% -- percentage of time spent doing a particular activity, ecdf -- empirical cumulative distribution function.}
    \label{tab:tstatistics}
    \resizebox{1.02\textwidth}{!}{%
    \begin{tabular}{l l r l r | l r l r}
     
     &
     \multicolumn{4}{c}{\cellcolor[HTML]{FFFFFF}\textbf{Activity-Level}} &
     \multicolumn{4}{c}{\cellcolor[HTML]{FFFFFF}\textbf{Statistical}} 
     \\
     
    \cmidrule{2-9}

    &
    \multicolumn{8}{c}{\cellcolor[HTML]{FFFFFF}\textbf{DPQ}}
    \\

    \cmidrule{2-9}

    & 
    &
    \textit{t-statistic} &
    &
    \textit{Cohen's-d} &
    &
    \textit{t-statistic} &
    &
    \textit{Cohen's-d} 
    \\

    %
    %
    
    \multirow{5}{1.9cm}{\textbf{Fearfulness}} &
    \textbf{sedentary \% (A)}        & \textbf{(-) 3.55*}    & 
    \textbf{sedentary \% (A)}        & \textbf{0.88*}        & 
    \textbf{gyro z histogram\_5 (N)} &  \textbf{(+) 17.92*} & 
    \textbf{gyro z histogram\_5 (N)} & \textbf{4.46*} 
    \\

    &
    light \% (N)        & (-) 3.12*    & 
    light \% (N)        & 0.78*        & 
    gyro z histogram\_5 (M) &  (+) 13.07* & 
    gyro z histogram\_5 (M) & 3.25* 
    \\
    
    &
    acceleration std (M) & (+) 2.46*    & 
    acceleration std (M) & 0.61*        & 
    gyro z histogram\_6 (N) &  (+) 12.86* & 
    gyro z histogram\_6 (N) & 3.21* 
    \\
    
    &
    sedentary \% (M)        & (-) 2.43*    & 
    sedentary \% (M)        & 0.60*        & 
    gyro z zero\_crossing\_rate (N) &  (+) 9.94* & 
    gyro z zero\_crossing\_rate (N) & 2.47* 
    \\
    
    &
    sleep \% (A)        & (+) 2.37*    & 
    acceleration std (A)        & 0.59*        & 
    gyro z histogram\_6 (M) &  (+) 9.56* & 
    gyro z histogram\_6 (M) & 2.38* 
    \\

    \arrayrulecolor{Gray}
    \cmidrule{2-9}

    %
    %
    
    \multirow{5}{1.9cm}{\textbf{Aggression Towards People}} &
    \textbf{light \% (M)}        & \textbf{(+) 1.74**}    & 
    \textbf{light \% (M)}        & \textbf{0.43}        & 
    \textbf{gyro y histogram\_8 (N)} &  \textbf{(-) 4.44*} & 
    \textbf{gyro y histogram\_8 (N)} & \textbf{1.11*} 
    \\
    
    &
    light \% (A)        & (+) 1.31**    & 
    light \% (A)        & 0.32        & 
    gyro y histogram\_4 (N) &  (+) 4.40* & 
    gyro y histogram\_4 (N) & 1.09* 
    \\
    
    &
    sleep \% (M)        & (-) 1.16***    & 
    sleep \% (M)        & 0.29        & 
    acc x histogram\_5 (M) &  (-) 4.38* & 
    gyro y histogram\_8 (M) & 1.09* 
    \\
    
    &
    moderate-vigorous \% (N)        & (-) 1.08***    & 
    moderate-vigorous \% (N)         & 0.27        & 
    gyro y histogram\_8 (M) &  (-) 4.35* & 
    acc x histogram\_5 (M) & 1.09* 
    \\
    
    &
    light \% (N)        & (+) 0.72***    & 
    light \% (N)         & 0.17*        & 
    gyro y histogram\_7 (M) &  (-) 4.33* & 
    gyro y histogram\_7 (M) & 1.08* 
    \\

    \arrayrulecolor{Gray}
    \cmidrule{2-9}

    %
    %
    
    \multirow{5}{1.9cm}{\textbf{Excitability}} &
    \textbf{sedentary \% (A)}        & \textbf{(+) 2.05*}    & 
    \textbf{sedentary \% (A)}        & \textbf{0.51*}        & 
    \textbf{gyro z histogram\_5 (M)} &  \textbf{(-)v5.86*} & 
    \textbf{gyro z histogram\_5 (M)} & \textbf{1.45*} 
    \\
    
    &
    acceleration std (A)     & (-) 1.50**    & 
    acceleration std (A)     & 0.37        & 
    gyro z histogram\_5 (N) &  (-) 5.85* & 
    gyro z histogram\_5 (N) & 1.45* 
    \\
    
    &
    light \% (M)        & (-) 1.44**    & 
    light \% (M)          & 0.36        & 
    gyro z zero\_crossing\_rate (M) &  (-) 5.67* & 
    gyro z zero\_crossing\_rate (M) & 1.41* 
    \\
    
    &
    sedentary \% (M)        & (+) 1.32**    & 
    sedentary \% (M)      & 0.33        & 
    gyro z zero\_crossing\_rate (N) &  (-) 5.63* & 
    gyro z zero\_crossing\_rate (N) & 1.40* 
    \\
    
    &
    acceleration max (A)        & (-) 1.27*    & 
    acceleration max (A)        & 0.32        & 
    gyro y zero\_crossing\_rate (N) &  (-) 5.54* & 
    gyro y zero\_crossing\_rate (N) & 1.37* 
    \\

    \arrayrulecolor{Gray}
    \cmidrule{2-9}

    %
    %
    
    \multirow{5}{1.9cm}{\textbf{Responsiveness to Training}} &
    \textbf{light \% (M)}        & \textbf{(-) 2.17*}    & 
    \textbf{light \% (M)}        & \textbf{0.52*}        & 
    \textbf{acc x mean (M)} &  \textbf{(+) 4.06*} & 
    \textbf{acc x mean (M)} & \textbf{0.99*} 
    \\
    
    &
    light \% (A)        & (-) 1.76**    & 
    light \% (A)        & 0.42        & 
    gyro x ecdf\_\_perc\_0 (M) &  (+) 3.87* & 
    gyro x histogram\_8 (A) & 0.97* 
    \\
    
    &
    sleep \% (M)        & (+) 1.52**    & 
    sleep \% (M)        & 0.38        & 
    gyro z ecdf\_\_perc\_0 (M) &  (+) 3.81* & 
    gyro x histogram\_7 (A) & 0.95* 
    \\
    
    &
    light \% (N)        & (-) 1.08**    & 
    acceleration std (A)        & 0.27        & 
    gyro z median (M) &  (+) 3.73* & 
    gyro z histogram\_8 (M) & 0.95* 
    \\
    
    &
    acceleration std (A)        & (+) 1.06**    & 
    light \% (N)        & 0.26        & 
    gyro z median (N) &  (+) 3.71** & 
    gyro x ecdf\_\_perc\_0 (N) & 0.94* 
    \\

    \arrayrulecolor{Gray}
    \cmidrule{2-9}

    %
    %
    
    \multirow{5}{1.9cm}{\textbf{Aggression Towards Animals}} &
    \textbf{sedentary \% (M)}        & \textbf{(+) 2.80*}    & 
    \textbf{sedentary \% (M)}       & \textbf{0.69*}        & 
    \textbf{acc y neighbourhood\_peaks (M)}        & \textbf{(+) 5.43*}  & 
    \textbf{acc y neighbourhood\_peaks (M)}   & \textbf{1.35*} 
    \\
    
    &
    sleep \% (M)        & (-) 2.61*    & 
    sleep \% (M)        & 0.65*        & 
    acc z neighbourhood\_peaks (M)  &  (+) 5.20* & 
    acc z neighbourhood\_peaks (M) & 1.29* 
    \\
    
    &
    acceleration mean (M)        & (+) 2.33*    & 
    acceleration mean (M)        & 0.58*        & 
    gyro y zero\_crossing\_rate (M) &  (+) 5.13* & 
    gyro y zero\_crossing\_rate (M) & 1.27* 
    \\
    
    &
    acceleration median (M)        & (+) 2.32*    & 
    acceleration median (M)        & 0.58*        & 
    gyro x neighbourhood\_peaks (M) &  (+) 5.06** & 
    gyro x neighbourhood\_peaks (M) & 1.25* 
    \\
    
    &
    acceleration median (N)        & (+) 2.32*    & 
    acceleration median (N)       & 0.58*        & 
    gyro x ecdf\_0 (M) &  (-) 4.99* & 
    gyro z ecdf\_percentile\_count\_0 (M) & 1.24* 
    \\

    
    \arrayrulecolor{black}
    \cmidrule{2-9}
    
      &
     \multicolumn{8}{c}{\cellcolor[HTML]{FFFFFF}\textbf{MCPQ-R}}
     \\

    \arrayrulecolor{black}
    \cmidrule{2-9}

    & 
    &
    \textit{t-statistic} &
    &
    \textit{Cohen's-d} &
    &
    \textit{t-statistic} &
    &
    \textit{Cohen's-d} 
    \\

    %
    %
    
    \multirow{5}{1.9cm}{\textbf{Extraversion}} &
    \textbf{acceleration min (M)}        & \textbf{(+) 2.34*}    & 
    \textbf{acceleration min (M)}        & \textbf{0.60*}        & 
    \textbf{gyro x histogram\_4 (N)} &  \textbf{(-) 4.86*} & 
    \textbf{gyro x histogram\_4 (N)} & \textbf{1.24*} 
    \\
    
    &
    acceleration max (A)           & (+) 2.19*    & 
    acceleration max (A)    & 0.56*        & 
    gyro x zero\_crossing\_rate (N) &  (-) 4.23* & 
    gyro x zero\_crossing\_rate (N) & 1.08* 
    \\
    
    &
    acceleration min (A)        & (+) 2.09*    & 
    acceleration min (A)        & 0.54*        & 
    acc y histogram\_8 (M) &  (-) 3.71* & 
    acc y histogram\_1 (A) & 0.94* 
    \\
    
    &
    acceleration min (N)        & (+) 1.96**    & 
    acceleration min (N)    & 0.50*        & 
    acc y histogram\_1 (A) &  (+) 3.68* & 
    acc y histogram\_8 (M) & 0.92* 
    \\
    
    &
    acceleration mean (N)        & (+) 1.95**    & 
    acceleration mean (N)    & 0.50*        & 
    gyro y auc (M) &  (-) 3.45* & 
    gyro y auc (M) (N) & 0.86* 
    \\
    
    \arrayrulecolor{Gray}
    \cmidrule{2-9}

    %
    %
    
    \multirow{5}{1.9cm}{\textbf{Motivation}} &
    \textbf{sedentary \% (M)}        & \textbf{(+) 3.22*}    & 
    \textbf{sedentary \% (M)}        & \textbf{0.80*}        & 
    \textbf{gyro z histogram\_7 (M)} &  \textbf{(-) 4.45*} & 
    \textbf{gyro z histogram\_7 (M)} & \textbf{1.10*} 
    \\
    
    &
    acceleration std (M)        & (-) 3.11*    & 
    acceleration std (M)        & 0.76*        & 
    gyro z histogram\_6 (M) &  (-) 4.25** & 
    gyro z histogram\_6 (M) & 1.06* 
    \\
    
    &
    sleep \% (A)        & (-) 2.58*    & 
    acceleration min (M)        & 0.65*        & 
    gyro z histogram\_9 (N) &  (-) 4.18** & 
    gyro z median (M) & 1.05* 
    \\
    
    &
    acceleration min (M)        & (+) 2.57*    & 
    sleep \% (A)        & 0.64*        & 
    gyro z median (M) &  (-) 4.18* & 
    gyro z median (N) & 1.05* 
    \\
    
    &
    sleep \% (M)        & (-) 2.40*    & 
    sleep \% (M)        & 0.60*        & 
    gyro z histogram\_7 (A) &  (-) 4.18* & 
    gyro z median (A) & 1.04* 
    \\

    \arrayrulecolor{Gray}
    \cmidrule{2-9}

    %
    %
    
    \multirow{5}{1.9cm}{\textbf{Training Focus}} &
    \textbf{light \% (A)}        & \textbf{(-) 1.88**}    & 
    \textbf{light \% (A)}        & \textbf{0.46}        & 
    \textbf{gyro x histogram\_4 (M)} &  \textbf{(+) 4.74*} & 
     \textbf{gyro x histogram\_4 (M)} & \textbf{1.17*} 
    \\
    
    &
    acceleration std (M)        & (-) 1.80**    & 
    acceleration std (M)        & 0.44        & 
    acc x histogram\_5 (M) &  (+) 4.71* & 
    acc x histogram\_5 (M) & 1.17* 
    \\
    
    &
    moderate-vigorous \% (M)        & (-) 1.69**    & 
    moderate-vigorous \% (M)        & 0.41        & 
    acc x mean (M) &  (+) 4.56* & 
    acc x mean (M) & 1.12* 
    \\
    
    &
    sedentary \% (A)        & (-) 1.59**    & 
    sedentary \% (A)        & 0.40        & 
    acc x median (M)        & (+) 3.98* & 
    acc x median (M) & 0.98* 
    \\
    
    &
    light \% (M)        & (-) 1.59**   & 
    light \% (M)        & 0.39        & 
    gyro y negative\_turning\_points (M) &  (+) 3.88* & 
    gyro y negative\_turning\_points (M) & 0.96* 
    \\

    \arrayrulecolor{Gray}
    \cmidrule{2-9}

    %
    %
    
    \multirow{5}{1.9cm}{\textbf{Amicability}} &
    \textbf{sleep \% (N)}        & \textbf{(-) 3.80*}    & 
    \textbf{sleep \% (N)}        & \textbf{0.97*}        & 
    \textbf{gyro y zero\_crossing\_rate (N)} &  \textbf{(-) 9.79*} & 
    \textbf{gyro y zero\_crossing\_rate (N)} & \textbf{2.42*} 
    \\
    
    &
    sleep \% (M)        & (-) 3.44*    & 
    sleep \% (M)        & 0.87*        & 
    gyro y zero\_crossing\_rate  (M) &  (-) 9.15* & 
    gyro y zero\_crossing\_rate (M) & 2.30* 
    \\
    
    &
    sedentary \% (M)        & (+) 3.29*    & 
    light \% (M)        & 0.84*        & 
    gyro y histogram\_4 (M) &  (-) 9.03* & 
    gyro y histogram\_4 (M) & 2.22* 
    \\
    
    &
    light \% (M)        & (+) 3.20*    & 
    sedentary \% (M)        & 0.83*        & 
    gyro y histogram\_4 (N) &  (-) 8.89* & 
    gyro y histogram\_4 (N) & 2.21* 
    \\
    
    &
    accleration min (M)        & (+) 3.17*    & 
    accleration min (M)        & 0.83*        & 
    gyro y histogram\_5 (M) &  (-) 7.82* & 
    gyro y histogram\_5 (M) & 1.94* 
    \\

    \arrayrulecolor{Gray}
    \cmidrule{2-9}

    %
    %
    
    \multirow{5}{1.9cm}{\textbf{Neuroticism}} &
    \textbf{acceleration std (M)}        & \textbf{(+) 3.02*}    & 
    \textbf{acceleration std (M)}      & \textbf{0.75*}        & 
    \textbf{gyro z histogram\_8 (N)} &  \textbf{(+) 6.19*} & 
    \textbf{gyro z histogram\_8 (N)} & \textbf{1.53*} 
    \\
    
    &
    moderate-vigorous \% (M)        & (+) 2.72*    & 
    acceleration min (M)       & 0.67*        & 
    gyro z histogram\_7 (N) &  (+) 5.80* & 
    gyro z histogram\_7 (N) & 1.44* 
    \\

    &
    acceleration min (M)        & (-) 2.69*    & 
    moderate-vigorous \% (M)     & 0.67*        & 
    gyro z histogram\_9 (N) &  (+) 5.78* & 
    gyro z histogram\_9 (N) & 1.43* 
    \\
    
    &
    acceleration max (A)        & (-) 2.49*    & 
    acceleration max (A)     & 0.62*        & 
    gyro z histogram\_9 (M) &  (+) 5.53* & 
    gyro z histogram\_9 (M) & 1.37* 
    \\
    
    &
    acceleration min (A)       & (-) 2.44*    & 
    acceleration min \% (A)       & 0.61*        & 
    gyro z ecdf\_percentile\_1 (M) &  (+) 5.43* & 
    gyro z ecdf\_percentile\_1 (M) & 1.35* 
    \\
    
    \arrayrulecolor{black}
    \hline 
    
    \end{tabular}
    }
\end{table*}

\begin{table*}
    \caption{{Inference results for each personality trait in DPQ and MCPQ-R for different types of models:} Mean ($\bar{S}$) and Standard Deviation ($S_\sigma$) area under the receiver operating characteristic curve (AUC) computed from five iterations. Results are presented as $\bar{S} (S_\sigma)$, where $S$ is AUC score, and the highest performing model is marked in bold text. SVM: Support Vector Machines; L-GBM: Light Gradient Boosting Machine; NB: Naive Bayes; RF: Random Forest.}
    \label{tab:accuracies_model_types}
    \resizebox{\textwidth}{!}{%
    \begin{tabular}{c c c c c c}

    &
    \multicolumn{5}{c}{\cellcolor[HTML]{FFFFFF}\textbf{DPQ}} 
     \\
    
    \arrayrulecolor{black}
    \cmidrule{2-6} 
    
    &
    \multicolumn{1}{c}{\cellcolor[HTML]{FFFFFF}\textbf{Fearfulness}} &
    \multicolumn{1}{c}{\cellcolor[HTML]{FFFFFF}\textbf{Aggression Towards People}} &
    \multicolumn{1}{c}{\cellcolor[HTML]{FFFFFF}\textbf{Excitability}} &
    \multicolumn{1}{c}{\cellcolor[HTML]{FFFFFF}\textbf{Responsiveness to Training}} &
    \multicolumn{1}{c}{\cellcolor[HTML]{FFFFFF}\textbf{Aggression Towards Animals}} 
    \\
    
    \arrayrulecolor{Gray}
    \hline

    \textbf{Baseline} &
    .50 (.00) & 
    .50 (.00) & 
    .50 (.00) & 
    .50 (.00) & 
    .50 (.00)  
    \\
    
    \arrayrulecolor{Gray}
    \hline 
    
    \textbf{SVM} &
    .71 (.02) & 
    .66 (.05) & 
    .59 (.03) & 
    \textbf{.72 (.06)} & 
    .63 (.09)  
    \\
    
    \textbf{L-GBM} &
    .76 (.09) & 
    \textbf{.68 (.06)} & 
    .61 (.08) & 
    .66 (.03) & 
    .59 (.10)  
    \\
    
    \textbf{NB} &
    .71 (.05) & 
    .63 (.08) & 
    \textbf{.62 (.02)} & 
    .65 (.03) & 
    .59 (.04)  
    \\
    
    \textbf{RF} &
    \textbf{.78 (.07)} & 
    .65 (.08) & 
    .62 (.08) & 
    .70 (.09f) & 
    \textbf{.68 (.06)}  
    \\
    
    \arrayrulecolor{Gray}
    \hline

    &
    \multicolumn{5}{c}{\cellcolor[HTML]{FFFFFF}\textbf{MCPQ-R}} 
    \\
    
    \arrayrulecolor{black}
    \cmidrule{2-6} 
    
    &
    \multicolumn{1}{c}{\cellcolor[HTML]{FFFFFF}\textbf{Extraversion}} &
    \multicolumn{1}{c}{\cellcolor[HTML]{FFFFFF}\textbf{Motivation}} &
    \multicolumn{1}{c}{\cellcolor[HTML]{FFFFFF}\textbf{Training Focus}} &
    \multicolumn{1}{c}{\cellcolor[HTML]{FFFFFF}\textbf{Amicability}} &
    \multicolumn{1}{c}{\cellcolor[HTML]{FFFFFF}\textbf{Neuroticism}} 
    \\
    
    \arrayrulecolor{Gray}
    \hline 
    
    \textbf{Baseline} &
    .50 (.00) & 
    .50 (.00) & 
    .50 (.00) & 
    .50 (.00) & 
    .50 (.00)  
    \\
    
    \arrayrulecolor{Gray}
    \hline 
    
    \textbf{SVM} &
    \textbf{.64 (.01)} & 
    .70 (.02) & 
    .73 (.06) & 
    .67 (.03) & 
    .62 (.09)  
    \\
    
    \textbf{L-GBM} &
    .62 (.07) & 
    .69 (.11) & 
    .70 (.08) & 
    .71 (.05) & 
    .70 (.08)  
    \\
    
    \textbf{NB} &
    .64 (.11) & 
    .72 (.05) & 
    .81 (.02) & 
    .70 (.08) & 
    .70 (.02)  
    \\
    
    \textbf{RF} &
    .59 (.07) & 
    \textbf{.76 (.07)} & 
    \textbf{.89 (.13)} & 
    \textbf{.74 (.08)} & 
    \textbf{.73 (.11)} 
    \\
    
    \arrayrulecolor{Gray}
    \hline 
    
    \end{tabular}%
    }
\end{table*}

\begin{table*}
    \caption{{Random Forest inference results for each personality trait in DPQ and MCPQ-R for different types of features:} Mean ($\bar{S}$) and Standard Deviation ($S_\sigma$) Area Under the receiver operating characteristic Curve (AUC) computed from five iterations. Results are presented as $\bar{S} (S_\sigma)$, where $S$ is AUC, and the highest performing model is marked in bold text. ACT: activity level features; STAT: statistical features; DEM: dog demographic attributes including its sex, weight, age, training rating, and whether neutered; O-INFO: dog owner's sex and personality traits.}
    \label{tab:accuracies_feature_types}
    \resizebox{\textwidth}{!}{%
    \begin{tabular}{c c c c c c}

    &
    \multicolumn{5}{c}{\cellcolor[HTML]{FFFFFF}\textbf{DPQ}}
    \\
    
    \cmidrule{2-6}

    
    &
    \multicolumn{1}{c}{\cellcolor[HTML]{FFFFFF}\textbf{Fearfulness}} &
    \multicolumn{1}{c}{\cellcolor[HTML]{FFFFFF}\textbf{Aggression Towards People}} &
    \multicolumn{1}{c}{\cellcolor[HTML]{FFFFFF}\textbf{Excitability}} &
    \multicolumn{1}{c}{\cellcolor[HTML]{FFFFFF}\textbf{Responsiveness to Training}} &
    \multicolumn{1}{c}{\cellcolor[HTML]{FFFFFF}\textbf{Aggression Towards Animals}} 
    \\
    
    \arrayrulecolor{Gray}
    \hline

    \textbf{B1} &
    .50 (.00) & 
    .50 (.00) & 
    .50 (.00) & 
    .50 (.00) & 
    .50 (.00)  
    \\
    
    \textbf{B2: O-INFO} &
    .54 (.10) & 
    .48 (.15) & 
    .40 (.18) & 
    .52 (.11) & 
    .49 (.09)  
    \\
    
    \textbf{B3: DEM} &
    .57 (.08) & 
    .49 (.06) & 
    .48 (.08) & 
    .55 (.07) & 
    .50 (.12)  
    \\
    
    \arrayrulecolor{Gray}
    \hline 
    
    \textbf{G1: ACT} &
    .67 (.05) & 
    .47 (.02) & 
    .60 (.06) & 
    .53 (.08) & 
    .56 (.05)  
    \\
    
    \textbf{G2: STAT} &
    .55 (.05) & 
    .47 (.13) & 
    .25 (.09) & 
    .58 (.07) & 
    .53 (.04)  
    \\
    
    \textbf{G3: ACT+DEM} &
    \textbf{.80 (.11)} & 
    .63 (.13) & 
    .47 (.12) & 
    .61 (.15) & 
    .67 (.05)  
    \\
    
    \textbf{G4: STAT+DEM} &
    .63 (.08) & 
    .57 (.11) & 
    .33 (.11) & 
    \textbf{.70 (.04)} & 
    .59 (.04)  
    \\
    
    \textbf{G5: ACT+STAT} &
    .78 (.07) & 
    .65 (.08) & 
    \textbf{.62 (.08)} & 
    .70 (.09) & 
    \textbf{.68 (.06)}  
    \\
    
    \textbf{G6: ACT+STAT+DEM} &
    .61 (.03) & 
    \textbf{.65 (.04)} & 
    .51 (.09) & 
    .61 (.09) & 
    .51 (.06)  
    \\
    
    \hline

    &
    \multicolumn{5}{c}{\cellcolor[HTML]{FFFFFF}\textbf{MCPQ-R}}
    \\
    
    \arrayrulecolor{black}
    \cmidrule{2-6} 
    
    &
    \multicolumn{1}{c}{\cellcolor[HTML]{FFFFFF}\textbf{Extraversion}} &
    \multicolumn{1}{c}{\cellcolor[HTML]{FFFFFF}\textbf{Motivation}} &
    \multicolumn{1}{c}{\cellcolor[HTML]{FFFFFF}\textbf{Training Focus}} &
    \multicolumn{1}{c}{\cellcolor[HTML]{FFFFFF}\textbf{Amicability}} &
    \multicolumn{1}{c}{\cellcolor[HTML]{FFFFFF}\textbf{Neuroticism}} 
    \\
    
    \arrayrulecolor{Gray}
    \hline 
    
    \textbf{B1} &
    .50 (.00) & 
    .50 (.00) & 
    .50 (.00) & 
    .50 (.00) & 
    .50 (.00)  
    \\
    
    \textbf{B2: O-INFO} &
    .43 (.02) & 
    .52 (.04) & 
    .56 (.03) & 
    .47 (.10) & 
    .51 (.11)  
    \\
    
    \textbf{B3: DEM} &
    .44 (.05) & 
    .54 (.08) & 
    .56 (.04) & 
    .49 (.09) & 
    .52 (.10)  
    \\
    
    \arrayrulecolor{Gray}
    \hline 
    
    \textbf{G1: ACT} &
    .41 (.03) & 
    .63 (.06) & 
    .47 (.06) & 
    .41 (.04) & 
    .57 (.02)  
    \\
    
    \textbf{G2: STAT} &
    .64 (.06) & 
    .36 (.11) & 
    .34 (.04) & 
    .71 (.05) & 
    .69 (.09)  
    \\
    
    \textbf{G3: ACT+DEM}  &
    .40 (.09) & 
    \textbf{.77 (.09)} & 
    \textbf{.89 (.08)} & 
    .38 (.07) & 
    .47 (.02)  
    \\
    
    \textbf{G4: STAT+DEM} &
    \textbf{.63 (.04)} & 
    .33 (.08) & 
    .43 (.06) & 
    .70 (.08) & 
    .67 (.09)  
    \\
    
    \textbf{G5: ACT+STAT} &
    .59 (.07) & 
    .76 (.07) & 
    .89 (.13) & 
    \textbf{.74 (.08)} & 
    \textbf{.73 (.11)}  
    \\
    
    \textbf{G6: ACT+STAT+DEM} &
    .62 (.04) & 
    .33 (.08) & 
    .43 (.07) & 
    .63 (.06) & 
    .70 (.09)  
    \\
    
    \arrayrulecolor{Gray}
    \hline 
    
    \end{tabular}%
    }
\end{table*}

\subsection{Activity Level and Statistical Features Discriminating Dog Personality}\label{subsec:statistical_analysis}
The statistical features showed high $t$-statistic, low $p$-values, and high Cohen's-$d$ values for a majority of dog personality traits on both questionnaires (Table~\ref{tab:tstatistics}). Most of these features were derived from gyroscope, and were captured during the morning. The activity-level features also showed high Cohen's-d values, but not as high as the statistical ones. 

\begin{itemize}[leftmargin=*,align=left]
\item DPQ: For \emph{Fearfulness}, the percentage of time spent in sedentary activity in the afternoon had a Cohen's-d of 0.88, and the amount time of time spent doing light activity at night had a Cohen's-d of 0.78; both of which have a large effect size. This translates into saying that a dog's sedentary activity in the afternoon or light activity at night are both informative of high vs. low levels of fearfulness.
These were in fact the highest Cohen's-d obtained for any activity-level features. In general, the lowest Cohen's-d values came for \emph{Aggression Towards People}. For that trait, light activity percentage at night, morning, and afternoon had Cohen's-d of 0.17, 0.43, and 0.32, respectively, while moderate-vigorous activity level during night had a Cohen's-d of 0.27. These features had above small effect size with low reliability (because the 95\% confidence interval was crossing zero in many cases), thus no association could have been drawn.

\item MCPQ-R: Across all traits, the statistical features had comparably similar values for $t$-statistic and Cohen's-$d$ to those previously obtained by DPQ. However, activity-level features had Cohen's-d for MCPQ-R higher compared to those for DPQ, showing a better discriminative capability of high class vs. low class for MCPQ-R traits compared to DPQ traits. In particular, statistical features capturing acceleration (i.e., dog movements throughout the day) had Cohen's-d values above 0.5, suggesting a link between Extraversion and high levels of activity; a finding in line with previous work~\cite{carrier2013exploring}.

\end{itemize}

These results suggest that both activity levels and the statistical features have discriminative power to various degrees, allowing us to draw conclusions on which features are associated with dog personality (\textbf{RQ\textsubscript{2}}).

\begin{figure*}
\captionsetup{labelfont=normalfont}
\centerline{\includegraphics[width=0.99\linewidth]{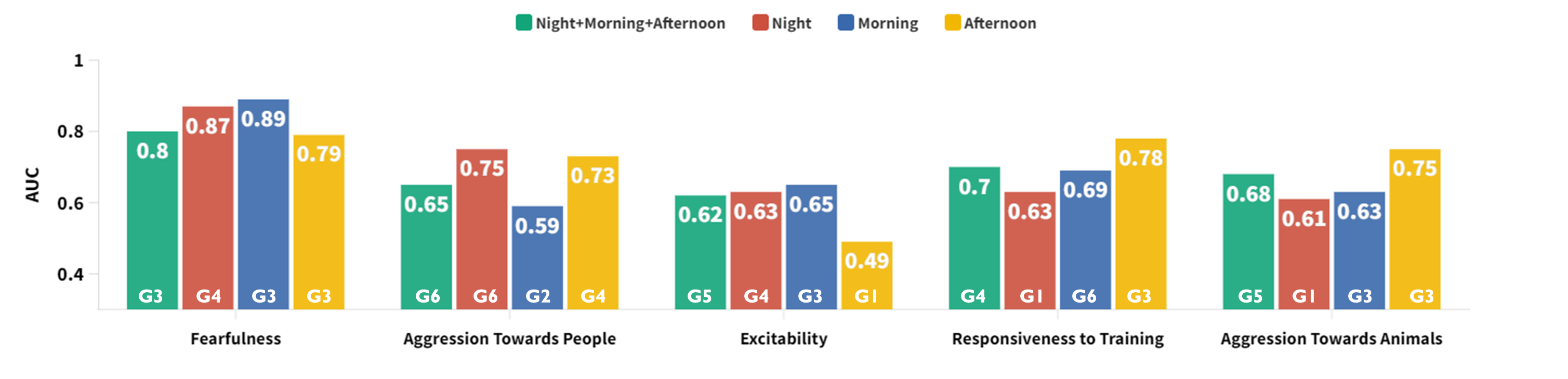}}
\caption{{Area Under the receiver operating characteristic Curve (AUC) score comparison for DPQ traits with models that used features from:} night (N); morning (M); afternoon (A); and all time periods (N+M+A). Feature type combinations (G1-G6 from Table~\ref{tab:accuracies_feature_types}) that provided the AUC score is marked in white color at the bottom of each bar.} 

\label{fig:nma_dpq_results}
\end{figure*}

\begin{figure*}
\captionsetup{labelfont=normalfont}
\centerline{\includegraphics[width=0.99\linewidth]{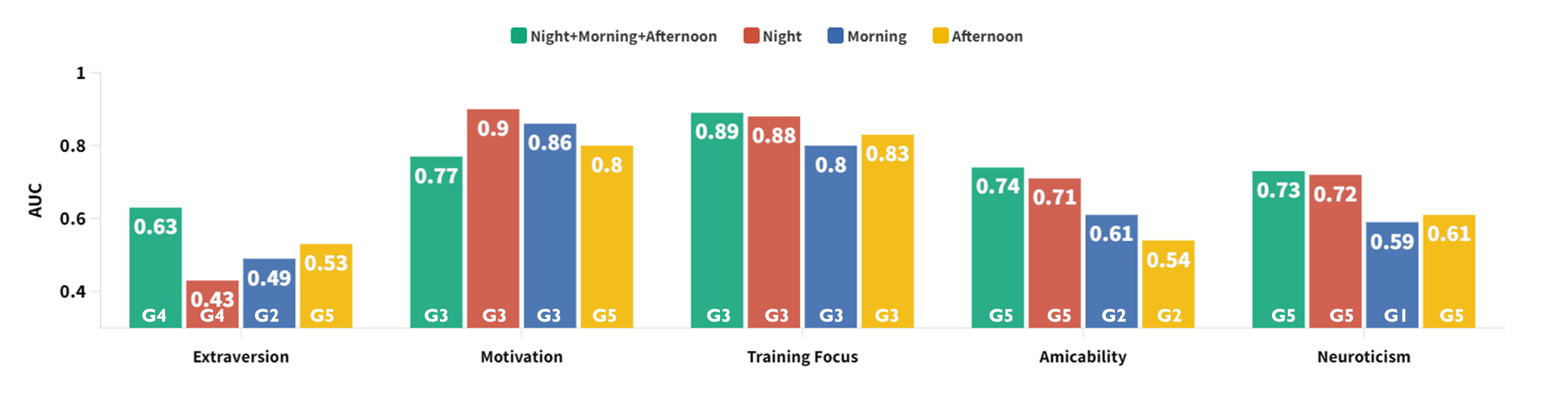}}
\caption{{Area Under the receiver operating characteristic Curve (AUC) score comparison for MCPQ-R traits with models that used features from:} night (N); morning (M); afternoon (A); and all time periods (N+M+A). Feature type combinations (=G1-G6 from Table~\ref{tab:accuracies_feature_types}) that provided the AUC score is marked in white color at the bottom of each bar.}
\label{fig:nma_mcpqr_results}
\end{figure*}
 
\subsection{Predicting Dog Personality} 
Table~\ref{tab:accuracies_model_types} shows the classification results for four model types predicting dog personality from both activity levels and statistical features. The columns show the performance for each personality trait, while the rows show model types. The baseline for all experiments is 0.5 as the testing sets were balanced~\cite{meegahapola2022sensing}. For DPQ, the best AUCs were obtained with Random Forest (in the range of 0.62-0.80), so for MCPQ-R (in the range of 0.63-0.89). AUC was highest for \emph{Training Focus} with a score of 0.89 using Random Forest, and lowest for \emph{Excitability} with a score of 0.59 using SVM. In general, the AUC scores of DPQ questionnaire were lower compared to those of MCPQ-R, meaning that our sensed features are better at classifying MCPQ-R traits (in line with the results in Section~\ref{subsec:statistical_analysis}). Given that Random Forest models showed the best performance for the majority of inferences, for brevity, we only present that model's results in the remainder of this section.

Table~\ref{tab:accuracies_feature_types} shows the performance for various feature type combinations, and two additional baseline models using: \emph{a)} dog owner's sex and personality (B2: O-INFO), and \emph{b)} dog demographics (i.e., sex, weight, age, training rating, and whether neutered) (B3: DEM). Models that used activity-level features alone performed with AUC scores in the range of 0.47-0.67 for DPQ, and 0.41-0.63 for MCPQ-R. Statistical features performed worse than activity-level features with AUC scores in the range of 0.25-0.58 for DPQ, and 0.34-0.71 for MCPQ-R. This suggests that, while being more interpretable, activity-level features offer higher predictive accuracies. When adding dog demographics to the models, their performance increased by considerable margins. The best performance for most dog personality traits was obtained when either one or both sensed feature types (activity-level and statistical features) were combined with demographic features. Overall, DPQ traits had AUC scores in the range of 0.62-0.80, with two traits above 0.70, while MCPQ-R traits had scores in the range of 0.63-0.89, with four traits having scores above 0.70. These results show that a combination of sensed features is predictive of dog personality traits with reasonable AUC scores above 0.70.

When the same features were computed at different times of day, they contributed differently to the predictive power. Figure~\ref{fig:nma_dpq_results} (DPQ) and Figure~\ref{fig:nma_mcpqr_results} (MCPQ-R) show varying performances for different time period-specific models (night, morning, or afternoon) compared to generic models that used the features computed throughout the whole day. For DPQ, models that used morning features were the best for predicting \emph{Fearfulness} and \emph{Excitability}, afternoon features were the best for \emph{Responsiveness to Training} and \emph{Aggressiveness Towards Animals}, and finally, night features were the best for \emph{Aggressiveness Towards People}. For MCPQ-R, models that used period-specific features did not yield better results---with the exception of \emph{Motivation} that yielded an AUC of 0.90 with night features. These results suggest that it would be better to use period-specific models for DPQ, and generic models for MCPQ-R.

Overall, these results suggest that a specific combination of features works best (i.e., using activity levels, statistical features, and demographics together), that the same feature computed at different times of day contributes differently to prediction power, yielding higher accuracy when predicting DPQ traits (\textbf{RQ\textsubscript{3}}).

\begin{figure*}
\captionsetup{labelfont=normalfont}
\centerline{\includegraphics[width=0.9\linewidth]{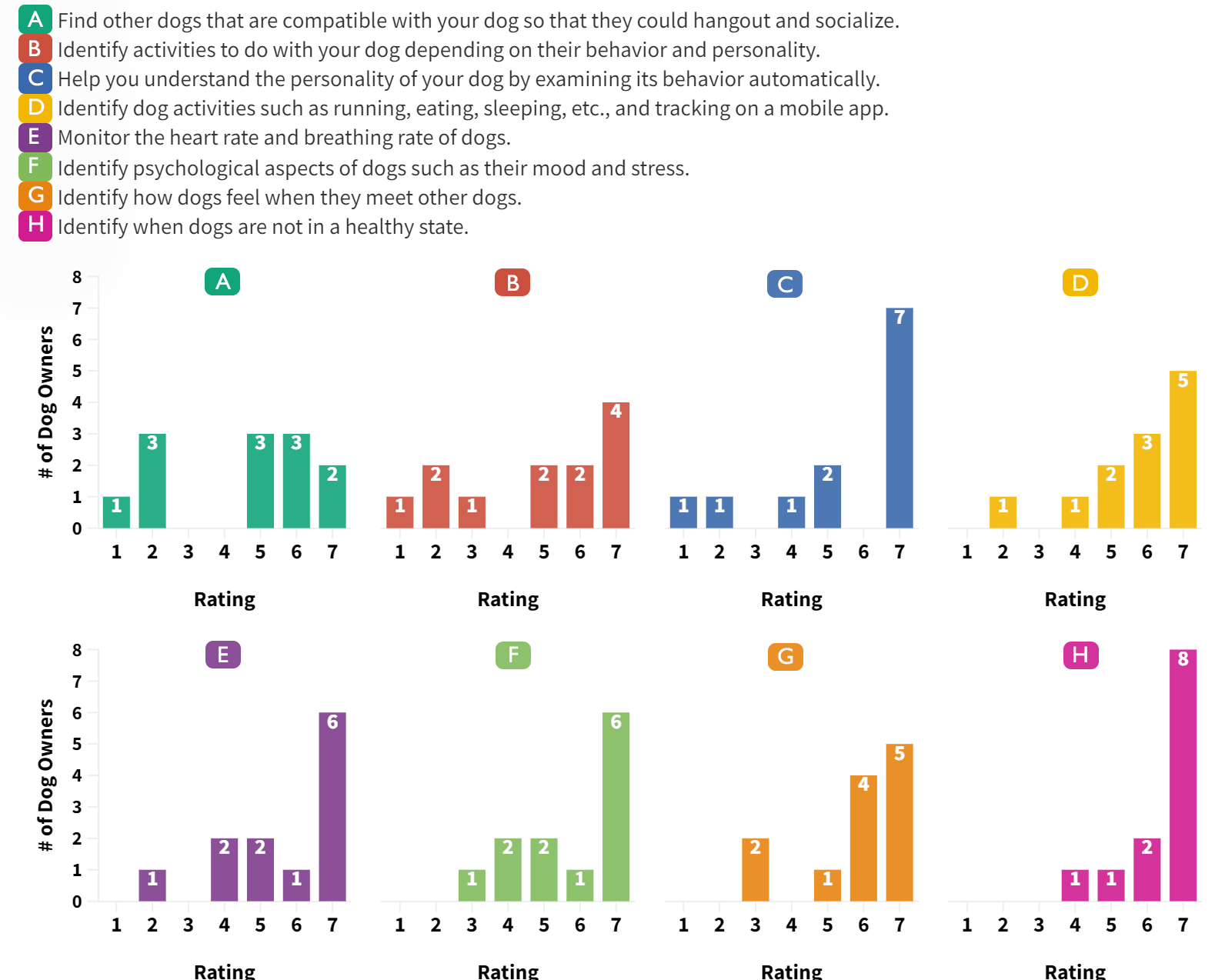}}
\caption{Response distribution from dog owners about features they would like to see in future dog wearables. Most of them mentioned measurements of aspects that are hard to quantify (e.g., psychological and behavioral states as proxies to a dog's health, feelings, mood, stress, and personality).}
\label{fig:owner_responses}
\end{figure*}

\subsection{\textbf{Follow-up Survey with Dog Owners}}\label{subsec:qualitative_analysis}
Dog owners answered a series of open-ended questions (in a free-text form) about the usability of Patchkeeper and a series of Likert-scale questions about the desirability of features in future dog monitoring wearables (Figure~\ref{fig:owner_responses}). We structured their responses into two sections: \emph{Usability of PatchKeeper} and \emph{Future dog monitoring wearables.} As a convention, for dog owners' quotes, we use the letter \textbf{P} (indicating a human participant) followed by the dog ID from Table~\ref{tab:participants} (e.g., \textbf{P1} is the owner of Dog\#1).

\noindent \paragraph{Usability of the PatchKeeper device.} A majority of dog owners had a good experience using the device. \textbf{P5} mentioned that \emph{`` it [PatchKeeper] was easy to use and charge''}, and \textbf{P7} stated that  \emph{``dog didn't mind wearing it''}. While the device was overall easy to use, owners had mixed feelings about its battery life. One said that \emph{``battery life is too short [...] charging it every day was a hassle''} (\textbf{P9}), while another stated that \textit{``battery lasted a full day without an issue''} (\textbf{P2}). 
The split opinions about battery life, however, are very subjective and might be driven by
the charging patterns of other owner's devices. For example, for a person who uses a smartwatch (e.g., Apple or Samsung Galaxy watches) that needs to be charged daily, the Patchkeeper's battery life might seem fair. However, for people who are accustomed to devices with longer battery lives (e.g.,  AmazFit, Fitbit or Garmin), PatchKeeper's battery life would be perceived to be much shorter. In terms of developing tools for dog monitoring, owners provided a range of answers, including health monitoring and understanding their pet better. For health monitoring, \textbf{P12} mentioned that \emph{``such devices can be useful to catch dog problems early [...] like maybe breathing too fast constantly, or cardiac problems''}. For understanding their pet better, \textbf{P3} put it nicely that \textit{``as dogs cannot speak [...] a device that allows my dog to `speak' and `express her feelings' is worth everything''}.

\noindent \paragraph{Future dog monitoring wearables.} Dog owners expressed their opinion about a set of features for future dog wearable devices. They provided answers on a Likert scale of 1 to 7 (1 – strongly not preferred; 3 – not preferred; 5 – preferred; 7 – strongly preferred), and their responses are summarized in Figure~\ref{fig:owner_responses}. Most dog owners (66.67\%) strongly preferred features that would allow them to monitor their dog's health. The second most desirable feature was to understand how dogs feel when they meet other dogs. Moreover, even though it was sixth in terms of the mean score, 58.33\% of dog owners (7 out of 12) gave a rating of seven out of seven for a feature that measured the personality of their dogs. This shows that dog owners were polarized regarding knowing dog personality. Another feature that received a very high rating was the ability of a wearable to track the mood and stress of dogs. In fact, most of the preferred features were about measures that are hard to quantify (e.g., psychological and behavioral states as proxies for a dog's health, feelings, mood, stress, and personality). 
\section{Discussion}
\label{sec:discussion}

\subsection{Summary of the Results}
In summary, our results showed that features captured from the inertial measurement unit along with dog demographic features are predictive of dog personality traits with reasonable AUC scores in the range of 0.62-0.89 (Table~\ref{tab:accuracies_feature_types}) in leave-K-dogs-out setting, with K = 4 (around 33\% of data was used for testing). In addition, as shown in Table~\ref{tab:accuracies_model_types}, we found that random forest classifiers performed best for the majority of inferences. However, results in Figure~\ref{fig:nma_dpq_results} and Figure~\ref{fig:nma_mcpqr_results} showed that, by using separate models for a particular time of day (i.e., night, morning, or afternoon), and by using different feature groups for each model (i.e., different combinations of activity level, statistical, and demographic features), led to increased performance. In particular, we observed an increased performance across all five personality traits under DPQ (AUC scores of 0.89 for \emph{Fearfulness}; 0.75 for \emph{Aggression towards People}; 0.65 for \emph{Excitability}; 0.78 for \emph{Responsiveness to Training}; and 0.75 for \emph{Aggression towards Animals}), while the performance gains for MCPQ-R (AUC scores of 0.63 for \emph{Extraversion}; 0.90 for \emph{Motivation}; 0.89 for \emph{Training Focus}; 0.74 for \emph{Amicability}; and 0.73 for \emph{Neuroticism}) was only visible for one personality trait, that is, Motivation. This highlights that capturing data from a certain period of the day provides better predictive power for certain personality traits, while for other traits, using all available features was a better option. 

\subsection{Implications}
Our work has both theoretical and practical implications. From a theoretical standpoint, our work adds empirical evidence to the growing body of research on animal personality~\cite{gosling2008personality}. We corroborated the previous findings suggesting that more extraverted dogs are associated with higher levels of activity (measured through our device's accelerometer sensor)~\cite{carrier2013exploring}, while amicable dogs engage in light activity (as previous work found~\cite{carrier2013exploring}). When it comes to aggression towards other animals, we found a moderate association with light activity. While previous work associated aggression with higher activity levels, they did so by studying Siberian Husky dogs~\cite{wan2013drd}; a breed not well represented in our sample. Our models also showed high performance for inferring fearfulness. Even though activity levels and fearfulness were not directly linked in previous literature, a possible explanation could be attributed to the relationship between activity levels and negative emotions (or stress), and negative emotions and stress have been associated with fearfulness~\cite{beaudet1994predictive, rayment2015applied, jones2014use}.
Overall, these results corroborate previous findings in the literature. At the same time, our study provides a fine-grained empirical analysis of the relationship between personality traits and activity levels, extending our theoretical understanding. Beyond animal personality research, our work contributes to the Animal-Computer Interaction literature, including recruitment techniques for in-the-wild studies and the use of pre-trained models for animal activity monitoring. In this work, we argue that dog monitoring could go beyond the typical activity level recognition to capture hard-to-quantify psychological aspects such as dog personality. Turning into our data processing pipeline, we showed that state-of-the-art wearable processing pipelines tailored to humans transferred, to a great extent, to our animal study. In a way, this might seem obvious because accelerometer data capture motion. However, only when data from dogs were properly scaled and processed (using a 1G scale) the pipeline and the inferred activities started to work (\S\ref{subsec:activity-level}). This finding holds great promise for future research in dog activity recognition. As for recruitment techniques, it was evident that dog recruitment required the element of trust. Word of mouth and recruiting in physical proximity through NextDoor turned out to be the best technique. Traditional techniques such as mailing lists, distributing leaflets, and posting on social media (Twitter, Facebook, Instagram) worked to a lesser extent. Another challenge was to retain participants. Given the battery life of 24 hours (even though it is similar to consumer-grade wearables such as Apple Watches\footnote{around 18 hours of battery: \href{https://www.apple.com/uk/apple-watch-series-7/}{https://www.apple.com/uk/apple-watch-series-7/}}), dog owners found it time-consuming and cumbersome to charge the device daily. 

From a practical perspective, our findings speak to both dog owners and shelters. A practical application would be a `dog health app' that tracks a dog's behavior patterns over time and detects its personality or psychological aspects, such as the pet's valence, arousal, and stress levels. Such an app could increase dog owners' awareness of their pet's health and allow them to take proactive actions (e.g., walk the pet to reduce its stress levels). In addition to dog health monitoring, our work can be used for dog socializing. Future platforms could offer owners the ability to receive personalized recommendations for their pets. For example, an owner could subscribe to a service wherein tailored pet social activities are recommended, or their pet is matched with another `like-minded' pet. Similar to how dating apps allow like-minded individuals to match, such a platform could offer the same experiences for a dog-dog social matching. Finally, dog shelters could benefit by developing platforms for matching dogs with prospective owners. Currently, matching dogs to owners based on personality is difficult for both small and large shelters because of the sheer amount of effort needed to characterize dog personality. On the one hand, using experts to do personality assessments needs specialized facilities and money. On the other hand, psychological scales are time-consuming, prone to biases, and require someone who knows the dog very well. In contrast, letting dogs wear a device like Patchkeeper for a week (or a few days) and obtaining a data-driven personality assessment could be immensely useful. 

\subsection{Limitations and Future Work}
Our work has several limitations that call for future research. First, we modeled dog personality by considering time-level (i.e., features were extracted by considering the signal for three time periods: night, morning, afternoon) and day-level (i.e., features were extracted by combining the signal of the three time periods) data. This allowed us to obtain several data points of the same dog on different days. In this way, we ensured robustness but also increased the relatively small (N=12 dogs) dataset size for inferences (up to 72 dog days). Second, while we obtained reasonable results with a small sample size, future studies could replicate our methodology with larger sample sizes. However, it is also worth noting that recruiting pet dogs for an in-the-wild study that runs for several days is challenging, and previous work had to resort to similar/lower sample sizes~\cite{hirskyj2021forming, weiss2013wagtag, ladha2013dog}. Additionally, while we enforced the same time schedule for the data collection to obtain comparable results, we acknowledge that different dogs might have different routines. Dogs with similar psychological readings (i.e., personality), but different routines might end up with different physiological readings (i.e., activity levels). Thus future studies could account for dog routines. 

Our findings are based on the assumption that dog activity levels serve as a good proxy for activity types. Given that the range of activity types in dogs is narrower (e.g., walking, eating, sleeping, running~\cite{hussain2022activity}) compared to that of humans, that was a reasonable assumption. While we prompted dog owners to share activity-type labels with supporting pictures and videos with us (e.g., eating, drinking, running, playing), not all owners were compliant, preventing any further analysis. Future studies could attempt to disentangle dog activity types from dog activity levels by building upon our results.
Third, while we resorted to previous literature to control for factors that might have influenced our results (e.g., dog demographics such as sex, age, and neutering~\cite{kubinyi2009dog,lofgren2014management}), future studies may well incorporate additional factors such as the size of the dog's living environment, or even the presence of other pets in that environment. Fourth, all the dogs in the study, are from the same city in the United Kingdom. Thus, whether these results replicate in other cities or countries remains a subject of future work, specially given that the generalization of mobile sensing-based models across countries is an important topic of interest \cite{meegahapola2023generalization, meegahapola2021smartphone}. Fifth, data collection occurred during the summer period and might not be generalizable. Therefore, future studies could explore whether our findings generalize to other seasons (i.e., winter, fall, spring), when dog behaviors vary (especially in countries closer to hemispheres, where weather drastically changes in different seasons\footnote{\url{https://www.pdsa.org.uk/pet-help-and-advice/pet-health-hub/conditions/seasons-in-dogs}}). Sixth, capturing personality could be done in many ways, and, in this study, we only focused on two commonly used personality measurement questionnaires that are filled by dog owners, which were specifically designed for shelter rehoming. Future work should evaluate these scales in the context of other facilities in the USA, and could also explore other dog personality measurement techniques (i.e., test batteries, experts).

\section{Conclusion}
\label{sec:conclusion}

We built a device called ``Patchkeeper'' that can be strapped on a dog's chest and that measures its activity through an accelerometer and a gyroscope. We experimented with the device on 12 dogs and collected sensor activity data for a week, along with dog personality test results. By matching these two datasets, we trained machine learning classifiers that predicted dog personality from activity data. We found that a combination of activity-level features (describing the activity as sleeping, sedentary, light, or moderate-vigorous) and statistical features (describing temporal and statistical aspects of the time-series accelerometer and gyroscope data) extracted from sensor data, together with dog demographics worked the best. We also found that the same feature computed at different times of day contributed differently to prediction power, with morning features being predictive of fearfulness and excitability, afternoon features being predictive of responsiveness to training and aggressiveness towards animals, and night features being predictive of aggressiveness towards people and motivation.


\bibliographystyle{ACM-Reference-Format}
\bibliography{main}


\begin{thebibliography}{100}


\ifx \showCODEN    \undefined \def \showCODEN     #1{\unskip}     \fi
\ifx \showDOI      \undefined \def \showDOI       #1{#1}\fi
\ifx \showISBNx    \undefined \def \showISBNx     #1{\unskip}     \fi
\ifx \showISBNxiii \undefined \def \showISBNxiii  #1{\unskip}     \fi
\ifx \showISSN     \undefined \def \showISSN      #1{\unskip}     \fi
\ifx \showLCCN     \undefined \def \showLCCN      #1{\unskip}     \fi
\ifx \shownote     \undefined \def \shownote      #1{#1}          \fi
\ifx \showarticletitle \undefined \def \showarticletitle #1{#1}   \fi
\ifx \showURL      \undefined \def \showURL       {\relax}        \fi
\providecommand\bibfield[2]{#2}
\providecommand\bibinfo[2]{#2}
\providecommand\natexlab[1]{#1}
\providecommand\showeprint[2][]{arXiv:#2}

\bibitem[\protect\citeauthoryear{Abdi and Williams}{Abdi and Williams}{2010}]%
        {abdi2010principal}
\bibfield{author}{\bibinfo{person}{Herv{\'e} Abdi} {and}
  \bibinfo{person}{Lynne~J Williams}.} \bibinfo{year}{2010}\natexlab{}.
\newblock \showarticletitle{Principal component analysis}.
\newblock \bibinfo{journal}{\emph{Wiley interdisciplinary reviews:
  computational statistics}} \bibinfo{volume}{2}, \bibinfo{number}{4}
  (\bibinfo{year}{2010}), \bibinfo{pages}{433--459}.
\newblock
\urldef\tempurl%
\url{https://doi.org/10.1002/wics.101}
\showDOI{\tempurl}
\showeprint{https://wires.onlinelibrary.wiley.com/doi/pdf/10.1002/wics.101}


\bibitem[\protect\citeauthoryear{Animals}{Animals}{2019}]%
        {personality_over_breed}
\bibfield{author}{\bibinfo{person}{Michelson~Found Animals}.}
  \bibinfo{year}{2019}\natexlab{}.
\newblock \bibinfo{booktitle}{\emph{{How to Select a Shelter Pet: Choose
  Personality Over Breed}}}.
\newblock
\urldef\tempurl%
\url{https://www.foundanimals.org/select-shelter-pet-choose-personality-breed/}
\showURL{%
Retrieved November 2022 from \tempurl}


\bibitem[\protect\citeauthoryear{Barandas, Folgado, Fernandes, Santos, Abreu,
  Bota, Liu, Schultz, and Gamboa}{Barandas et~al\mbox{.}}{2020}]%
        {barandas2020tsfel}
\bibfield{author}{\bibinfo{person}{Mar{\'\i}lia Barandas},
  \bibinfo{person}{Duarte Folgado}, \bibinfo{person}{Let{\'\i}cia Fernandes},
  \bibinfo{person}{Sara Santos}, \bibinfo{person}{Mariana Abreu},
  \bibinfo{person}{Patr{\'\i}cia Bota}, \bibinfo{person}{Hui Liu},
  \bibinfo{person}{Tanja Schultz}, {and} \bibinfo{person}{Hugo Gamboa}.}
  \bibinfo{year}{2020}\natexlab{}.
\newblock \showarticletitle{TSFEL: Time Series Feature Extraction Library}.
\newblock \bibinfo{journal}{\emph{SoftwareX}}  \bibinfo{volume}{11}
  (\bibinfo{year}{2020}), \bibinfo{pages}{100456}.
\newblock
\urldef\tempurl%
\url{https://doi.org/10.1016/j.softx.2020.100456}
\showDOI{\tempurl}


\bibitem[\protect\citeauthoryear{Beaudet, Chalifoux, and Dallaire}{Beaudet
  et~al\mbox{.}}{1994}]%
        {beaudet1994predictive}
\bibfield{author}{\bibinfo{person}{R Beaudet}, \bibinfo{person}{A Chalifoux},
  {and} \bibinfo{person}{A Dallaire}.} \bibinfo{year}{1994}\natexlab{}.
\newblock \showarticletitle{Predictive value of activity level and behavioral
  evaluation on future dominance in puppies}.
\newblock \bibinfo{journal}{\emph{Applied Animal Behaviour Science}}
  \bibinfo{volume}{40}, \bibinfo{number}{3-4} (\bibinfo{year}{1994}),
  \bibinfo{pages}{273--284}.
\newblock


\bibitem[\protect\citeauthoryear{B{\'o}dizs, Kis, G{\'a}csi, and
  Top{\'a}l}{B{\'o}dizs et~al\mbox{.}}{2020}]%
        {bodizs2020sleep}
\bibfield{author}{\bibinfo{person}{R{\'o}bert B{\'o}dizs},
  \bibinfo{person}{Anna Kis}, \bibinfo{person}{M{\'a}rta G{\'a}csi}, {and}
  \bibinfo{person}{J{\'o}zsef Top{\'a}l}.} \bibinfo{year}{2020}\natexlab{}.
\newblock \showarticletitle{Sleep in the dog: comparative, behavioral and
  translational relevance}.
\newblock \bibinfo{journal}{\emph{Current Opinion in Behavioral Sciences}}
  \bibinfo{volume}{33} (\bibinfo{year}{2020}), \bibinfo{pages}{25--33}.
\newblock


\bibitem[\protect\citeauthoryear{BRACKIN}{BRACKIN}{2022}]%
        {caroline2022do}
\bibfield{author}{\bibinfo{person}{CAROLINE BRACKIN}.}
  \bibinfo{year}{2022}\natexlab{}.
\newblock \bibinfo{booktitle}{\emph{Do Dogs Need to Socialize With Other Dogs?
  (Why Socialization Is Key)}}.
\newblock
\urldef\tempurl%
\url{https://pethelpful.com/dogs/Why-Socializing-With-Other-Dogs-Is-More-Important-Than-You-Think}
\showURL{%
Retrieved August 19, 2022 from \tempurl}


\bibitem[\protect\citeauthoryear{Bradley}{Bradley}{1997}]%
        {bradley1997use}
\bibfield{author}{\bibinfo{person}{Andrew~P Bradley}.}
  \bibinfo{year}{1997}\natexlab{}.
\newblock \showarticletitle{The use of the area under the ROC curve in the
  evaluation of machine learning algorithms}.
\newblock \bibinfo{journal}{\emph{Pattern recognition}} \bibinfo{volume}{30},
  \bibinfo{number}{7} (\bibinfo{year}{1997}), \bibinfo{pages}{1145--1159}.
\newblock
\urldef\tempurl%
\url{https://doi.org/10.1016/S0031-3203(96)00142-2}
\showDOI{\tempurl}


\bibitem[\protect\citeauthoryear{Bradshaw}{Bradshaw}{2021}]%
        {bradshaw2021impact}
\bibfield{author}{\bibinfo{person}{Yolonda~F Bradshaw}.}
  \bibinfo{year}{2021}\natexlab{}.
\newblock \emph{\bibinfo{title}{The Impact of Breed Identification, Potential
  Adopter Perceptions and Demographics, and Dog Behavior on Shelter Dog
  Adoptability}}.
\newblock \bibinfo{thesistype}{Ph.D. Dissertation}. \bibinfo{school}{The Ohio
  State University}.
\newblock


\bibitem[\protect\citeauthoryear{Brugarolas, Latif, Dieffenderfer, Walker,
  Yuschak, Sherman, Roberts, and Bozkurt}{Brugarolas et~al\mbox{.}}{2015}]%
        {brugarolas2015wearable}
\bibfield{author}{\bibinfo{person}{Rita Brugarolas}, \bibinfo{person}{Tahmid
  Latif}, \bibinfo{person}{James Dieffenderfer}, \bibinfo{person}{Katherine
  Walker}, \bibinfo{person}{Sherrie Yuschak}, \bibinfo{person}{Barbara~L
  Sherman}, \bibinfo{person}{David~L Roberts}, {and} \bibinfo{person}{Alper
  Bozkurt}.} \bibinfo{year}{2015}\natexlab{}.
\newblock \showarticletitle{Wearable heart rate sensor systems for wireless
  canine health monitoring}.
\newblock \bibinfo{journal}{\emph{IEEE Sensors Journal}} \bibinfo{volume}{16},
  \bibinfo{number}{10} (\bibinfo{year}{2015}), \bibinfo{pages}{3454--3464}.
\newblock


\bibitem[\protect\citeauthoryear{burgesspetcare}{burgesspetcare}{2021}]%
        {burgesspetcare2021tailor}
\bibfield{author}{\bibinfo{person}{burgesspetcare}.}
  \bibinfo{year}{2021}\natexlab{}.
\newblock \bibinfo{booktitle}{\emph{Tailor your training to your dog’s
  personality}}.
\newblock
\urldef\tempurl%
\url{https://www.burgesspetcare.com/blog/dogs/tailor-your-training-to-your-dogs-personality/}
\showURL{%
Retrieved August 19, 2022 from \tempurl}


\bibitem[\protect\citeauthoryear{Byrne, Zuerndorfer, Freil, Han, Sirolly,
  Cilliland, Starner, and Jackson}{Byrne et~al\mbox{.}}{2018}]%
        {byrne2018predicting}
\bibfield{author}{\bibinfo{person}{Ceara Byrne}, \bibinfo{person}{Jay
  Zuerndorfer}, \bibinfo{person}{Larry Freil}, \bibinfo{person}{Xiaochuang
  Han}, \bibinfo{person}{Andrew Sirolly}, \bibinfo{person}{Scott Cilliland},
  \bibinfo{person}{Thad Starner}, {and} \bibinfo{person}{Melody Jackson}.}
  \bibinfo{year}{2018}\natexlab{}.
\newblock \showarticletitle{Predicting the Suitability of Service Animals Using
  Instrumented Dog Toys}.
\newblock \bibinfo{journal}{\emph{Proc. ACM Interact. Mob. Wearable Ubiquitous
  Technol.}} \bibinfo{volume}{1}, \bibinfo{number}{4}, Article
  \bibinfo{articleno}{127} (\bibinfo{date}{jan} \bibinfo{year}{2018}),
  \bibinfo{numpages}{20}~pages.
\newblock
\urldef\tempurl%
\url{https://doi.org/10.1145/3161184}
\showDOI{\tempurl}


\bibitem[\protect\citeauthoryear{Carrier, Cyr, Anderson, and Walsh}{Carrier
  et~al\mbox{.}}{2013}]%
        {carrier2013exploring}
\bibfield{author}{\bibinfo{person}{Lydia~Ottenheimer Carrier},
  \bibinfo{person}{Amanda Cyr}, \bibinfo{person}{Rita~E Anderson}, {and}
  \bibinfo{person}{Carolyn~J Walsh}.} \bibinfo{year}{2013}\natexlab{}.
\newblock \showarticletitle{Exploring the dog park: Relationships between
  social behaviours, personality and cortisol in companion dogs}.
\newblock \bibinfo{journal}{\emph{Applied Animal Behaviour Science}}
  \bibinfo{volume}{146}, \bibinfo{number}{1-4} (\bibinfo{year}{2013}),
  \bibinfo{pages}{96--106}.
\newblock


\bibitem[\protect\citeauthoryear{Chambers, Yoder, Carson, Junge, Allen,
  Prescott, Bradley, Wymore, Lloyd, and Lyle}{Chambers et~al\mbox{.}}{2021}]%
        {chambers2021deep}
\bibfield{author}{\bibinfo{person}{Robert~D. Chambers},
  \bibinfo{person}{Nathanael~C. Yoder}, \bibinfo{person}{Aletha~B. Carson},
  \bibinfo{person}{Christian Junge}, \bibinfo{person}{David~E. Allen},
  \bibinfo{person}{Laura~M. Prescott}, \bibinfo{person}{Sophie Bradley},
  \bibinfo{person}{Garrett Wymore}, \bibinfo{person}{Kevin Lloyd}, {and}
  \bibinfo{person}{Scott Lyle}.} \bibinfo{year}{2021}\natexlab{}.
\newblock \showarticletitle{Deep Learning Classification of Canine Behavior
  Using a Single Collar-Mounted Accelerometer: Real-World Validation}.
\newblock \bibinfo{journal}{\emph{Animals}} \bibinfo{volume}{11},
  \bibinfo{number}{6} (\bibinfo{year}{2021}).
\newblock
\showISSN{2076-2615}
\urldef\tempurl%
\url{https://doi.org/10.3390/ani11061549}
\showDOI{\tempurl}


\bibitem[\protect\citeauthoryear{Chersini, Hall, and Wynne}{Chersini
  et~al\mbox{.}}{2018}]%
        {chersini2018dog}
\bibfield{author}{\bibinfo{person}{Nadine Chersini}, \bibinfo{person}{Nathan~J
  Hall}, {and} \bibinfo{person}{Clive~DL Wynne}.}
  \bibinfo{year}{2018}\natexlab{}.
\newblock \showarticletitle{Dog pups’ attractiveness to humans peaks at
  weaning age}.
\newblock \bibinfo{journal}{\emph{Anthrozo{\"o}s}} \bibinfo{volume}{31},
  \bibinfo{number}{3} (\bibinfo{year}{2018}), \bibinfo{pages}{309--318}.
\newblock


\bibitem[\protect\citeauthoryear{Chollet et~al\mbox{.}}{Chollet
  et~al\mbox{.}}{2015}]%
        {chollet2015keras}
\bibfield{author}{\bibinfo{person}{Fran{\c{c}}ois Chollet} {et~al\mbox{.}}}
  \bibinfo{year}{2015}\natexlab{}.
\newblock \bibinfo{title}{keras}.
\newblock
\newblock


\bibitem[\protect\citeauthoryear{Constantinides, Busk, Matic, Faurholt-Jepsen,
  Kessing, and Bardram}{Constantinides et~al\mbox{.}}{2018}]%
        {constantinides2018personalized}
\bibfield{author}{\bibinfo{person}{Marios Constantinides},
  \bibinfo{person}{Jonas Busk}, \bibinfo{person}{Aleksandar Matic},
  \bibinfo{person}{Maria Faurholt-Jepsen}, \bibinfo{person}{Lars~Vedel
  Kessing}, {and} \bibinfo{person}{Jakob~E Bardram}.}
  \bibinfo{year}{2018}\natexlab{}.
\newblock \showarticletitle{Personalized versus generic mood prediction models
  in bipolar disorder}. In \bibinfo{booktitle}{\emph{Proceedings of the 2018
  ACM International Joint Conference and 2018 International Symposium on
  Pervasive and Ubiquitous Computing and Wearable Computers}}.
  \bibinfo{pages}{1700--1707}.
\newblock
\urldef\tempurl%
\url{https://doi.org/10.1145/3267305.3267536}
\showDOI{\tempurl}


\bibitem[\protect\citeauthoryear{Corsetti, Pimpolari, and Natoli}{Corsetti
  et~al\mbox{.}}{2021}]%
        {corsetti2021different}
\bibfield{author}{\bibinfo{person}{Sara Corsetti}, \bibinfo{person}{Luisa
  Pimpolari}, {and} \bibinfo{person}{Eugenia Natoli}.}
  \bibinfo{year}{2021}\natexlab{}.
\newblock \showarticletitle{How Different Personalities Affect the Reaction to
  Adoption of Dogs Adopted from a Shelter}.
\newblock \bibinfo{journal}{\emph{Animals}} \bibinfo{volume}{11},
  \bibinfo{number}{6} (\bibinfo{year}{2021}), \bibinfo{pages}{1816}.
\newblock
\urldef\tempurl%
\url{https://doi.org/10.3390/ani11061816}
\showDOI{\tempurl}


\bibitem[\protect\citeauthoryear{Costa~Jr and McCrae}{Costa~Jr and
  McCrae}{1992}]%
        {costa1992four}
\bibfield{author}{\bibinfo{person}{Paul~T Costa~Jr} {and}
  \bibinfo{person}{Robert~R McCrae}.} \bibinfo{year}{1992}\natexlab{}.
\newblock \showarticletitle{Four ways five factors are basic}.
\newblock \bibinfo{journal}{\emph{Personality and individual differences}}
  \bibinfo{volume}{13}, \bibinfo{number}{6} (\bibinfo{year}{1992}),
  \bibinfo{pages}{653--665}.
\newblock
\showISSN{0191-8869}
\urldef\tempurl%
\url{https://doi.org/10.1016/0191-8869(92)90236-I}
\showDOI{\tempurl}


\bibitem[\protect\citeauthoryear{Cotur, Olenik, Asfour, Bruyns-Haylett,
  Kasimatis, Tanriverdi, Gonzalez-Macia, Lee, Kozlov, and G{\"u}der}{Cotur
  et~al\mbox{.}}{2022}]%
        {cotur2022bioinspired}
\bibfield{author}{\bibinfo{person}{Yasin Cotur}, \bibinfo{person}{Selin
  Olenik}, \bibinfo{person}{Tarek Asfour}, \bibinfo{person}{Michael
  Bruyns-Haylett}, \bibinfo{person}{Michael Kasimatis}, \bibinfo{person}{Ugur
  Tanriverdi}, \bibinfo{person}{Laura Gonzalez-Macia},
  \bibinfo{person}{Hong~Seok Lee}, \bibinfo{person}{Andrei~S Kozlov}, {and}
  \bibinfo{person}{Firat G{\"u}der}.} \bibinfo{year}{2022}\natexlab{}.
\newblock \showarticletitle{Bioinspired Stretchable Transducer for Wearable
  Continuous Monitoring of Respiratory Patterns in Humans and Animals}.
\newblock \bibinfo{journal}{\emph{Advanced Materials}} (\bibinfo{year}{2022}),
  \bibinfo{pages}{2203310}.
\newblock
\urldef\tempurl%
\url{https://doi.org/10.1002/adma.202203310}
\showDOI{\tempurl}
\showeprint{https://onlinelibrary.wiley.com/doi/pdf/10.1002/adma.202203310}


\bibitem[\protect\citeauthoryear{Cox, Mancini, and Ruge}{Cox
  et~al\mbox{.}}{2020}]%
        {cox2020understanding}
\bibfield{author}{\bibinfo{person}{Elizabeth Cox}, \bibinfo{person}{Clara
  Mancini}, {and} \bibinfo{person}{Luisa Ruge}.}
  \bibinfo{year}{2020}\natexlab{}.
\newblock \showarticletitle{Understanding Dogs' Engagement with Interactive
  Games: Interaction Style, Behaviour and Personality}. In
  \bibinfo{booktitle}{\emph{Proceedings of the Seventh International Conference
  on Animal-Computer Interaction}}. \bibinfo{pages}{1--12}.
\newblock
\urldef\tempurl%
\url{https://doi.org/10.1145/3446002.3446122}
\showDOI{\tempurl}


\bibitem[\protect\citeauthoryear{Cutler, Cutler, and Stevens}{Cutler
  et~al\mbox{.}}{2011}]%
        {cutler2011random}
\bibfield{author}{\bibinfo{person}{Adele Cutler}, \bibinfo{person}{David
  Cutler}, {and} \bibinfo{person}{John Stevens}.}
  \bibinfo{year}{2011}\natexlab{}.
\newblock \bibinfo{booktitle}{\emph{Random Forests}}.
  Vol.~\bibinfo{volume}{45}.
\newblock \bibinfo{pages}{157--176}.
\newblock
\showISBNx{978-1-4419-9325-0}
\urldef\tempurl%
\url{https://doi.org/10.1007/978-1-4419-9326-7_5}
\showDOI{\tempurl}


\bibitem[\protect\citeauthoryear{Doherty, Jackson, Hammerla, Pl{\"o}tz,
  Olivier, Granat, White, Van~Hees, Trenell, Owen, et~al\mbox{.}}{Doherty
  et~al\mbox{.}}{2017}]%
        {doherty2017large}
\bibfield{author}{\bibinfo{person}{Aiden Doherty}, \bibinfo{person}{Dan
  Jackson}, \bibinfo{person}{Nils Hammerla}, \bibinfo{person}{Thomas
  Pl{\"o}tz}, \bibinfo{person}{Patrick Olivier}, \bibinfo{person}{Malcolm~H
  Granat}, \bibinfo{person}{Tom White}, \bibinfo{person}{Vincent~T Van~Hees},
  \bibinfo{person}{Michael~I Trenell}, \bibinfo{person}{Christoper~G Owen},
  {et~al\mbox{.}}} \bibinfo{year}{2017}\natexlab{}.
\newblock \showarticletitle{Large scale population assessment of physical
  activity using wrist worn accelerometers: the UK biobank study}.
\newblock \bibinfo{journal}{\emph{PloS one}} \bibinfo{volume}{12},
  \bibinfo{number}{2} (\bibinfo{year}{2017}), \bibinfo{pages}{e0169649}.
\newblock
\urldef\tempurl%
\url{https://doi.org/10.1371/journal.pone.0169649}
\showURL{%
\tempurl}


\bibitem[\protect\citeauthoryear{Doherty, Smith-Byrne, Ferreira, Holmes,
  Holmes, Pulit, and Lindgren}{Doherty et~al\mbox{.}}{2018}]%
        {doherty2018gwas}
\bibfield{author}{\bibinfo{person}{Aiden Doherty}, \bibinfo{person}{Karl
  Smith-Byrne}, \bibinfo{person}{Teresa Ferreira}, \bibinfo{person}{Michael~V
  Holmes}, \bibinfo{person}{Chris Holmes}, \bibinfo{person}{Sara~L Pulit},
  {and} \bibinfo{person}{Cecilia~M Lindgren}.} \bibinfo{year}{2018}\natexlab{}.
\newblock \showarticletitle{GWAS identifies 14 loci for device-measured
  physical activity and sleep duration}.
\newblock \bibinfo{journal}{\emph{Nature communications}} \bibinfo{volume}{9},
  \bibinfo{number}{1} (\bibinfo{year}{2018}), \bibinfo{pages}{1--8}.
\newblock
\urldef\tempurl%
\url{https://doi.org/10.1038/s41467-018-07743-4}
\showDOI{\tempurl}


\bibitem[\protect\citeauthoryear{Ellard-Gray, Jeffrey, Choubak, and
  Crann}{Ellard-Gray et~al\mbox{.}}{2015}]%
        {ellard2015finding}
\bibfield{author}{\bibinfo{person}{Amy Ellard-Gray}, \bibinfo{person}{Nicole~K
  Jeffrey}, \bibinfo{person}{Melisa Choubak}, {and} \bibinfo{person}{Sara~E
  Crann}.} \bibinfo{year}{2015}\natexlab{}.
\newblock \showarticletitle{Finding the hidden participant: Solutions for
  recruiting hidden, hard-to-reach, and vulnerable populations}.
\newblock \bibinfo{journal}{\emph{International Journal of Qualitative
  Methods}} \bibinfo{volume}{14}, \bibinfo{number}{5} (\bibinfo{year}{2015}),
  \bibinfo{pages}{1609406915621420}.
\newblock
\urldef\tempurl%
\url{https://doi.org/10.1177/1609406915621420}
\showDOI{\tempurl}
\showeprint{https://doi.org/10.1177/1609406915621420}


\bibitem[\protect\citeauthoryear{Foster, Brugarolas, Walker, Mealin, Cleghern,
  Yuschak, Clark, Adin, Russenberger, Gruen, et~al\mbox{.}}{Foster
  et~al\mbox{.}}{2020}]%
        {foster2020preliminary}
\bibfield{author}{\bibinfo{person}{Marc Foster}, \bibinfo{person}{Rita
  Brugarolas}, \bibinfo{person}{Katherine Walker}, \bibinfo{person}{Sean
  Mealin}, \bibinfo{person}{Zach Cleghern}, \bibinfo{person}{Sherrie Yuschak},
  \bibinfo{person}{Julia~Condit Clark}, \bibinfo{person}{Darcy Adin},
  \bibinfo{person}{Jane Russenberger}, \bibinfo{person}{Margaret Gruen},
  {et~al\mbox{.}}} \bibinfo{year}{2020}\natexlab{}.
\newblock \showarticletitle{Preliminary evaluation of a wearable sensor system
  for heart rate assessment in guide dog puppies}.
\newblock \bibinfo{journal}{\emph{IEEE Sensors Journal}} \bibinfo{volume}{20},
  \bibinfo{number}{16} (\bibinfo{year}{2020}), \bibinfo{pages}{9449--9459}.
\newblock


\bibitem[\protect\citeauthoryear{Foundation}{Foundation}{2020}]%
        {animal_foundation}
\bibfield{author}{\bibinfo{person}{Animal~Farm Foundation}.}
  \bibinfo{year}{2020}\natexlab{}.
\newblock \bibinfo{booktitle}{\emph{{Does breed really matter when choosing a
  pet dog?}}}
\newblock
\urldef\tempurl%
\url{https://www.animalfarmfoundation.org/dont-make-this-mistake-when-adopting-a-dog/}
\showURL{%
Retrieved November 2022 from \tempurl}


\bibitem[\protect\citeauthoryear{Gartner}{Gartner}{2015}]%
        {gartner2015pet}
\bibfield{author}{\bibinfo{person}{Marieke~Cassia Gartner}.}
  \bibinfo{year}{2015}\natexlab{}.
\newblock \showarticletitle{Pet personality: A review}.
\newblock \bibinfo{journal}{\emph{Personality and Individual Differences}}
  \bibinfo{volume}{75} (\bibinfo{year}{2015}), \bibinfo{pages}{102--113}.
\newblock
\showISSN{0191-8869}
\urldef\tempurl%
\url{https://doi.org/10.1016/j.paid.2014.10.042}
\showDOI{\tempurl}


\bibitem[\protect\citeauthoryear{Goldsmith, Buss, Plomin, Rothbart, Thomas,
  Chess, Hinde, and McCall}{Goldsmith et~al\mbox{.}}{1987}]%
        {goldsmith1987roundtable}
\bibfield{author}{\bibinfo{person}{H~Hill Goldsmith}, \bibinfo{person}{Arnold~H
  Buss}, \bibinfo{person}{Robert Plomin}, \bibinfo{person}{Mary~Klevjord
  Rothbart}, \bibinfo{person}{Alexander Thomas}, \bibinfo{person}{Stella
  Chess}, \bibinfo{person}{Robert~A Hinde}, {and} \bibinfo{person}{Robert~B
  McCall}.} \bibinfo{year}{1987}\natexlab{}.
\newblock \showarticletitle{Roundtable: What is temperament? Four approaches}.
\newblock \bibinfo{journal}{\emph{Child development}} (\bibinfo{year}{1987}),
  \bibinfo{pages}{505--529}.
\newblock


\bibitem[\protect\citeauthoryear{Gosling}{Gosling}{2001}]%
        {gosling2001mice}
\bibfield{author}{\bibinfo{person}{Samuel~D Gosling}.}
  \bibinfo{year}{2001}\natexlab{}.
\newblock \showarticletitle{From mice to men: what can we learn about
  personality from animal research?}
\newblock \bibinfo{journal}{\emph{Psychological bulletin}}
  \bibinfo{volume}{127}, \bibinfo{number}{1} (\bibinfo{year}{2001}),
  \bibinfo{pages}{45}.
\newblock


\bibitem[\protect\citeauthoryear{Gosling}{Gosling}{2008}]%
        {gosling2008personality}
\bibfield{author}{\bibinfo{person}{Samuel~D Gosling}.}
  \bibinfo{year}{2008}\natexlab{}.
\newblock \showarticletitle{Personality in non-human animals}.
\newblock \bibinfo{journal}{\emph{Social and Personality Psychology Compass}}
  \bibinfo{volume}{2}, \bibinfo{number}{2} (\bibinfo{year}{2008}),
  \bibinfo{pages}{985--1001}.
\newblock
\urldef\tempurl%
\url{https://doi.org/10.1111/j.1751-9004.2008.00087.x}
\showDOI{\tempurl}
\showeprint{https://compass.onlinelibrary.wiley.com/doi/pdf/10.1111/j.1751-9004.2008.00087.x}


\bibitem[\protect\citeauthoryear{Gosling, Kwan, and John}{Gosling
  et~al\mbox{.}}{2003a}]%
        {gosling2003dog}
\bibfield{author}{\bibinfo{person}{Samuel~D Gosling},
  \bibinfo{person}{Virginia~SY Kwan}, {and} \bibinfo{person}{Oliver~P John}.}
  \bibinfo{year}{2003}\natexlab{a}.
\newblock \showarticletitle{A dog's got personality: a cross-species
  comparative approach to personality judgments in dogs and humans.}
\newblock \bibinfo{journal}{\emph{Journal of personality and social
  psychology}} \bibinfo{volume}{85}, \bibinfo{number}{6}
  (\bibinfo{year}{2003}), \bibinfo{pages}{1161}.
\newblock


\bibitem[\protect\citeauthoryear{Gosling, Rentfrow, and Swann~Jr}{Gosling
  et~al\mbox{.}}{2003b}]%
        {gosling2003very}
\bibfield{author}{\bibinfo{person}{Samuel~D Gosling}, \bibinfo{person}{Peter~J
  Rentfrow}, {and} \bibinfo{person}{William~B Swann~Jr}.}
  \bibinfo{year}{2003}\natexlab{b}.
\newblock \showarticletitle{A very brief measure of the Big-Five personality
  domains}.
\newblock \bibinfo{journal}{\emph{Journal of Research in personality}}
  \bibinfo{volume}{37}, \bibinfo{number}{6} (\bibinfo{year}{2003}),
  \bibinfo{pages}{504--528}.
\newblock
\urldef\tempurl%
\url{https://doi.org/10.1016/S0092-6566(03)00046-1}
\showDOI{\tempurl}


\bibitem[\protect\citeauthoryear{{Greenland Sander}, {Senn Stephen J.},
  {Rothman Kenneth J.}, {Carlin John B.}, {Poole Charles}, {Goodman Steven N.},
  and {Altman Douglas G.}}{{Greenland Sander} et~al\mbox{.}}{2016}]%
        {Greenland2016}
\bibfield{author}{\bibinfo{person}{{Greenland Sander}}, \bibinfo{person}{{Senn
  Stephen J.}}, \bibinfo{person}{{Rothman Kenneth J.}},
  \bibinfo{person}{{Carlin John B.}}, \bibinfo{person}{{Poole Charles}},
  \bibinfo{person}{{Goodman Steven N.}}, {and} \bibinfo{person}{{Altman Douglas
  G.}}} \bibinfo{year}{2016}\natexlab{}.
\newblock \showarticletitle{Statistical tests, P values, confidence intervals,
  and power: a guide to misinterpretations}.
\newblock \bibinfo{journal}{\emph{European Journal of Epidemiology}}
  \bibinfo{volume}{31}, \bibinfo{number}{4} (\bibinfo{year}{2016}),
  \bibinfo{pages}{337--350}.
\newblock
\urldef\tempurl%
\url{https://doi.org/10.1007/s10654-016-0149-3}
\showDOI{\tempurl}


\bibitem[\protect\citeauthoryear{Griffies, Zutty, Sarzen, and
  Soorholtz}{Griffies et~al\mbox{.}}{2018}]%
        {griffies2018wearable}
\bibfield{author}{\bibinfo{person}{Joel Griffies}, \bibinfo{person}{Jason
  Zutty}, \bibinfo{person}{Marcel Sarzen}, {and} \bibinfo{person}{Stuart
  Soorholtz}.} \bibinfo{year}{2018}\natexlab{}.
\newblock \showarticletitle{Wearable sensor shown to specifically quantify
  pruritic behaviors in dogs}.
\newblock \bibinfo{journal}{\emph{BMC Veterinary Research}}
  \bibinfo{volume}{14} (\bibinfo{date}{04} \bibinfo{year}{2018}),
  \bibinfo{pages}{124}.
\newblock
\urldef\tempurl%
\url{https://doi.org/10.1186/s12917-018-1428-x}
\showDOI{\tempurl}


\bibitem[\protect\citeauthoryear{Hecht and Horowitz}{Hecht and
  Horowitz}{2015}]%
        {hecht2015seeing}
\bibfield{author}{\bibinfo{person}{Julie Hecht} {and}
  \bibinfo{person}{Alexandra Horowitz}.} \bibinfo{year}{2015}\natexlab{}.
\newblock \showarticletitle{Seeing dogs: Human preferences for dog physical
  attributes}.
\newblock \bibinfo{journal}{\emph{Anthrozo{\"o}s}} \bibinfo{volume}{28},
  \bibinfo{number}{1} (\bibinfo{year}{2015}), \bibinfo{pages}{153--163}.
\newblock


\bibitem[\protect\citeauthoryear{Hirskyj-Douglas and Lucero}{Hirskyj-Douglas
  and Lucero}{2019}]%
        {hirskyj2019internet}
\bibfield{author}{\bibinfo{person}{Ilyena Hirskyj-Douglas} {and}
  \bibinfo{person}{Andr{\'e}s Lucero}.} \bibinfo{year}{2019}\natexlab{}.
\newblock \showarticletitle{On the Internet, Nobody Knows You're a Dog...
  Unless You're Another Dog}. In \bibinfo{booktitle}{\emph{Proceedings of the
  2019 CHI Conference on Human Factors in Computing Systems}}.
  \bibinfo{pages}{1--12}.
\newblock
\urldef\tempurl%
\url{https://doi.org/10.1145/3290605.3300347}
\showDOI{\tempurl}


\bibitem[\protect\citeauthoryear{Hirskyj-Douglas, Piitulainen, Lucero,
  et~al\mbox{.}}{Hirskyj-Douglas et~al\mbox{.}}{2021}]%
        {hirskyj2021forming}
\bibfield{author}{\bibinfo{person}{Ilyena Hirskyj-Douglas},
  \bibinfo{person}{Roosa Piitulainen}, \bibinfo{person}{Andr{\'e}s Lucero},
  {et~al\mbox{.}}} \bibinfo{year}{2021}\natexlab{}.
\newblock \showarticletitle{Forming the Dog Internet: Prototyping a
  Dog-to-Human Video Call Device.}
\newblock \bibinfo{journal}{\emph{Proc. ACM Hum. Comput. Interact.}}
  \bibinfo{volume}{5}, \bibinfo{number}{ISS} (\bibinfo{year}{2021}),
  \bibinfo{pages}{1--20}.
\newblock
\urldef\tempurl%
\url{https://doi.org/10.1145/3488539}
\showDOI{\tempurl}


\bibitem[\protect\citeauthoryear{H{\"o}glin, Van~Poucke, Katajamaa, Jensen,
  Theodorsson, and Roth}{H{\"o}glin et~al\mbox{.}}{2021}]%
        {hoglin2021long}
\bibfield{author}{\bibinfo{person}{Amanda H{\"o}glin}, \bibinfo{person}{Enya
  Van~Poucke}, \bibinfo{person}{Rebecca Katajamaa}, \bibinfo{person}{Per
  Jensen}, \bibinfo{person}{Elvar Theodorsson}, {and} \bibinfo{person}{Lina~SV
  Roth}.} \bibinfo{year}{2021}\natexlab{}.
\newblock \showarticletitle{Long-term stress in dogs is related to the
  human--dog relationship and personality traits}.
\newblock \bibinfo{journal}{\emph{Scientific Reports}} \bibinfo{volume}{11},
  \bibinfo{number}{1} (\bibinfo{year}{2021}), \bibinfo{pages}{1--9}.
\newblock


\bibitem[\protect\citeauthoryear{Hsu and Serpell}{Hsu and Serpell}{2003}]%
        {hsu2003development}
\bibfield{author}{\bibinfo{person}{Yuying Hsu} {and} \bibinfo{person}{James~A
  Serpell}.} \bibinfo{year}{2003}\natexlab{}.
\newblock \showarticletitle{Development and validation of a questionnaire for
  measuring behavior and temperament traits in pet dogs}.
\newblock \bibinfo{journal}{\emph{Journal of the American Veterinary Medical
  Association}} \bibinfo{volume}{223}, \bibinfo{number}{9}
  (\bibinfo{year}{2003}), \bibinfo{pages}{1293--1300}.
\newblock
\urldef\tempurl%
\url{https://doi.org/10.2460/javma.2003.223.1293}
\showDOI{\tempurl}


\bibitem[\protect\citeauthoryear{HUMPHREY}{HUMPHREY}{1934}]%
        {humphrey1934mental}
\bibfield{author}{\bibinfo{person}{ELLIOTT~S HUMPHREY}.}
  \bibinfo{year}{1934}\natexlab{}.
\newblock \showarticletitle{``Mental tests'' for shepherd dogs: An attempted
  classification and evaluation of the various traits that go to make up
  “temperament” in the german shepherd dog}.
\newblock \bibinfo{journal}{\emph{Journal of Heredity}} \bibinfo{volume}{25},
  \bibinfo{number}{4} (\bibinfo{year}{1934}), \bibinfo{pages}{129--136}.
\newblock
\showISSN{0022-1503}
\urldef\tempurl%
\url{https://doi.org/10.1093/oxfordjournals.jhered.a103899}
\showDOI{\tempurl}
\showeprint{https://academic.oup.com/jhered/article-pdf/25/4/129/2591195/25-4-129.pdf}


\bibitem[\protect\citeauthoryear{Hussain, Ali, Kim, et~al\mbox{.}}{Hussain
  et~al\mbox{.}}{2022}]%
        {hussain2022activity}
\bibfield{author}{\bibinfo{person}{Ali Hussain}, \bibinfo{person}{Sikandar
  Ali}, \bibinfo{person}{Hee-Cheol Kim}, {et~al\mbox{.}}}
  \bibinfo{year}{2022}\natexlab{}.
\newblock \showarticletitle{Activity Detection for the Wellbeing of Dogs Using
  Wearable Sensors Based on Deep Learning}.
\newblock \bibinfo{journal}{\emph{IEEE Access}}  \bibinfo{volume}{10}
  (\bibinfo{year}{2022}), \bibinfo{pages}{53153--53163}.
\newblock


\bibitem[\protect\citeauthoryear{Isgate and Couchman}{Isgate and
  Couchman}{2018}]%
        {isgate2018makes}
\bibfield{author}{\bibinfo{person}{Sara Isgate} {and} \bibinfo{person}{Justin~J
  Couchman}.} \bibinfo{year}{2018}\natexlab{}.
\newblock \showarticletitle{What makes a dog adoptable? An eye-tracking
  Investigation}.
\newblock \bibinfo{journal}{\emph{Journal of applied animal welfare science}}
  \bibinfo{volume}{21}, \bibinfo{number}{1} (\bibinfo{year}{2018}),
  \bibinfo{pages}{69--81}.
\newblock


\bibitem[\protect\citeauthoryear{John}{John}{1990}]%
        {john1990big}
\bibfield{author}{\bibinfo{person}{Oliver~P John}.}
  \bibinfo{year}{1990}\natexlab{}.
\newblock \showarticletitle{The" Big Five" factor taxonomy: Dimensions of
  personality in the natural language and in questionnaires}.
\newblock \bibinfo{journal}{\emph{Handbook of personality: Theory and
  research}} (\bibinfo{year}{1990}).
\newblock


\bibitem[\protect\citeauthoryear{John and Gosling}{John and Gosling}{2000}]%
        {john2000personality}
\bibfield{author}{\bibinfo{person}{Oliver~P John} {and}
  \bibinfo{person}{Samuel~D Gosling}.} \bibinfo{year}{2000}\natexlab{}.
\newblock \showarticletitle{Personality traits}.
\newblock  (\bibinfo{year}{2000}).
\newblock


\bibitem[\protect\citeauthoryear{Jones}{Jones}{2008}]%
        {jones2008development}
\bibfield{author}{\bibinfo{person}{Amanda~Claire Jones}.}
  \bibinfo{year}{2008}\natexlab{}.
\newblock \bibinfo{booktitle}{\emph{Development and* validation of a Dog
  Personality Questionnaire}}.
\newblock \bibinfo{publisher}{The University of Texas at Austin}.
\newblock


\bibitem[\protect\citeauthoryear{Jones and Gosling}{Jones and Gosling}{2005}]%
        {jones2005temperament}
\bibfield{author}{\bibinfo{person}{Amanda~C Jones} {and}
  \bibinfo{person}{Samuel~D Gosling}.} \bibinfo{year}{2005}\natexlab{}.
\newblock \showarticletitle{Temperament and personality in dogs (Canis
  familiaris): A review and evaluation of past research}.
\newblock \bibinfo{journal}{\emph{Applied Animal Behaviour Science}}
  \bibinfo{volume}{95}, \bibinfo{number}{1-2} (\bibinfo{year}{2005}),
  \bibinfo{pages}{1--53}.
\newblock
\showISSN{0168-1591}
\urldef\tempurl%
\url{https://doi.org/10.1016/j.applanim.2005.04.008}
\showDOI{\tempurl}


\bibitem[\protect\citeauthoryear{Jones, Dowling-Guyer, Patronek, Marder,
  Segurson~D'Arpino, and McCobb}{Jones et~al\mbox{.}}{2014}]%
        {jones2014use}
\bibfield{author}{\bibinfo{person}{Sarah Jones}, \bibinfo{person}{Seana
  Dowling-Guyer}, \bibinfo{person}{Gary~J Patronek}, \bibinfo{person}{Amy~R
  Marder}, \bibinfo{person}{Sheila Segurson~D'Arpino}, {and}
  \bibinfo{person}{Emily McCobb}.} \bibinfo{year}{2014}\natexlab{}.
\newblock \showarticletitle{Use of accelerometers to measure stress levels in
  shelter dogs}.
\newblock \bibinfo{journal}{\emph{Journal of Applied Animal Welfare Science}}
  \bibinfo{volume}{17}, \bibinfo{number}{1} (\bibinfo{year}{2014}),
  \bibinfo{pages}{18--28}.
\newblock


\bibitem[\protect\citeauthoryear{Ke, Meng, Finley, Wang, Chen, Ma, Ye, and
  Liu}{Ke et~al\mbox{.}}{2017}]%
        {ke2017lightgbm}
\bibfield{author}{\bibinfo{person}{Guolin Ke}, \bibinfo{person}{Qi Meng},
  \bibinfo{person}{Thomas Finley}, \bibinfo{person}{Taifeng Wang},
  \bibinfo{person}{Wei Chen}, \bibinfo{person}{Weidong Ma},
  \bibinfo{person}{Qiwei Ye}, {and} \bibinfo{person}{Tie-Yan Liu}.}
  \bibinfo{year}{2017}\natexlab{}.
\newblock \showarticletitle{Lightgbm: A highly efficient gradient boosting
  decision tree}.
\newblock \bibinfo{journal}{\emph{Advances in neural information processing
  systems}}  \bibinfo{volume}{30} (\bibinfo{year}{2017}).
\newblock
\urldef\tempurl%
\url{https://doi.org/10.5555/3294996.3295074}
\showDOI{\tempurl}


\bibitem[\protect\citeauthoryear{Khwaja, Vaid, Zannone, Harari, Faisal, and
  Matic}{Khwaja et~al\mbox{.}}{2019}]%
        {khwaja2019modeling}
\bibfield{author}{\bibinfo{person}{Mohammed Khwaja}, \bibinfo{person}{Sumer~S
  Vaid}, \bibinfo{person}{Sara Zannone}, \bibinfo{person}{Gabriella~M Harari},
  \bibinfo{person}{A~Aldo Faisal}, {and} \bibinfo{person}{Aleksandar Matic}.}
  \bibinfo{year}{2019}\natexlab{}.
\newblock \showarticletitle{Modeling personality vs. modeling personalidad:
  In-the-wild mobile data analysis in five countries suggests cultural impact
  on personality models}.
\newblock \bibinfo{journal}{\emph{Proceedings of the ACM on Interactive,
  Mobile, Wearable and Ubiquitous Technologies}} \bibinfo{volume}{3},
  \bibinfo{number}{3} (\bibinfo{year}{2019}), \bibinfo{pages}{1--24}.
\newblock
\urldef\tempurl%
\url{https://doi.org/10.1145/3351246}
\showDOI{\tempurl}


\bibitem[\protect\citeauthoryear{Kim}{Kim}{2015a}]%
        {kim2015statistical}
\bibfield{author}{\bibinfo{person}{Hae-Young Kim}.}
  \bibinfo{year}{2015}\natexlab{a}.
\newblock \showarticletitle{Statistical notes for clinical researchers: effect
  size}.
\newblock \bibinfo{journal}{\emph{Restorative dentistry \& endodontics}}
  \bibinfo{volume}{40}, \bibinfo{number}{4} (\bibinfo{year}{2015}),
  \bibinfo{pages}{328--331}.
\newblock
\urldef\tempurl%
\url{https://doi.org/10.5395/rde.2015.40.4.328}
\showDOI{\tempurl}


\bibitem[\protect\citeauthoryear{Kim}{Kim}{2015b}]%
        {Kim2015}
\bibfield{author}{\bibinfo{person}{Tae Kim}.} \bibinfo{year}{2015}\natexlab{b}.
\newblock \showarticletitle{T test as a parametric statistic}.
\newblock \bibinfo{journal}{\emph{Korean Journal of Anesthesiology}}
  \bibinfo{volume}{68} (\bibinfo{date}{11} \bibinfo{year}{2015}),
  \bibinfo{pages}{540}.
\newblock
\urldef\tempurl%
\url{https://doi.org/10.4097/kjae.2015.68.6.540}
\showDOI{\tempurl}


\bibitem[\protect\citeauthoryear{Knowsley}{Knowsley}{2021}]%
        {knowsley2021meet}
\bibfield{author}{\bibinfo{person}{Jo Knowsley}.}
  \bibinfo{year}{2021}\natexlab{}.
\newblock \bibinfo{booktitle}{\emph{Meet the dog charity that matches up
  pooches to new owners by personality}}.
\newblock
\urldef\tempurl%
\url{https://metro.co.uk/2021/06/15/meet-the-dog-charity-that-matches-up-pets-to-new-owners-by-personality-14768607/}
\showURL{%
Retrieved August 19, 2022 from \tempurl}


\bibitem[\protect\citeauthoryear{Kubinyi, Turcs{\'a}n, and Mikl{\'o}si}{Kubinyi
  et~al\mbox{.}}{2009}]%
        {kubinyi2009dog}
\bibfield{author}{\bibinfo{person}{Enik{\H{o}} Kubinyi},
  \bibinfo{person}{Borb{\'a}la Turcs{\'a}n}, {and}
  \bibinfo{person}{{\'A}d{\'a}m Mikl{\'o}si}.} \bibinfo{year}{2009}\natexlab{}.
\newblock \showarticletitle{Dog and owner demographic characteristics and dog
  personality trait associations}.
\newblock \bibinfo{journal}{\emph{Behavioural processes}} \bibinfo{volume}{81},
  \bibinfo{number}{3} (\bibinfo{year}{2009}), \bibinfo{pages}{392--401}.
\newblock
\urldef\tempurl%
\url{https://doi.org/10.1016/j.beproc.2009.04.004}
\showDOI{\tempurl}


\bibitem[\protect\citeauthoryear{Ladha, Hammerla, Hughes, Olivier, and
  Ploetz}{Ladha et~al\mbox{.}}{2013}]%
        {ladha2013dog}
\bibfield{author}{\bibinfo{person}{Cassim Ladha}, \bibinfo{person}{Nils
  Hammerla}, \bibinfo{person}{Emma Hughes}, \bibinfo{person}{Patrick Olivier},
  {and} \bibinfo{person}{Thomas Ploetz}.} \bibinfo{year}{2013}\natexlab{}.
\newblock \showarticletitle{Dog's Life: Wearable Activity Recognition for
  Dogs}. In \bibinfo{booktitle}{\emph{Proceedings of the 2013 ACM International
  Joint Conference on Pervasive and Ubiquitous Computing}} (Zurich,
  Switzerland) \emph{(\bibinfo{series}{UbiComp '13})}.
  \bibinfo{publisher}{Association for Computing Machinery},
  \bibinfo{address}{New York, NY, USA}, \bibinfo{pages}{415–418}.
\newblock
\showISBNx{9781450317702}
\urldef\tempurl%
\url{https://doi.org/10.1145/2493432.2493519}
\showDOI{\tempurl}


\bibitem[\protect\citeauthoryear{Lakens}{Lakens}{2013}]%
        {Lakens2013}
\bibfield{author}{\bibinfo{person}{Daniel Lakens}.}
  \bibinfo{year}{2013}\natexlab{}.
\newblock \showarticletitle{Calculating and reporting effect sizes to
  facilitate cumulative science: a practical primer for t-tests and ANOVAs}.
\newblock \bibinfo{journal}{\emph{Frontiers in Psychology}}
  \bibinfo{volume}{4} (\bibinfo{year}{2013}).
\newblock
\showISSN{1664-1078}
\urldef\tempurl%
\url{https://doi.org/10.3389/fpsyg.2013.00863}
\showDOI{\tempurl}


\bibitem[\protect\citeauthoryear{Lamb, Andrukonis, and Protopopova}{Lamb
  et~al\mbox{.}}{2021}]%
        {lamb2021role}
\bibfield{author}{\bibinfo{person}{Fiona Lamb}, \bibinfo{person}{Allison
  Andrukonis}, {and} \bibinfo{person}{Alexandra Protopopova}.}
  \bibinfo{year}{2021}\natexlab{}.
\newblock \showarticletitle{The role of artificial photo backgrounds of shelter
  dogs on pet profile clicking and the perception of sociability}.
\newblock \bibinfo{journal}{\emph{PloS one}} \bibinfo{volume}{16},
  \bibinfo{number}{12} (\bibinfo{year}{2021}), \bibinfo{pages}{e0255551}.
\newblock


\bibitem[\protect\citeauthoryear{Ley, Bennett, and Coleman}{Ley
  et~al\mbox{.}}{2008}]%
        {ley2008personality}
\bibfield{author}{\bibinfo{person}{Jacqueline Ley}, \bibinfo{person}{Pauleen
  Bennett}, {and} \bibinfo{person}{Grahame Coleman}.}
  \bibinfo{year}{2008}\natexlab{}.
\newblock \showarticletitle{Personality dimensions that emerge in companion
  canines}.
\newblock \bibinfo{journal}{\emph{Applied Animal Behaviour Science}}
  \bibinfo{volume}{110}, \bibinfo{number}{3-4} (\bibinfo{year}{2008}),
  \bibinfo{pages}{305--317}.
\newblock
\urldef\tempurl%
\url{https://doi.org/10.1016/j.applanim.2007.04.016}
\showDOI{\tempurl}


\bibitem[\protect\citeauthoryear{Ley, Bennett, and Coleman}{Ley
  et~al\mbox{.}}{2009}]%
        {ley2009refinement}
\bibfield{author}{\bibinfo{person}{Jacqui~M Ley}, \bibinfo{person}{Pauleen~C
  Bennett}, {and} \bibinfo{person}{Grahame~J Coleman}.}
  \bibinfo{year}{2009}\natexlab{}.
\newblock \showarticletitle{A refinement and validation of the Monash Canine
  Personality Questionnaire (MCPQ)}.
\newblock \bibinfo{journal}{\emph{Applied Animal Behaviour Science}}
  \bibinfo{volume}{116}, \bibinfo{number}{2-4} (\bibinfo{year}{2009}),
  \bibinfo{pages}{220--227}.
\newblock
\urldef\tempurl%
\url{https://doi.org/10.1016/j.applanim.2008.09.009}
\showDOI{\tempurl}


\bibitem[\protect\citeauthoryear{Lofgren, Wiener, Blott, Sanchez-Molano,
  Woolliams, Clements, and Haskell}{Lofgren et~al\mbox{.}}{2014}]%
        {lofgren2014management}
\bibfield{author}{\bibinfo{person}{Sarah~E. Lofgren}, \bibinfo{person}{Pamela
  Wiener}, \bibinfo{person}{Sarah~C. Blott}, \bibinfo{person}{Enrique
  Sanchez-Molano}, \bibinfo{person}{John~A. Woolliams},
  \bibinfo{person}{Dylan~N. Clements}, {and} \bibinfo{person}{Marie~J.
  Haskell}.} \bibinfo{year}{2014}\natexlab{}.
\newblock \showarticletitle{Management and personality in Labrador Retriever
  dogs}.
\newblock \bibinfo{journal}{\emph{Applied Animal Behaviour Science}}
  \bibinfo{volume}{156} (\bibinfo{year}{2014}), \bibinfo{pages}{44--53}.
\newblock
\showISSN{0168-1591}
\urldef\tempurl%
\url{https://doi.org/10.1016/j.applanim.2014.04.006}
\showDOI{\tempurl}


\bibitem[\protect\citeauthoryear{McCrae, Costa~Jr, Ostendorf, Angleitner,
  H{\v{r}}eb{\'\i}{\v{c}}kov{\'a}, Avia, Sanz, Sanchez-Bernardos, Kusdil,
  Woodfield, et~al\mbox{.}}{McCrae et~al\mbox{.}}{2000}]%
        {mccrae2000nature}
\bibfield{author}{\bibinfo{person}{Robert~R McCrae}, \bibinfo{person}{Paul~T
  Costa~Jr}, \bibinfo{person}{Fritz Ostendorf}, \bibinfo{person}{Alois
  Angleitner}, \bibinfo{person}{Martina H{\v{r}}eb{\'\i}{\v{c}}kov{\'a}},
  \bibinfo{person}{Maria~D Avia}, \bibinfo{person}{Jes{\'u}s Sanz},
  \bibinfo{person}{Maria~L Sanchez-Bernardos}, \bibinfo{person}{M~Ersin
  Kusdil}, \bibinfo{person}{Ruth Woodfield}, {et~al\mbox{.}}}
  \bibinfo{year}{2000}\natexlab{}.
\newblock \showarticletitle{Nature over nurture: temperament, personality, and
  life span development.}
\newblock \bibinfo{journal}{\emph{Journal of personality and social
  psychology}} \bibinfo{volume}{78}, \bibinfo{number}{1}
  (\bibinfo{year}{2000}), \bibinfo{pages}{173}.
\newblock


\bibitem[\protect\citeauthoryear{Meegahapola, Bangamuarachchi, Chamantha,
  Ruiz-Correa, Perera, and Gatica-Perez}{Meegahapola et~al\mbox{.}}{2022}]%
        {meegahapola2022sensing}
\bibfield{author}{\bibinfo{person}{Lakmal Meegahapola},
  \bibinfo{person}{Wageesha Bangamuarachchi}, \bibinfo{person}{Anju Chamantha},
  \bibinfo{person}{Salvador Ruiz-Correa}, \bibinfo{person}{Indika Perera},
  {and} \bibinfo{person}{Daniel Gatica-Perez}.}
  \bibinfo{year}{2022}\natexlab{}.
\newblock \showarticletitle{Sensing Eating Events in Context: A Smartphone-Only
  Approach}.
\newblock \bibinfo{journal}{\emph{IEEE Access}} \bibinfo{volume}{10},
  \bibinfo{number}{ARTICLE} (\bibinfo{year}{2022}).
\newblock
\urldef\tempurl%
\url{https://doi.org/10.1109/ACCESS.2022.3179702}
\showDOI{\tempurl}


\bibitem[\protect\citeauthoryear{Meegahapola, Droz, Kun, de~G\"{o}tzen,
  Nutakki, Diwakar, Correa, Song, Xu, Bidoglia, Gaskell, Chagnaa, Ganbold,
  Zundui, Caprini, Miorandi, Hume, Zarza, Cernuzzi, Bison, Britez, Busso,
  Chenu-Abente, G\"{u}nel, Giunchiglia, Schelenz, and Gatica-Perez}{Meegahapola
  et~al\mbox{.}}{2023}]%
        {meegahapola2023generalization}
\bibfield{author}{\bibinfo{person}{Lakmal Meegahapola},
  \bibinfo{person}{William Droz}, \bibinfo{person}{Peter Kun},
  \bibinfo{person}{Amalia de G\"{o}tzen}, \bibinfo{person}{Chaitanya Nutakki},
  \bibinfo{person}{Shyam Diwakar}, \bibinfo{person}{Salvador~Ruiz Correa},
  \bibinfo{person}{Donglei Song}, \bibinfo{person}{Hao Xu},
  \bibinfo{person}{Miriam Bidoglia}, \bibinfo{person}{George Gaskell},
  \bibinfo{person}{Altangerel Chagnaa}, \bibinfo{person}{Amarsanaa Ganbold},
  \bibinfo{person}{Tsolmon Zundui}, \bibinfo{person}{Carlo Caprini},
  \bibinfo{person}{Daniele Miorandi}, \bibinfo{person}{Alethia Hume},
  \bibinfo{person}{Jose~Luis Zarza}, \bibinfo{person}{Luca Cernuzzi},
  \bibinfo{person}{Ivano Bison}, \bibinfo{person}{Marcelo~Rodas Britez},
  \bibinfo{person}{Matteo Busso}, \bibinfo{person}{Ronald Chenu-Abente},
  \bibinfo{person}{Can G\"{u}nel}, \bibinfo{person}{Fausto Giunchiglia},
  \bibinfo{person}{Laura Schelenz}, {and} \bibinfo{person}{Daniel
  Gatica-Perez}.} \bibinfo{year}{2023}\natexlab{}.
\newblock \showarticletitle{Generalization and Personalization of Mobile
  Sensing-Based Mood Inference Models: An Analysis of College Students in Eight
  Countries}.
\newblock \bibinfo{journal}{\emph{Proc. ACM Interact. Mob. Wearable Ubiquitous
  Technol.}} \bibinfo{volume}{6}, \bibinfo{number}{4}, Article
  \bibinfo{articleno}{176} (\bibinfo{date}{jan} \bibinfo{year}{2023}),
  \bibinfo{numpages}{32}~pages.
\newblock
\urldef\tempurl%
\url{https://doi.org/10.1145/3569483}
\showDOI{\tempurl}


\bibitem[\protect\citeauthoryear{Meegahapola, Ruiz-Correa, Robledo-Valero,
  Hernandez-Huerfano, Alvarez-Rivera, Chenu-Abente, and
  Gatica-Perez}{Meegahapola et~al\mbox{.}}{2021}]%
        {meegahapola2021one}
\bibfield{author}{\bibinfo{person}{Lakmal Meegahapola},
  \bibinfo{person}{Salvador Ruiz-Correa}, \bibinfo{person}{Viridiana del~Carmen
  Robledo-Valero}, \bibinfo{person}{Emilio~Ernesto Hernandez-Huerfano},
  \bibinfo{person}{Leonardo Alvarez-Rivera}, \bibinfo{person}{Ronald
  Chenu-Abente}, {and} \bibinfo{person}{Daniel Gatica-Perez}.}
  \bibinfo{year}{2021}\natexlab{}.
\newblock \showarticletitle{One More Bite? Inferring Food Consumption Level of
  College Students Using Smartphone Sensing and Self-Reports}.
\newblock \bibinfo{journal}{\emph{Proc. ACM Interact. Mob. Wearable Ubiquitous
  Technol.}} \bibinfo{volume}{5}, \bibinfo{number}{1}, Article
  \bibinfo{articleno}{26} (\bibinfo{date}{mar} \bibinfo{year}{2021}),
  \bibinfo{numpages}{28}~pages.
\newblock
\urldef\tempurl%
\url{https://doi.org/10.1145/3448120}
\showDOI{\tempurl}


\bibitem[\protect\citeauthoryear{Mirk{\'o}, Kubinyi, G{\'a}csi, and
  Mikl{\'o}si}{Mirk{\'o} et~al\mbox{.}}{2012}]%
        {mirko2012preliminary}
\bibfield{author}{\bibinfo{person}{Erika Mirk{\'o}},
  \bibinfo{person}{Enik{\H{o}} Kubinyi}, \bibinfo{person}{M{\'a}rta G{\'a}csi},
  {and} \bibinfo{person}{{\'A}d{\'a}m Mikl{\'o}si}.}
  \bibinfo{year}{2012}\natexlab{}.
\newblock \showarticletitle{Preliminary analysis of an adjective-based dog
  personality questionnaire developed to measure some aspects of personality in
  the domestic dog (Canis familiaris)}.
\newblock \bibinfo{journal}{\emph{Applied Animal Behaviour Science}}
  \bibinfo{volume}{138}, \bibinfo{number}{1-2} (\bibinfo{year}{2012}),
  \bibinfo{pages}{88--98}.
\newblock
\showISSN{0168-1591}
\urldef\tempurl%
\url{https://doi.org/10.1016/j.applanim.2012.02.016}
\showDOI{\tempurl}


\bibitem[\protect\citeauthoryear{Morrison, Penpraze, Beber, Reilly, and
  Yam}{Morrison et~al\mbox{.}}{2013}]%
        {morrison2013associations}
\bibfield{author}{\bibinfo{person}{R Morrison}, \bibinfo{person}{V Penpraze},
  \bibinfo{person}{A Beber}, \bibinfo{person}{JJ Reilly}, {and}
  \bibinfo{person}{PS Yam}.} \bibinfo{year}{2013}\natexlab{}.
\newblock \showarticletitle{Associations between obesity and physical activity
  in dogs: a preliminary investigation}.
\newblock \bibinfo{journal}{\emph{Journal of Small Animal Practice}}
  \bibinfo{volume}{54}, \bibinfo{number}{11} (\bibinfo{year}{2013}),
  \bibinfo{pages}{570--574}.
\newblock


\bibitem[\protect\citeauthoryear{Nepal, Mirjafari, Martinez, Audia, Striegel,
  and Campbell}{Nepal et~al\mbox{.}}{2020}]%
        {nepal2020detecting}
\bibfield{author}{\bibinfo{person}{Subigya Nepal}, \bibinfo{person}{Shayan
  Mirjafari}, \bibinfo{person}{Gonzalo~J Martinez}, \bibinfo{person}{Pino
  Audia}, \bibinfo{person}{Aaron Striegel}, {and} \bibinfo{person}{Andrew~T
  Campbell}.} \bibinfo{year}{2020}\natexlab{}.
\newblock \showarticletitle{Detecting job promotion in information workers
  using mobile sensing}.
\newblock \bibinfo{journal}{\emph{Proceedings of the ACM on Interactive,
  Mobile, Wearable and Ubiquitous Technologies}} \bibinfo{volume}{4},
  \bibinfo{number}{3} (\bibinfo{year}{2020}), \bibinfo{pages}{1--28}.
\newblock
\urldef\tempurl%
\url{https://doi.org/10.1145/3414118}
\showDOI{\tempurl}


\bibitem[\protect\citeauthoryear{Noble}{Noble}{2006}]%
        {noble2006support}
\bibfield{author}{\bibinfo{person}{William~S Noble}.}
  \bibinfo{year}{2006}\natexlab{}.
\newblock \showarticletitle{What is a support vector machine?}
\newblock \bibinfo{journal}{\emph{Nature biotechnology}} \bibinfo{volume}{24},
  \bibinfo{number}{12} (\bibinfo{year}{2006}), \bibinfo{pages}{1565--1567}.
\newblock
\urldef\tempurl%
\url{https://doi.org/10.1038/nbt1206-1565}
\showDOI{\tempurl}


\bibitem[\protect\citeauthoryear{Obuchi, Huckins, Wang, daSilva, Rogers,
  Murphy, Hedlund, Holtzheimer, Mirjafari, and Campbell}{Obuchi
  et~al\mbox{.}}{2020}]%
        {obuchi2020predicting}
\bibfield{author}{\bibinfo{person}{Mikio Obuchi}, \bibinfo{person}{Jeremy~F
  Huckins}, \bibinfo{person}{Weichen Wang}, \bibinfo{person}{Alex daSilva},
  \bibinfo{person}{Courtney Rogers}, \bibinfo{person}{Eilis Murphy},
  \bibinfo{person}{Elin Hedlund}, \bibinfo{person}{Paul Holtzheimer},
  \bibinfo{person}{Shayan Mirjafari}, {and} \bibinfo{person}{Andrew Campbell}.}
  \bibinfo{year}{2020}\natexlab{}.
\newblock \showarticletitle{Predicting brain functional connectivity using
  mobile sensing}.
\newblock \bibinfo{journal}{\emph{Proceedings of the ACM on Interactive,
  Mobile, Wearable and Ubiquitous Technologies}} \bibinfo{volume}{4},
  \bibinfo{number}{1} (\bibinfo{year}{2020}), \bibinfo{pages}{1--22}.
\newblock
\urldef\tempurl%
\url{https://doi.org/10.1145/3381001}
\showDOI{\tempurl}


\bibitem[\protect\citeauthoryear{Ortmeyer, Robey, and McDonald}{Ortmeyer
  et~al\mbox{.}}{2018}]%
        {ortmeyer2018combining}
\bibfield{author}{\bibinfo{person}{Heidi~K Ortmeyer}, \bibinfo{person}{Lynda
  Robey}, {and} \bibinfo{person}{Tara McDonald}.}
  \bibinfo{year}{2018}\natexlab{}.
\newblock \showarticletitle{Combining actigraph link and PetPace collar data to
  measure activity, proximity, and physiological responses in freely moving
  dogs in a natural environment}.
\newblock \bibinfo{journal}{\emph{Animals}} \bibinfo{volume}{8},
  \bibinfo{number}{12} (\bibinfo{year}{2018}), \bibinfo{pages}{230}.
\newblock


\bibitem[\protect\citeauthoryear{Pedregosa, Varoquaux, Gramfort, Michel,
  Thirion, Grisel, Blondel, Prettenhofer, Weiss, Dubourg,
  et~al\mbox{.}}{Pedregosa et~al\mbox{.}}{2011}]%
        {pedregosa2011scikit}
\bibfield{author}{\bibinfo{person}{Fabian Pedregosa}, \bibinfo{person}{Ga{\"e}l
  Varoquaux}, \bibinfo{person}{Alexandre Gramfort}, \bibinfo{person}{Vincent
  Michel}, \bibinfo{person}{Bertrand Thirion}, \bibinfo{person}{Olivier
  Grisel}, \bibinfo{person}{Mathieu Blondel}, \bibinfo{person}{Peter
  Prettenhofer}, \bibinfo{person}{Ron Weiss}, \bibinfo{person}{Vincent
  Dubourg}, {et~al\mbox{.}}} \bibinfo{year}{2011}\natexlab{}.
\newblock \showarticletitle{Scikit-Learn: Machine learning in Python}.
\newblock \bibinfo{journal}{\emph{the Journal of machine Learning research}}
  \bibinfo{volume}{12} (\bibinfo{year}{2011}), \bibinfo{pages}{2825--2830}.
\newblock
\urldef\tempurl%
\url{http://arxiv.org/abs/1201.0490}
\showURL{%
\tempurl}


\bibitem[\protect\citeauthoryear{Peremans, Audenaert, Coopman, Blanckaert,
  Jacobs, Otte, Verschooten, van Bree, van Heeringen, Mertens,
  et~al\mbox{.}}{Peremans et~al\mbox{.}}{2003}]%
        {peremans2003estimates}
\bibfield{author}{\bibinfo{person}{Kathelijne Peremans}, \bibinfo{person}{Kurt
  Audenaert}, \bibinfo{person}{Frank Coopman}, \bibinfo{person}{Peter
  Blanckaert}, \bibinfo{person}{Filip Jacobs}, \bibinfo{person}{Andreas Otte},
  \bibinfo{person}{Francis Verschooten}, \bibinfo{person}{Henri van Bree},
  \bibinfo{person}{Kees van Heeringen}, \bibinfo{person}{John Mertens},
  {et~al\mbox{.}}} \bibinfo{year}{2003}\natexlab{}.
\newblock \showarticletitle{Estimates of regional cerebral blood flow and
  5-HT2A receptor density in impulsive, aggressive dogs with 99mTc-ECD and
  123I-5-I-R91150}.
\newblock \bibinfo{journal}{\emph{European journal of nuclear medicine and
  molecular imaging}} \bibinfo{volume}{30}, \bibinfo{number}{11}
  (\bibinfo{year}{2003}), \bibinfo{pages}{1538--1546}.
\newblock


\bibitem[\protect\citeauthoryear{Pervin and John}{Pervin and John}{1997}]%
        {pervin1997}
\bibfield{author}{\bibinfo{person}{Lawrence~A Pervin} {and}
  \bibinfo{person}{Oliver~P. John}.} \bibinfo{year}{1997}\natexlab{}.
\newblock \bibinfo{booktitle}{\emph{Personality: Theory and research, 7th ed.}}
\newblock \bibinfo{publisher}{John Wiley \& Sons}.
\newblock


\bibitem[\protect\citeauthoryear{Posluns, Anderson, and Walsh}{Posluns
  et~al\mbox{.}}{2017}]%
        {posluns2017comparing}
\bibfield{author}{\bibinfo{person}{Julie~A Posluns}, \bibinfo{person}{Rita~E
  Anderson}, {and} \bibinfo{person}{Carolyn~J Walsh}.}
  \bibinfo{year}{2017}\natexlab{}.
\newblock \showarticletitle{Comparing two canine personality assessments:
  Convergence of the MCPQ-R and DPQ and consensus between dog owners and dog
  walkers}.
\newblock \bibinfo{journal}{\emph{Applied Animal Behaviour Science}}
  \bibinfo{volume}{188} (\bibinfo{year}{2017}), \bibinfo{pages}{68--76}.
\newblock
\urldef\tempurl%
\url{https://doi.org/10.1016/j.applanim.2016.12.013}
\showDOI{\tempurl}


\bibitem[\protect\citeauthoryear{Rayment, De~Groef, Peters, and
  Marston}{Rayment et~al\mbox{.}}{2015}]%
        {rayment2015applied}
\bibfield{author}{\bibinfo{person}{Diana~J Rayment}, \bibinfo{person}{Bert
  De~Groef}, \bibinfo{person}{Richard~A Peters}, {and} \bibinfo{person}{Linda~C
  Marston}.} \bibinfo{year}{2015}\natexlab{}.
\newblock \showarticletitle{Applied personality assessment in domestic dogs:
  Limitations and caveats}.
\newblock \bibinfo{journal}{\emph{Applied Animal Behaviour Science}}
  \bibinfo{volume}{163} (\bibinfo{year}{2015}), \bibinfo{pages}{1--18}.
\newblock


\bibitem[\protect\citeauthoryear{Rice and Harris}{Rice and Harris}{2005}]%
        {Rice2005}
\bibfield{author}{\bibinfo{person}{Marnie Rice} {and} \bibinfo{person}{Grant
  Harris}.} \bibinfo{year}{2005}\natexlab{}.
\newblock \showarticletitle{Comparing effect sizes in follow-up studies: ROC
  Area, Cohen's d}.
\newblock \bibinfo{journal}{\emph{Law and human behavior}}
  \bibinfo{volume}{29} (\bibinfo{date}{11} \bibinfo{year}{2005}),
  \bibinfo{pages}{615--20}.
\newblock
\urldef\tempurl%
\url{https://doi.org/10.1007/s10979-005-6832-7}
\showDOI{\tempurl}


\bibitem[\protect\citeauthoryear{Rowan}{Rowan}{1992}]%
        {rowan1992shelters}
\bibfield{author}{\bibinfo{person}{Andrew~N Rowan}.}
  \bibinfo{year}{1992}\natexlab{}.
\newblock \bibinfo{title}{Shelters and pet overpopulation: A statistical black
  hole}.
\newblock , \bibinfo{numpages}{140--143}~pages.
\newblock
\urldef\tempurl%
\url{https://doi.org/10.2752/089279392787011430}
\showDOI{\tempurl}


\bibitem[\protect\citeauthoryear{Salonen, Mikkola, Hakanen, Sulkama, Puurunen,
  and Lohi}{Salonen et~al\mbox{.}}{2021}]%
        {ani11051234}
\bibfield{author}{\bibinfo{person}{Milla Salonen}, \bibinfo{person}{Salla
  Mikkola}, \bibinfo{person}{Emma Hakanen}, \bibinfo{person}{Sini Sulkama},
  \bibinfo{person}{Jenni Puurunen}, {and} \bibinfo{person}{Hannes Lohi}.}
  \bibinfo{year}{2021}\natexlab{}.
\newblock \showarticletitle{Reliability and Validity of a Dog Personality and
  Unwanted Behavior Survey}.
\newblock \bibinfo{journal}{\emph{Animals}} \bibinfo{volume}{11},
  \bibinfo{number}{5} (\bibinfo{year}{2021}).
\newblock
\showISSN{2076-2615}
\urldef\tempurl%
\url{https://doi.org/10.3390/ani11051234}
\showDOI{\tempurl}


\bibitem[\protect\citeauthoryear{Schneider, Delfabbro, and Burns}{Schneider
  et~al\mbox{.}}{2013}]%
        {schneider2013temperament}
\bibfield{author}{\bibinfo{person}{Luke~A Schneider}, \bibinfo{person}{Paul~H
  Delfabbro}, {and} \bibinfo{person}{Nicholas~R Burns}.}
  \bibinfo{year}{2013}\natexlab{}.
\newblock \showarticletitle{Temperament and lateralization in the domestic dog
  (Canis familiaris)}.
\newblock \bibinfo{journal}{\emph{Journal of Veterinary Behavior}}
  \bibinfo{volume}{8}, \bibinfo{number}{3} (\bibinfo{year}{2013}),
  \bibinfo{pages}{124--134}.
\newblock
\showISSN{1558-7878}
\urldef\tempurl%
\url{https://doi.org/10.1016/j.jveb.2012.06.004}
\showDOI{\tempurl}


\bibitem[\protect\citeauthoryear{Schork, Manzo, De~Oliveira, da~Costa, Young,
  and de~Azevedo}{Schork et~al\mbox{.}}{2022}]%
        {schork2022cyclic}
\bibfield{author}{\bibinfo{person}{Ivana~Gabriela Schork},
  \bibinfo{person}{Isabele~Aparecida Manzo}, \bibinfo{person}{Marcos
  Roberto~Beiral De~Oliveira}, \bibinfo{person}{Fernanda~Vieira da Costa},
  \bibinfo{person}{Robert~John Young}, {and}
  \bibinfo{person}{Cristiano~Schetini de Azevedo}.}
  \bibinfo{year}{2022}\natexlab{}.
\newblock \showarticletitle{The cyclic interaction between daytime behavior and
  the sleep behavior of laboratory dogs}.
\newblock \bibinfo{journal}{\emph{Scientific reports}} \bibinfo{volume}{12},
  \bibinfo{number}{1} (\bibinfo{year}{2022}), \bibinfo{pages}{1--9}.
\newblock


\bibitem[\protect\citeauthoryear{shibashake}{shibashake}{2017}]%
        {shibashake2017do}
\bibfield{author}{\bibinfo{person}{shibashake}.}
  \bibinfo{year}{2017}\natexlab{}.
\newblock \bibinfo{booktitle}{\emph{Do Dogs Need Other Dogs?}}
\newblock
\urldef\tempurl%
\url{https://shibashake.com/dog/do-dogs-need-other-dogs/#:~:text=Dogs\%20do\%20not\%20need\%20to,companion\%2C\%20so\%20much\%20the\%20better.}
\showURL{%
Retrieved August 19, 2022 from \tempurl}


\bibitem[\protect\citeauthoryear{Sibona and Walczak}{Sibona and
  Walczak}{2012}]%
        {sibona2012purposive}
\bibfield{author}{\bibinfo{person}{Christopher Sibona} {and}
  \bibinfo{person}{Steven Walczak}.} \bibinfo{year}{2012}\natexlab{}.
\newblock \showarticletitle{Purposive sampling on Twitter: A case study}. In
  \bibinfo{booktitle}{\emph{2012 45th Hawaii international conference on system
  sciences}}. IEEE, \bibinfo{pages}{3510--3519}.
\newblock
\urldef\tempurl%
\url{https://doi.org/10.1109/HICSS.2012.493}
\showDOI{\tempurl}


\bibitem[\protect\citeauthoryear{Sinn, Gosling, and Hilliard}{Sinn
  et~al\mbox{.}}{2010}]%
        {sinn2010personality}
\bibfield{author}{\bibinfo{person}{David~L Sinn}, \bibinfo{person}{Samuel~D
  Gosling}, {and} \bibinfo{person}{Stewart Hilliard}.}
  \bibinfo{year}{2010}\natexlab{}.
\newblock \showarticletitle{Personality and performance in military working
  dogs: Reliability and predictive validity of behavioral tests}.
\newblock \bibinfo{journal}{\emph{Applied Animal Behaviour Science}}
  \bibinfo{volume}{127}, \bibinfo{number}{1-2} (\bibinfo{year}{2010}),
  \bibinfo{pages}{51--65}.
\newblock
\urldef\tempurl%
\url{https://doi.org/10.1016/j.applanim.2010.08.007}
\showDOI{\tempurl}


\bibitem[\protect\citeauthoryear{Society}{Society}{2022}]%
        {nhhumane2022}
\bibfield{author}{\bibinfo{person}{NH~Humane Society}.}
  \bibinfo{year}{2022}\natexlab{}.
\newblock \bibinfo{booktitle}{\emph{Canine Personalities}}.
\newblock
\urldef\tempurl%
\url{https://nhhumane.org/adopt/meet-your-match/dogs}
\showURL{%
Retrieved August 19, 2022 from \tempurl}


\bibitem[\protect\citeauthoryear{Svartberg}{Svartberg}{2006}]%
        {svartberg2006breed}
\bibfield{author}{\bibinfo{person}{Kenth Svartberg}.}
  \bibinfo{year}{2006}\natexlab{}.
\newblock \showarticletitle{Breed-typical behaviour in dogs—historical
  remnants or recent constructs?}
\newblock \bibinfo{journal}{\emph{Applied Animal Behaviour Science}}
  \bibinfo{volume}{96}, \bibinfo{number}{3-4} (\bibinfo{year}{2006}),
  \bibinfo{pages}{293--313}.
\newblock
\urldef\tempurl%
\url{https://doi.org/10.1016/j.applanim.2005.06.014}
\showDOI{\tempurl}


\bibitem[\protect\citeauthoryear{Theerawatanasirikul, Suriyaphol,
  Thanawongnuwech, and Sailasuta}{Theerawatanasirikul et~al\mbox{.}}{2012}]%
        {theerawatanasirikul2012histologic}
\bibfield{author}{\bibinfo{person}{Sirin Theerawatanasirikul},
  \bibinfo{person}{Gunnaporn Suriyaphol}, \bibinfo{person}{Roongroje
  Thanawongnuwech}, {and} \bibinfo{person}{Achariya Sailasuta}.}
  \bibinfo{year}{2012}\natexlab{}.
\newblock \showarticletitle{Histologic morphology and involucrin, filaggrin,
  and keratin expression in normal canine skin from dogs of different breeds
  and coat types}.
\newblock \bibinfo{journal}{\emph{Journal of Veterinary Science}}
  \bibinfo{volume}{13}, \bibinfo{number}{2} (\bibinfo{year}{2012}),
  \bibinfo{pages}{163--170}.
\newblock


\bibitem[\protect\citeauthoryear{Vas, Top{\'a}l, P{\'e}ch, and Mikl{\'o}si}{Vas
  et~al\mbox{.}}{2007}]%
        {vas2007measuring}
\bibfield{author}{\bibinfo{person}{Judit Vas}, \bibinfo{person}{J{\'o}zsef
  Top{\'a}l}, \bibinfo{person}{Eva P{\'e}ch}, {and} \bibinfo{person}{Ad{\'a}m
  Mikl{\'o}si}.} \bibinfo{year}{2007}\natexlab{}.
\newblock \showarticletitle{Measuring attention deficit and activity in dogs: a
  new application and validation of a human ADHD questionnaire}.
\newblock \bibinfo{journal}{\emph{Applied Animal Behaviour Science}}
  \bibinfo{volume}{103}, \bibinfo{number}{1-2} (\bibinfo{year}{2007}),
  \bibinfo{pages}{105--117}.
\newblock


\bibitem[\protect\citeauthoryear{Vinciarelli and Mohammadi}{Vinciarelli and
  Mohammadi}{2014}]%
        {vinciarelli2014survey}
\bibfield{author}{\bibinfo{person}{Alessandro Vinciarelli} {and}
  \bibinfo{person}{Gelareh Mohammadi}.} \bibinfo{year}{2014}\natexlab{}.
\newblock \showarticletitle{A survey of personality computing}.
\newblock \bibinfo{journal}{\emph{IEEE Transactions on Affective Computing}}
  \bibinfo{volume}{5}, \bibinfo{number}{3} (\bibinfo{year}{2014}),
  \bibinfo{pages}{273--291}.
\newblock
\urldef\tempurl%
\url{https://doi.org/10.1109/TAFFC.2014.2330816}
\showDOI{\tempurl}


\bibitem[\protect\citeauthoryear{Walmsley, Chan, Smith-Byrne, Ramakrishnan,
  Woodward, Rahimi, Dwyer, Bennett, and Doherty}{Walmsley
  et~al\mbox{.}}{2021}]%
        {walmsley2021reallocation}
\bibfield{author}{\bibinfo{person}{Rosemary Walmsley}, \bibinfo{person}{Shing
  Chan}, \bibinfo{person}{Karl Smith-Byrne}, \bibinfo{person}{Rema
  Ramakrishnan}, \bibinfo{person}{Mark Woodward}, \bibinfo{person}{Kazem
  Rahimi}, \bibinfo{person}{Terence Dwyer}, \bibinfo{person}{Derrick Bennett},
  {and} \bibinfo{person}{Aiden Doherty}.} \bibinfo{year}{2021}\natexlab{}.
\newblock \showarticletitle{Reallocation of time between device-measured
  movement behaviours and risk of incident cardiovascular disease}.
\newblock \bibinfo{journal}{\emph{British journal of sports medicine}}
  (\bibinfo{year}{2021}).
\newblock
\urldef\tempurl%
\url{https://doi.org/10.1136/bjsports-2021-104050}
\showDOI{\tempurl}


\bibitem[\protect\citeauthoryear{Walmsley, Chan, Smith-Byrne, Ramakrishnan,
  Woodward, Rahimi, Dwyer, Bennett, and Doherty}{Walmsley
  et~al\mbox{.}}{2022}]%
        {walmsley2022reallocation}
\bibfield{author}{\bibinfo{person}{Rosemary Walmsley}, \bibinfo{person}{Shing
  Chan}, \bibinfo{person}{Karl Smith-Byrne}, \bibinfo{person}{Rema
  Ramakrishnan}, \bibinfo{person}{Mark Woodward}, \bibinfo{person}{Kazem
  Rahimi}, \bibinfo{person}{Terence Dwyer}, \bibinfo{person}{Derrick Bennett},
  {and} \bibinfo{person}{Aiden Doherty}.} \bibinfo{year}{2022}\natexlab{}.
\newblock \showarticletitle{Reallocation of time between device-measured
  movement behaviours and risk of incident cardiovascular disease}.
\newblock \bibinfo{journal}{\emph{British journal of sports medicine}}
  \bibinfo{volume}{56}, \bibinfo{number}{18} (\bibinfo{year}{2022}),
  \bibinfo{pages}{1008--1017}.
\newblock


\bibitem[\protect\citeauthoryear{Wan, Hejjas, Ronai, Elek, Sasvari-Szekely,
  Champagne, Mikl{\'o}si, and Kubinyi}{Wan et~al\mbox{.}}{2013}]%
        {wan2013drd}
\bibfield{author}{\bibinfo{person}{Michele Wan}, \bibinfo{person}{Krisztina
  Hejjas}, \bibinfo{person}{Zsolt Ronai}, \bibinfo{person}{Zsuzsanna Elek},
  \bibinfo{person}{Maria Sasvari-Szekely}, \bibinfo{person}{Frances~A
  Champagne}, \bibinfo{person}{{\'A}d{\'a}m Mikl{\'o}si}, {and}
  \bibinfo{person}{Enik{\H{o}} Kubinyi}.} \bibinfo{year}{2013}\natexlab{}.
\newblock \showarticletitle{DRD 4 and TH gene polymorphisms are associated with
  activity, impulsivity and inattention in Siberian Husky dogs}.
\newblock \bibinfo{journal}{\emph{Animal genetics}} \bibinfo{volume}{44},
  \bibinfo{number}{6} (\bibinfo{year}{2013}), \bibinfo{pages}{717--727}.
\newblock


\bibitem[\protect\citeauthoryear{Wang, Wang, Aung, Ben-Zeev, Brian, Campbell,
  Choudhury, Hauser, Kane, Scherer, et~al\mbox{.}}{Wang et~al\mbox{.}}{2017}]%
        {wang2017predicting}
\bibfield{author}{\bibinfo{person}{Rui Wang}, \bibinfo{person}{Weichen Wang},
  \bibinfo{person}{Min~SH Aung}, \bibinfo{person}{Dror Ben-Zeev},
  \bibinfo{person}{Rachel Brian}, \bibinfo{person}{Andrew~T Campbell},
  \bibinfo{person}{Tanzeem Choudhury}, \bibinfo{person}{Marta Hauser},
  \bibinfo{person}{John Kane}, \bibinfo{person}{Emily~A Scherer},
  {et~al\mbox{.}}} \bibinfo{year}{2017}\natexlab{}.
\newblock \showarticletitle{Predicting symptom trajectories of schizophrenia
  using mobile sensing}.
\newblock \bibinfo{journal}{\emph{Proceedings of the ACM on Interactive,
  Mobile, Wearable and Ubiquitous Technologies}} \bibinfo{volume}{1},
  \bibinfo{number}{3} (\bibinfo{year}{2017}), \bibinfo{pages}{1--24}.
\newblock
\urldef\tempurl%
\url{https://doi.org/10.1145/3130976}
\showDOI{\tempurl}


\bibitem[\protect\citeauthoryear{Wang, Wang, Obuchi, Scherer, Brian, Ben-Zeev,
  Choudhury, Kane, Hauser, Walsh, et~al\mbox{.}}{Wang et~al\mbox{.}}{2020}]%
        {wang2020predicting}
\bibfield{author}{\bibinfo{person}{Rui Wang}, \bibinfo{person}{Weichen Wang},
  \bibinfo{person}{Mikio Obuchi}, \bibinfo{person}{Emily Scherer},
  \bibinfo{person}{Rachel Brian}, \bibinfo{person}{Dror Ben-Zeev},
  \bibinfo{person}{Tanzeem Choudhury}, \bibinfo{person}{John Kane},
  \bibinfo{person}{Martar Hauser}, \bibinfo{person}{Megan Walsh},
  {et~al\mbox{.}}} \bibinfo{year}{2020}\natexlab{}.
\newblock \showarticletitle{On predicting relapse in schizophrenia using mobile
  sensing in a randomized control trial}. In \bibinfo{booktitle}{\emph{2020
  IEEE International Conference on Pervasive Computing and Communications
  (PerCom)}}. IEEE, \bibinfo{pages}{1--8}.
\newblock
\urldef\tempurl%
\url{https://doi.org/10.1109/PerCom45495.2020.9127365}
\showDOI{\tempurl}


\bibitem[\protect\citeauthoryear{Wang, Nepal, Huckins, Hernandez, Vojdanovski,
  Mack, Plomp, Pillai, Obuchi, daSilva, et~al\mbox{.}}{Wang
  et~al\mbox{.}}{2022}]%
        {wang2022first}
\bibfield{author}{\bibinfo{person}{Weichen Wang}, \bibinfo{person}{Subigya
  Nepal}, \bibinfo{person}{Jeremy~F Huckins}, \bibinfo{person}{Lessley
  Hernandez}, \bibinfo{person}{Vlado Vojdanovski}, \bibinfo{person}{Dante
  Mack}, \bibinfo{person}{Jane Plomp}, \bibinfo{person}{Arvind Pillai},
  \bibinfo{person}{Mikio Obuchi}, \bibinfo{person}{Alex daSilva},
  {et~al\mbox{.}}} \bibinfo{year}{2022}\natexlab{}.
\newblock \showarticletitle{First-Gen Lens: Assessing Mental Health of
  First-Generation Students across Their First Year at College Using Mobile
  Sensing}.
\newblock \bibinfo{journal}{\emph{Proceedings of the ACM on Interactive,
  Mobile, Wearable and Ubiquitous Technologies}} \bibinfo{volume}{6},
  \bibinfo{number}{2} (\bibinfo{year}{2022}), \bibinfo{pages}{1--32}.
\newblock
\urldef\tempurl%
\url{https://doi.org/10.1145/3543194}
\showDOI{\tempurl}


\bibitem[\protect\citeauthoryear{Webb, Keogh, and Miikkulainen}{Webb
  et~al\mbox{.}}{2010}]%
        {webb2010naive}
\bibfield{author}{\bibinfo{person}{Geoffrey~I Webb}, \bibinfo{person}{Eamonn
  Keogh}, {and} \bibinfo{person}{Risto Miikkulainen}.}
  \bibinfo{year}{2010}\natexlab{}.
\newblock \showarticletitle{Na{\"\i}ve Bayes.}
\newblock \bibinfo{journal}{\emph{Encyclopedia of machine learning}}
  \bibinfo{volume}{15} (\bibinfo{year}{2010}), \bibinfo{pages}{713--714}.
\newblock
\urldef\tempurl%
\url{https://doi.org/10.1007/978-1-4899-7502-7_581-1}
\showDOI{\tempurl}


\bibitem[\protect\citeauthoryear{Weiss}{Weiss}{2002}]%
        {weiss2002selecting}
\bibfield{author}{\bibinfo{person}{Emily Weiss}.}
  \bibinfo{year}{2002}\natexlab{}.
\newblock \showarticletitle{Selecting shelter dogs for service dog training}.
\newblock \bibinfo{journal}{\emph{Journal of Applied Animal Welfare Science}}
  \bibinfo{volume}{5}, \bibinfo{number}{1} (\bibinfo{year}{2002}),
  \bibinfo{pages}{43--62}.
\newblock


\bibitem[\protect\citeauthoryear{Weiss, Nathan, Kropp, and Lockhart}{Weiss
  et~al\mbox{.}}{2013}]%
        {weiss2013wagtag}
\bibfield{author}{\bibinfo{person}{Gary~M. Weiss}, \bibinfo{person}{Ashwin
  Nathan}, \bibinfo{person}{J.B. Kropp}, {and} \bibinfo{person}{Jeffrey~W.
  Lockhart}.} \bibinfo{year}{2013}\natexlab{}.
\newblock \showarticletitle{WagTag: A Dog Collar Accessory for Monitoring
  Canine Activity Levels}. In \bibinfo{booktitle}{\emph{Proceedings of the 2013
  ACM Conference on Pervasive and Ubiquitous Computing Adjunct Publication}}
  (Zurich, Switzerland) \emph{(\bibinfo{series}{UbiComp '13 Adjunct})}.
  \bibinfo{publisher}{Association for Computing Machinery},
  \bibinfo{address}{New York, NY, USA}, \bibinfo{pages}{405–414}.
\newblock
\showISBNx{9781450322157}
\urldef\tempurl%
\url{https://doi.org/10.1145/2494091.2495972}
\showDOI{\tempurl}


\bibitem[\protect\citeauthoryear{Weisstein}{Weisstein}{2004}]%
        {weisstein2004bonferroni}
\bibfield{author}{\bibinfo{person}{Eric~W Weisstein}.}
  \bibinfo{year}{2004}\natexlab{}.
\newblock \showarticletitle{Bonferroni correction}.
\newblock \bibinfo{journal}{\emph{https://mathworld. wolfram. com/}}
  (\bibinfo{year}{2004}).
\newblock


\bibitem[\protect\citeauthoryear{Whitaker, Stevelink, Fear,
  et~al\mbox{.}}{Whitaker et~al\mbox{.}}{2017}]%
        {whitaker2017use}
\bibfield{author}{\bibinfo{person}{Christopher Whitaker},
  \bibinfo{person}{Sharon Stevelink}, \bibinfo{person}{Nicola Fear},
  {et~al\mbox{.}}} \bibinfo{year}{2017}\natexlab{}.
\newblock \showarticletitle{The use of Facebook in recruiting participants for
  health research purposes: a systematic review}.
\newblock \bibinfo{journal}{\emph{Journal of medical Internet research}}
  \bibinfo{volume}{19}, \bibinfo{number}{8} (\bibinfo{year}{2017}),
  \bibinfo{pages}{e7071}.
\newblock
\urldef\tempurl%
\url{https://doi.org/10.2196/jmir.7071}
\showDOI{\tempurl}


\bibitem[\protect\citeauthoryear{Willetts, Hollowell, Aslett, Holmes, and
  Doherty}{Willetts et~al\mbox{.}}{2018}]%
        {willetts2018statistical}
\bibfield{author}{\bibinfo{person}{Matthew Willetts}, \bibinfo{person}{Sven
  Hollowell}, \bibinfo{person}{Louis Aslett}, \bibinfo{person}{Chris Holmes},
  {and} \bibinfo{person}{Aiden Doherty}.} \bibinfo{year}{2018}\natexlab{}.
\newblock \showarticletitle{Statistical machine learning of sleep and physical
  activity phenotypes from sensor data in 96,220 UK Biobank participants}.
\newblock \bibinfo{journal}{\emph{Scientific reports}} \bibinfo{volume}{8},
  \bibinfo{number}{1} (\bibinfo{year}{2018}), \bibinfo{pages}{1--10}.
\newblock
\urldef\tempurl%
\url{https://doi.org/10.1038/s41598-018-26174-1}
\showDOI{\tempurl}


\bibitem[\protect\citeauthoryear{Zamansky, Van Der~Linden, Hadar, and
  Bleuer-Elsner}{Zamansky et~al\mbox{.}}{2019}]%
        {zamansky2019log}
\bibfield{author}{\bibinfo{person}{Anna Zamansky}, \bibinfo{person}{Dirk Van
  Der~Linden}, \bibinfo{person}{Irit Hadar}, {and} \bibinfo{person}{Stephane
  Bleuer-Elsner}.} \bibinfo{year}{2019}\natexlab{}.
\newblock \showarticletitle{Log my dog: perceived impact of dog activity
  tracking}.
\newblock \bibinfo{journal}{\emph{Computer}} \bibinfo{volume}{52},
  \bibinfo{number}{9} (\bibinfo{year}{2019}), \bibinfo{pages}{35--43}.
\newblock
\urldef\tempurl%
\url{https://doi.org/10.1109/MC.2018.2889637}
\showDOI{\tempurl}


\end{thebibliography}

\end{document}